\documentclass[prb,twocolumn,superscriptaddress,showpacs,floatfix]{revtex4-1}  
\usepackage{amsmath}
\usepackage{color}
\usepackage{txfonts}
\usepackage{mathrsfs}
\usepackage{dcolumn}
\usepackage{bm}
\usepackage{graphicx}
\usepackage{multirow}
\usepackage{sidecap}

\allowdisplaybreaks

\begin{document}

\title{{
Rigorous \em ab initio} quantum embedding for quantum chemistry using Green's function theory: screened interaction, non-local self-energy relaxation, orbital basis, and chemical accuracy}
\date{\today}
\author{Tran Nguyen Lan}
\email{latran@umich.edu}
\affiliation{Department of Chemistry, University of Michigan, Ann Arbor, Michigan, 48109, USA}
\affiliation{Department of Physics, University of Michigan, Ann Arbor, Michigan, 48109, USA}
\altaffiliation{On leave from: Ho Chi Minh City Institute of Physics, VAST, Ho Chi Minh City, Vietnam}
\author{Alexei A. Kananenka}
\affiliation{Department of Chemistry, University of Michigan, Ann Arbor, Michigan, 48109, USA}
\author{Dominika Zgid}
\affiliation{Department of Chemistry, University of Michigan, Ann Arbor, Michigan, 48109, USA}

\begin{abstract}
 We present a detailed discussion of self-energy embedding theory (SEET) which is a quantum embedding scheme allowing us to describe a chosen subsystem very accurately while keeping the description of the environment at a lower cost. We apply SEET to molecular examples where commonly our chosen subsystem is made out of a set of  strongly correlated orbitals while the weakly correlated orbitals constitute an environment. Such a self-energy separation is very general and to make this procedure applicable to multiple systems a detailed and practical procedure for the evaluation of the system and environment self-energy is necessary. We list all the intricacies for one of the possible procedures while focusing our discussion on many practical implementation aspects such as the choice of best orbital basis, impurity solver, and many steps necessary to reach chemical accuracy.
Finally, on a set of carefully chosen molecular examples, we demonstrate that SEET which is a controlled, systematically improvable Green's function method can be as accurate as established wavefunction quantum chemistry methods.
\end{abstract}

\maketitle
 
\section{Introduction}
In molecular systems the interplay between the localized (strongly correlated) and delocalized (weakly correlated) electrons is the chief reason causing a difficulty in describing these systems 
since a robust quantum chemistry method has to be able to treat both types of electrons simultaneously with a comparable effectiveness and yield  molecular energies reaching chemical accuracy. Similar challenges are faced by condensed matter methods where both itinerant and localized electrons present in a solid have to be accounted for by a computational method that yields thermodynamic quantities and spectra.

In quantum chemistry, complete active space second order perturbation theory (CASPT2)~\cite{Ghigo:cpl/396/142,Kerstin:jpc/94/5483}, $n$-electron valence state second-order perturbation theory (NEVPT2)~\cite{Angeli:jcp/114/10252,Angeli:jcp/117/9138}, multi-reference coupled cluster (MRCC) \cite{Bartlett:rmp/79/291,Jeziorski:pra/24/1668,Jeziorski:jcp/88/5673,Musial:jcp/134/114108,Lindgren:ijqc/14/1097,Mukherjee:mp/30/1861}, multi-reference configuration interaction (MRCI)~\cite{Szalay:cr/112/108}, and multi-configurational density functional theory (MC-DFT)~\cite{Grafenstein:cpl/316/569,Grafenstein:mp/103/279,Pollet:jcp/116/1250,Limanni:jctc/10/3669,Carlson:jctc/11/4077} belong to a class of methods capable of addressing the problem of strongly correlated electrons in the presence of weakly correlated ones. These are primarily wavefunction methods that are at least in principle systematically improvable and aim to produce very accurate energies for the ground state and couple of low-lying excited states.

In condensed matter community when solids are described by a realistic  Hamiltonian, dynamical mean-field theory (DMFT)~\cite{Georges:rmp/68/13,Kotliar:rmp/78/865,Maier:rmp/77/1027} on top of GW~\cite{Hedin:pr/139/A796} method (GW+DMFT)~\cite{Biermann:prl/90/086402,Tomczak:epl/100/67001} formulated in many-body Green's function language is the primary tool to describe the intricate nature of realistic systems and give information about the spectral function and thermodynamic properties.

A natural direction is to find a common ground between approaches present in different communities and create a method that could yield not only thermodynamic quantities and spectra but also a satisfactory chemical accuracy while using the Green's function language. We have developed the self-energy embedding theory (SEET)~\cite{Kananenka:prb/91/121111,Lan:jcp/143/241102} where the self-energy describing couple of strongly correlated  orbitals is embedded into the self-energy coming from the weakly correlated orbitals present in the physical problem.  
Consequently, SEET is a hybrid approach where we use two Green's function methods, a weekly correlated and strongly correlated ones. The second order Green's function (GF2)~\cite{Phillips:jcp/140/241101,Dahlen:jcp/122/164102} method is used as a weakly correlated approach presenting a useful compromise between the computational cost and accuracy. We have demonstrated that GF2 can describe moderately correlated systems well~\cite{Phillips:jcp/140/241101}, can be successfully used to find strongly correlated orbitals, and can deliver information about periodic systems~\cite{Rusakov:jcp/144/054106} and their thermodynamic properties~\cite{Welden_arxiv}. The accurate correlated method that is used to describe a chosen set of strongly correlated orbitals is in our case a quantum impurity solver based on configuration interaction (CI) with various levels of wavefunction truncation~\cite{Zgid:jcp/134/094115, Zgid:prb/86/165128}.

We have applied SEET to the 2D Hubbard model~\cite{Kananenka:prb/91/121111} and small molecular problems~\cite{Lan:jcp/143/241102} obtaining in both cases excellent results when compared against other established methods. For the 2D Hubbard model at half- and away from half-filling, SEET self-energies matched very well the self-energies obtained from the continuous time auxiliary field quantum Monte Carlo (CT-QMC)~\cite{Gull:el/82/57003,Gull:prb/83/075122} calculations at all interaction strengths analyzed. In Ref.~\onlinecite{Lan:jcp/143/241102} using molecular examples, we have demonstrated that SEET, which is a perturb and diagonalize scheme, can reach a level of accuracy that is comparable to other quantum chemistry methods such as NEVPT2 and CASPT2. Moreover, we have shown that SEET has potential to be extended to periodic systems since in contrast to NEVPT2/CASPT2 it does not require preparing one-, two-, three- or four-body reduced density matrices and only one-body Green's function and self-energy are sufficient to perform all calculations. Additionally, SEET in the CI-in-GF2 variant does not suffer from the intruder state problem resulting in denominator divergences present in CASPT2.

In this paper, we are further examining all the aspects of SEET and we are focusing on describing all the requirements that have to be fulfilled to make SEET systematically improvable and applicable to a wide variety of systems. We base our discussion on the molecular systems where the near exact (or very accurate) solutions at zero temperature are known from variety of methods such as density matrix renormalization group (DMRG)~\cite{White:prl/69/2863,White:jcp/110/4127,Chan_dmrg_jcp2003,Chan_dmrg_jcp2004,Zgid:jcp/128/144116,Kurashige:jcp/130/234114,Marti:pccp/13/6750} or NEVPT2/CASPT2. These simple benchmark systems allow us to establish the best way of treating molecular systems using Green's function methods. 

This paper is organized in the following way. In Sec.~\ref{sec:seet_theory}, we present a theoretical motivation behind SEET. In Sec.~\ref{GF2_non_local}, we describe the GF2 theory that allows us to produce the self-energy for the weakly correlated orbitals. Subsequently, in Sec.~\ref{screening_seet}, we proceed to explaining how SEET is related to DMFT-type approaches that use on-site effective interactions. SEET can be executed using various strongly correlated orbital selection schemes either based on energy or spatial criteria as discussed in Sec.~\ref{orb_basis}.
To ensure that very high accuracy and systematic improbability is maintained during SEET calculations, it is essential to use Green's function quantum impurity solvers that  can describe full realistic Hamiltonian in the impurity orbitals. We discuss possible implication of this fact in Sec.~\ref{realistic_impurity_solvers}.
Finally, we present numerical results in Sec.~\ref{results} and conclude this paper in Sec.~\ref{conclusions}.

\section{Theoretical description}

\subsection{Self-energy embedding theory (SEET)}\label{sec:seet_theory}

Let us assume that we would like to calculate properties of a realistic system described by a Hamiltonian 
\begin{align}
H =  \sum_{ij} t_{ij}{a_i^\dagger a_j}  + \frac{1}{2}\sum_{ijkl} v_{ijkl} a_i^\dagger a_k^\dagger a_j a_l,
\end{align} 
where $t_{ij}$ and $v_{ijkl}$ are the one- and two-body integrals in an orbital basis that can be either orthogonal or non-orthogonal. Many realistic systems are correlated enough that low-level many-body methods cannot describe them with sufficient accuracy. While high-level many-body methods can deliver accurate answers, for many interesting systems the total number of orbitals $N$ may be too large to compute the whole problem using a high-level method. However, for most realistic cases only few orbitals contribute significantly to the physics or chemistry of the total problem. Consequently, these physically or chemically important orbitals, which we will call active or strongly correlated orbitals, can be described by a high-level method, while all the other inactive or weakly correlated orbitals can be described by a lower level method. 

The separation of the orbital space into active and inactive or strongly and weakly correlated orbitals implies in the Green's function language that we will express the self-energy of the strongly correlated orbitals $u$ and $v$ as
\begin{align}\label{eq:se_separation}
[\mathbf{\Sigma}]_{uv}=[\mathbf{\Sigma}^{low-level}_{non-local}]_{uv}+[\mathbf{\Sigma}^{high-level}_{local}]_{uv},
\end{align}
where a given self-energy matrix element $[\mathbf{\Sigma}]_{uv}$ is composed out of the local self-energy described by a high-level theory embedded into the self-energy described by a low-level theory. We assume that the self-energy contains both the frequency dependent and independent parts, $\mathbf{\Sigma}=\mathbf{\Sigma_{\infty}}+\mathbf{\Sigma}(i\omega)$.
Note that the separation of the self-energy presented in Eq.~\ref{eq:se_separation} is characteristic for embedding methods and thus is general without necessarily specifying how the low- and high-level self-energies are evaluated in practice. Here, few major routes exist {\bf (i)} either by a variational minimization of the free energy functional (that depends on the self-energy and Green's function), {\bf (ii)} by an explicit construction of the diagrammatic series necessary to represent different parts of the self-energy, {\bf (iii)} or by perturbative construction that allows for updating the hybridization between the embedded system and the environment.
Clearly while multitude of approximations are possible, the choice of a particular route should depend on the physics of the system under study.

In our two previous papers~\cite{Kananenka:prb/91/121111,Lan:jcp/143/241102} and in this paper, we use the third option and construct an impurity model for the strongly correlated orbitals that then has its self-energy evaluated in DMFT-type iterations; however, we would like to stress that the self-energy partitioning is general and our choice of evaluating the two parts of self-energy is one of many possible choices.
The separation of the self-energy contributions allows us to evaluate the non-local part of the self-energy at a perturbative, inexpensive level, while the local, strongly correlated part is expressed as an impurity problem and is evaluated with a more accurate method.

Consequently, one of the most important questions is how to choose orbitals that are strongly correlated and ought to be treated by a higher level method. To make such a choice based on the physics and not only our intuition, we first perform a low-level perturbative calculation and analyze occupations, energy, or spatial domains of the orbitals involved. The details of such a procedure will be discussed in subsection~\ref{orb_basis}.

For now, let us assume that using one of the possible criteria (orbital occupations, energies, or spatial domains), we have chosen the active (strongly correlated) orbitals that we denote $u, v, t, w, ...$, while $\mu,\lambda$ are the inactive (weakly correlated) orbitals, and $b$ stands for a bath orbital index.
We define the active space Green's function as 
 \begin{align} \label{eq:gfimp}
\textbf{G}_{act}(i\omega) = \left[ i\omega\textbf{1} - \textbf{f}^{\ nodc}_{act} -\boldsymbol{\Delta}(i\omega)-\boldsymbol{\Sigma}_{act}(i\omega) \right]^{-1},
 \end{align}
where subscript $act$ stands for the active space.
A molecular Fock matrix without double counting is denoted as $\textbf{f}^{\ nodc}$. The double counting between the mean-field and high-level treatments in the active space is removed since the mean-field Coulomb interaction in the active space is exactly subtracted.
Explicitly, the above operation is given by
\begin{align}
 f_{uv}^{nodc} &= h_{uv} + \sum_{\mu \lambda} P_{\mu \lambda} \left(v_{u v \mu \lambda} - \frac{1}{2}v_{u \lambda \mu v}\right)
  - \sum_{tw} P_{tw} \left(v_{uvtw} - \frac{1}{2}v_{tvuw}\right),
\end{align}
where $\textbf{P}$ is the one-body density matrix obtained from a perturbative method. 
Note that the chemical potential has been included in the one-body electron integral matrix, $\mathbf{h}$, for convenience.

SEET is general and multiple perturbative methods such as GF2, GW, or FLEX~\cite{Bickers:ap/193/206,Bickers:prl/62/961}, etc. can be used to describe the inactive orbitals and deliver initial density and Fock matrices describing all the orbitals. We have chosen to use GF2 since it contains an exchange diagram that is important for reaching the chemical accuracy in molecular systems. The details of the GF2 procedure will be listed in subsection~\ref{GF2_non_local}.
In GF2, the one-body density matrix $\mathbf{P}$ is obtained using Eq.~\ref{eq:gf2dm}.

The hybridization $\boldsymbol{\Delta}(i\omega)$ from Eq.~\ref{eq:gfimp} describes the coupling of the active orbitals to the remaining weakly correlated ones.
While at later iterations the active space Green's function and self-energy are evaluated by a more accurate many-body methods, at the first iteration of SEET, to initialize these matrices, we express them as sub-matrices of the GF2 Green's function and self-energy in the subset of active orbitals,
\begin{align}
  \textbf{G}_{act}(i\omega)=[\textbf{G}^{GF2}(i\omega)]_{act} \label{eq:gf2actgf}\\
  \boldsymbol{\Sigma}_{act}(i\omega)=[\boldsymbol{\Sigma}^{GF2}(i\omega)]_{act} \label{eq:gf2actse} \ .
\end{align}
The initial hybridization is then computed by substituting $\textbf{G}_{act}(i\omega)$ and $\boldsymbol{\Sigma}_{act}(i\omega)$ into Eq.~\ref{eq:gfimp}.

The infinite, continuum bath that describes the hybridization can be approximated by a finite, discrete one.
A finite number of bath orbital energies $\epsilon$ and the impurity-bath couplings $V$ are then fitted to the hybridization $\boldsymbol{\Delta}(i\omega)$, 
\begin{align}
  \Delta_{uv}(i\omega) = \sum_b \frac{V_{ub}^*V_{vb}^{}}{i\omega-\epsilon_b}
\end{align}
in order to produce the impurity Hamiltonian.

To compute the active space Green's function from Eq.~\ref{eq:gfimp} using quantum impurity solvers that need an explicit Hamiltonian formulation, one needs to formulate the Anderson impurity Hamiltonian as follows~\cite{Anderson:pr/124/41}
\begin{align}
  H_{act+bath} = H_{act} + H_{coupling} + H_{bath},
  \label{eq:impH}
\end{align}
where $H_{act}$ is the full active space Hamiltonian 
\begin{align}
H_{act} &=  \sum_{uv} f_{uv}^{nodc}{a_u^\dagger a_v}  + \frac{1}{2}\sum_{uvtw} v_{uvtw} a_u^\dagger a_t^\dagger a_v a_w,
\end{align}
where $H_{bath}$ describes the non-interacting bath, and $H_{coupling}$ stands for the coupling between the active space and non-interacting bath,
\begin{align}
  H_{bath} &= \sum_{b} \epsilon_b a_b^\dagger a_b, \\
  H_{coupling}  &= \sum_{ub} V_{ub}\left(a_u^\dagger a_b + a_b^\dagger a_u\right).
\end{align}

Multiple numerical methods have been developed to solve the  impurity Hamiltonian from Eq.~\ref{eq:impH}.
Quantum chemistry approaches that use an explicit Hamiltonian formulation and  properly capture strong correlations in the active space can be extended
to become Green's function impurity solvers. A discussion concerning the exact diagonalization (ED) or full configuration interaction (FCI) solvers and a later class of solvers based on 
a truncated configuration interaction (CI) or restricted active space configuration interaction (RASCI) approaches can be found in Refs.~\onlinecite{Davidson:jcp/17/87,Capone:prb/76/245116,Lin:cp/7/400,Liebsch:jpcm/24/053201,Zgid:prb/86/165128,Zgid:jcp/134/094115}. We employ a RASCI solver in SEET if the total number of orbitals (impurity + bath) in the impurity Hamiltonian exceeds 16 orbitals and is intractable with FCI.
This quantum chemistry RASCI solver that works with Hamiltonians containing general two-body interactions is important since it allows us to achieve chemical accuracy in molecular {\it ab initio} SEET.

As an alternative to the CI solvers, the continuous time quantum Monte Carlo (CT-QMC) methods, which introduce an explicit temperature-dependence can also be used with SEET. 
In subsection~\ref{realistic_impurity_solvers}, we will discuss in more details how realistic interactions impact the impurity problem and solvers.

After obtaining the active space Green's function using the impurity Hamiltonian from Eq.~\ref{eq:impH}, the active space self-energy is evaluated using the Dyson equation.
Combining the non-local self-energy obtained from a perturbative method (e.g. GF2, see Eq.~\ref{eq:gf2nl}) and the active space self-energy obtained from a high level many-body method, we construct the molecular self-energy for strongly correlated orbitals as
\begin{align}
  [\boldsymbol{\Sigma}_{mol}(i\omega)]_{uv} = [\boldsymbol{\Sigma}^{GF2}_{non-local}(i\omega)]_{uv} + [\boldsymbol{\Sigma}_{local}^{high-level}(i\omega)]_{uv}, \label{eq:semol1}
\end{align}
where the non-local part of the self-energy coming from the perturbative method does not contain any double counting of diagrams, for details see subsection~\ref{GF2_non_local}.

The inactive orbitals are described using only the self-energy obtained from a perturbative method and consequently
\begin{align}
  [\boldsymbol{\Sigma}_{mol}(i\omega)]_{\mu\lambda} = [\boldsymbol{\Sigma}^{GF2}(i\omega)]_{\mu\lambda}. \label{eq:semol_inactive}
\end{align}

In SEET, the impurity Hamiltonian is solved using bare interactions since the frequency dependent field coming from all the other orbitals is described by $\boldsymbol{\Sigma}^{GF2}_{non-local}(i\omega)$. However, to connect to the condensed matter and materials science community that uses screened interactions in the impurity Hamiltonian, we present a discussion of how these interactions can be evaluated in SEET in subsection~\ref{screening_seet}.

The new molecular Green's function, which contains the GF2 self-energy coming from all the inactive, weakly correlated orbitals as well as the strongly correlated self-energy from the active space orbitals, is reconstructed as
\begin{align} \label{eq:gfmol}
  \textbf{G}_{mol}(i\omega) = \left[ i\omega\textbf{1} - \textbf{f}^{\ nodc} -\boldsymbol{\Sigma}_{mol}(i\omega) \right]^{-1}.
\end{align}
This $\textbf{G}_{mol}(i\omega)$ is an $N\times N \times w_{max}$ matrix and has elements for all the orbitals present in the molecular problem $N=N_{inactive}+N_{active}$.

The hybridization $\boldsymbol{\Delta}(i\omega)$ present in Eq.~\ref{eq:gfimp} is then updated using the new Green's function and self-energy, and subsequent iterations are performed until convergence is achieved.

In SEET, the self-consistency procedure is made out of two loops. The inner loop has DMFT-like iterations and in this loop the matrices such as  $\textbf{G}_{mol}$, $\boldsymbol{\Sigma}_{mol}(i\omega)$, and $\boldsymbol{\Delta}(i\omega)$ are determined self-consistently.
At the convergence of the DMFT-like loop, the quantities of interest, such as the density matrix, molecular energy, and density of states, can be evaluated using the converged Green's function and self-energy. 

In SEET outer loop, we use the converged Green's function that contains the self-energy obtained using an accurate many-body method and we pass it back to the perturbative method as a zeroth-order Green's function.
Note that in the case of GF2 only a single iteration is performed in the SEET outer loop.
(Otherwise, due to the self-consistent nature of GF2, the converged GF2 Green's function  will be the same as the one obtained previously.)

An overall scheme of SEET self-consistency is summarized as follows.

\begin{enumerate}
\item Generate Hartree--Fock (HF) or density functional theory (DFT) Green's function as an initial guess.
\item Perform a self-consistent GF2 calculation for the whole molecule, see Ref.~\onlinecite{Phillips:jcp/140/241101} for details of the GF2 iterative loop.
\item \label{dmeval_iter} Evaluate the one-body density matrix $\textbf{P}$ using Eq.~\ref{eq:gf2dm} and construct a desired orbital basis such as natural orbitals, localized orbitals, or orthogonal atomic orbitals.
\item Transform all quantities from the atomic orbital (AO) basis to the new desired orbital basis.
\item Construct the impurity Hamiltonian in which two-electron term is either a subset of bare Coulomb  or screened interactions in the active space.
\item Perform DMFT-like loop:
  \begin{enumerate}
  \item Use an impurity solver (RASCI/FCI or CT-HYB) to obtain the active space Green's function $\textbf{G}_{act}(i\omega)$ and extract the active space self-energy $\boldsymbol{\Sigma}_{act}(i\omega)$. At the first iteration the hybridization $\boldsymbol{\Delta}(i\omega)$ is initialized using GF2 quantities. 
  \item Set up the molecular self-energy according to Eqs.~\ref{eq:semol1}-\ref{eq:semol_inactive} if a non-local self-energy is used or Eq.~\ref{eq:screened_u}  if screened interactions are used.
  \item Reconstruct the molecular Green's function via the Dyson equation and adjust the chemical potential to obtain a correct electron number for the  whole molecule.
  \item Update the hybridization $\boldsymbol{\Delta}(i\omega)$ using the new molecular Green's function and self-energy.
  \item Go back to step 6(a) and iterate until convergence is reached.
  \end{enumerate}
\item \label{outloop1} Pass the converged molecular Green's function to a GF2 calculation and perform only a single GF2 iteration.
\item \label{outloop2} Go back to step~\ref{dmeval_iter} and iterate until outer loop convergence is reached.
\end{enumerate}

Steps \ref{outloop1} and \ref{outloop2} are optional. In the past, we have investigated a single-shot scheme of DMFT-like iterations on top of GF2 without any further self-consistent iterations (steps \ref{outloop1} and \ref{outloop2}).
In practical calculations, we found that when the embedding was done in the energy domain then performing single-shot calculations was almost always sufficient. The outer loop was necessary to relax the non-local self-energy when the embedding construction was executed using spatial fragments.

\subsection{GF2 non-local self-energy}\label{GF2_non_local}

In the first step of SEET, a low-level Green's function method is used to obtain a non-local self-energy.
In our work, we employ the {\it ab initio} self-consistent GF2 method using either HF or DFT Green's function as an initial guess.
The GF2 self-energy in the imaginary time domain reads
\begin{align}
  \left[\Sigma^{GF2}_{mol}(\tau)\right]_{ij} = &- \sum_{klmnpq}\left[G^{GF2}_{mol}(\tau)\right]_{kl} \
                                                \left[G^{GF2}_{mol}(\tau)\right]_{mn} \left[G^{GF2}_{mol}(-\tau)\right]_{pq} \nonumber \\
                                             &\times v_{ikmq} \left( 2v_{ljpn} - v_{pjln} \right),  \label{eq:siggf2}
\end{align}
where $\textbf{G}^{GF2}_{mol}(\tau)$ is the imaginary time GF2 Green's function and
\begin{align}
  v_{ijkl} = \int d\textbf{r}_1 d\textbf{r}_2 \phi^*_i(\textbf{r}_1)\phi_j(\textbf{r}_1) v(\textbf{r}_1-\textbf{r}_2) \phi^*_k(\textbf{r}_2)\phi_l(\textbf{r}_2),
\end{align}
are two-electron integrals in the AO basis.
The resulting $\boldsymbol{\Sigma}^{GF2}_{mol}(\tau)$ matrix is then transformed to the imaginary frequency domain $\boldsymbol{\Sigma}^{GF2}_{mol}(i\omega)$ using either a traditional Fourier transform (FT) or a Fourier transform in the basis of orthogonal polynomials~\cite{Kananenka:jctc/12/564}.
The non-local part of the GF2 self-energy without the double counting is directly defined as a difference between the GF2 self-energy obtained from Eq.~\ref{eq:siggf2} while summing over all the molecular orbitals and that obtained by summing only over the local (active) part 
\begin{align}\label{eq:gf2nl}
  \boldsymbol{\Sigma}^{GF2}_{non-local}(i\omega) = \boldsymbol{\Sigma}^{GF2}_{mol}(i\omega) - \boldsymbol{\Sigma}^{GF2}_{act}(i\omega),
\end{align}
where $\boldsymbol{\Sigma}^{GF2}_{act}(i\omega)$ is evaluated using local interactions within the active space.
The inclusion of this non-local self-energy into the DMFT-like iterations of SEET eliminates the need for screened, effective interactions in the active space since the non-local self-energy term contains all the non-local interactions between the active and remaining orbitals. We will elaborate on this point in subsection~\ref{screening_seet}.

Recently, a comprehensive comparison between approximate diagrammatic schemes including second-order perturbation theory ($\Sigma^{(2)}$), GW, FLEX, and $T-$matrix approximation (TMA)~\cite{Galitskii:zetp/34/151} was performed for the 2D Hubbard model\cite{Gukelberger:prb/91/235114} where formally exact results are known for multiple regimes.
This study showed that the GW, FLEX, and TMA methods consisting of partial summations of bubble and/or ladder diagrams can yield worse results than the second-order perturbation theory.
More importantly, when combined with DMFT, $\Sigma^{(2)}$ non-local self-energy is more reliable than that from other methods including partial diagrammatic summations \cite{Gukelberger:prb/91/235114}.
We have chosen GF2 as the low-level perturbative method since we expect that similar conclusions will hold for realistic systems, particularly for molecular systems, where it is widely acknowledged that an exchange diagram is important if chemical accuracy is required.

The main bottleneck of GF2 is the evaluation of the second-order self-energy from Eq.~\ref{eq:siggf2}.
However, a highly parallel scheme consisting of factorizations and multiplications can easily make the GF2 calculation possible for a couple of hundred orbitals.

The other bottleneck in the GF2 calculation is the size of the imaginary frequency and time grids necessary to express the self-energy and Matsubara Green's function.
To reach chemical accuracy for realistic systems, these grids can have up to ten or even hundred thousand of points, leading to computations that are both processor and memory demanding.
Recently, we have optimized these numerical grids.
First, a Fourier transform in the basis of orthogonal polynomials~\cite{Kananenka:jctc/12/564} was used to transform the second-order self-energy from the imaginary time to the Matsubara frequency domain instead of a conventional Fourier transformation.
The imaginary time grid can be therefore reduced to a couple of hundred points, while preserving a micro-Hartree accuracy in the energy evaluation.
Later, the cubic spline interpolation was implemented to approximate the equidistant Matsubara frequency grid~\cite{Kananenka:jctc/12/2250}.
We showed that the chemical accuracy can be maintained with a very sparse subset of imaginary frequency points (only a few percent of the full imaginary frequency grid).
These new grids significantly lower the required computation time and memory storage for the the self-energy and Matsubara Green's function.

\subsection{Screened interactions in SEET}\label{screening_seet}

In a molecular system, the active space self-energy obtained from an impurity described by the bare interactions $v_{bare}$ combined with the  the self-energy $[\boldsymbol{\Sigma}^{GF2}_{non-local}(\tau,v_{bare})]_{uv}$
evaluated by summing over all weakly correlated (inactive) orbitals 
is equivalent to the self-energy evaluated for an impurity problem described by screened interactions $U(\tau)$.
Numerically, this amounts to requiring that the impurity self-energy evaluated using imaginary time dependent screened interactions recovers the full second-order self-energy at every $\tau$ point
\begin{align}\label{eq:screened_u}
  [\boldsymbol{\Sigma}_{mol}(\tau)]_{uv} &= [\boldsymbol{\Sigma}^{GF2}_{non-local}(\tau,v_{bare})]_{uv} + [\boldsymbol{\Sigma}_{act}(\tau, v_{bare})]_{uv} \nonumber \\ 
&  =[\boldsymbol{\Sigma}_{act}(\tau,U(\tau))]_{uv}
.
\end{align}
Consequently, in multiple theories present in condensed matter and materials science the non-local field coming from all the other electrons is described by $U(\tau)$, while in SEET this non-local field is described by $[\boldsymbol{\Sigma}^{GF2}_{non-local}(\tau,v_{bare})]_{uv}$.
Typically, screened interactions are obtained by a {\it downfolding procedure} of the full realistic band structure or infinite lattice Hamiltonian to an effective impurity model with only few correlated orbitals~\cite{Cococcioni:prb/71/035105,Aryasetiawan:prb/70195104,werner:prl2010,werner2016dynamical}. 
In SEET, we avoid the explicit downfolding procedure by including the $\boldsymbol{\Sigma}^{GF2}_{non-local}(\tau,v_{bare})$ matrix in the DMFT-like self-consistency.

However, in order to connect to the many-body condensed matter and materials science community, we demonstrate how to evaluate on-site screened interactions in molecular problems and 
by numerical examples (subsection~\ref{numres:screening}) show that both the procedures, SEET and DMFT with screened interactions, give similar numerical answers.

In our previous work, we proposed a procedure for finding screened interactions that reached satisfactory chemical accuracy for molecular systems~\cite{Rusakov:jcp/141/194105}. 
In this work, we capitalize on the previous procedure, and we require that for a single strongly correlated on-site orbital we fulfill the following relationship
\begin{align}\label{eqn:screened_u_se}
[\Sigma^{GF2}_{mol}(\tau)]_{ii} &=[\Sigma^{GF2}_{act}(\tau,U(\tau))]_{ii}\nonumber \\ 
&=-\left[G^{GF2}_{mol}(\tau)\right]_{ii}^2\left[G^{GF2}_{mol}(-\tau)\right]_{ii}[U(\tau)]^2,
\end{align}
for the on-site GF2 imaginary time self-energy (Eq.~\ref{eq:siggf2}).
The on-site screened interaction $U(\tau)$ is then obtained as follows
\begin{align}
  U(\tau) = - \sqrt{\frac{\left[\Sigma^{GF2}_{mol}(\tau)\right]_{ii}} {\left[G^{GF2}_{mol}(\tau)\right]_{ii}^2\left[G^{GF2}_{mol}(-\tau)\right]_{ii}} }.
  \label{eq:Uscr}
\end{align}

It is worth mentioning that while the non-local self-energy in Eq.~\ref{eq:gf2nl} for multiple-orbital active spaces can be straightforwardly evaluated, the generalization of the on-site screened interaction to treat multiple orbitals, unfortunately, is non-trivial.
This is due to the multiple choices of screened interactions $U_{ijkl}(\tau)$ and the parametrization not always being unique~\cite{Rusakov:jcp/141/194105}.
Additional advantages of using the non-local self-energy as compared to using the screened interaction in SEET will be numerically demonstrated in subsection~\ref{numres:screening}.

Now, let us shortly discuss the main distinction between SEET using the GF2 non-local self-energy and GW+DMFT.
First, in GW+DMFT, the GW non-local self-energy is given by the product of Green's function and the dynamically screened interaction $W$.
Thus, in GW+DMFT, when the realistic band structure or infinite lattice Hamiltonian is mapped onto an effective low energy model, it is mandatory to reconstruct the local components of both $G$ and screened interaction $W$ \cite{Biermann:prl/90/086402}.
This leads to ``the {\it double} embedding in both Green's function $G$ and dynamically screened interaction $W$" \cite{werner2016dynamical}.
In SEET, the GF2 non-local self-energy is computed using bare Coulomb interactions (Eq.~\ref{eq:siggf2}), it is therefore unnecessary to reconstruct the local component for the interaction in the active space, which can be directly taken from the bare interaction matrix.
Second, because of the dynamical nature of local screened interactions, the rigorous solution the impurity problem without neglecting the $\tau$-dependence of the interactions in GW+DMFT can be only obtained by CT-QMC solver. In SEET, any existing quantum impurity solver can be used to tackle the impurity problem.
GW+DMFT is known to yield good results for realistic materials such as NiO, it remains to be established if SEET can yield accurate results for these systems.
However, for molecular systems,  SEET with the GF2 non-local self-energy is a fully {\it ab initio}, systematically improvable, and self-consistent procedure able to give results with chemical accuracy.

\subsection{Orbital basis}\label{orb_basis}

\begin{figure} [h]
  \includegraphics[width=\columnwidth]{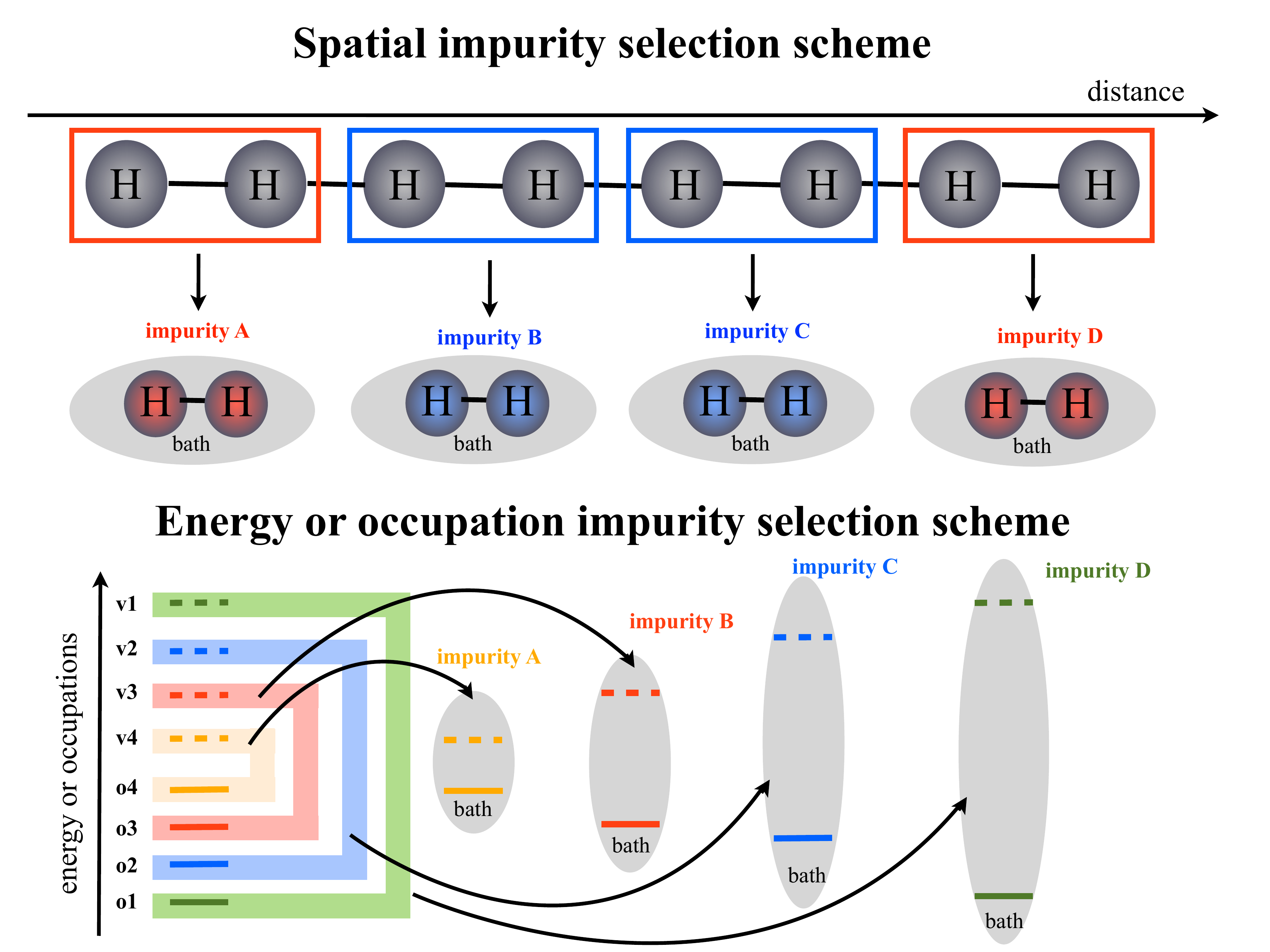} 
  \caption{\normalsize Two schemes for selecting orbitals used to construct impurities for the H$_8$ chain. Note that in the spatial selection scheme there are two pairs of degenerate impurities (A, D) and (B, C). In the energy or occupation scheme, there is no such degeneracy and impurities are build using bonding and corresponding antibonding orbitals.}
  \label{fig:seet_selection}
\end{figure}

After a self-consistent GF2 calculation is performed, the next step in the SEET procedure is to construct an orbital basis in which active orbitals are chosen.
Here, as we have mentioned earlier, two general schemes are possible: either to perform the selection based on the orbital energies or spatial domains criteria, depending on system studied. 
As an example,  in Fig.~\ref{fig:seet_selection}, we present the selection of impurities using the aforementioned criteria for the H$_8$ chain.

To construct the orbital basis in the energy domain, the one-body density matrix $\textbf{P}$ is first evaluated using the converged GF2 Green's function,
\begin{align}
  \textbf{P} = - 2 \textbf{G}^{GF2}_{mol}(\tau = \beta), \label{eq:gf2dm}
\end{align}
where $\beta=1/(k_BT)$ is the inverse temperature.
Then,  this whole molecular one-body density matrix is diagonalized to obtain the natural orbitals (NOs) and occupation numbers.
The active orbitals are then chosen from this set of NOs based on their occupations, as it is done in traditional CAS type methods.
The strongly correlated orbitals have occupations significantly different from 0 or 2, while the weakly correlated orbitals are mostly empty or doubly occupied.
In SEET performed in the NO basis, if the number of active orbitals is too large to be included in a single impurity, the orbitals can be easily split into different groups (impurities) belonging to different fragments or different symmetries without any further implementation as shown in Fig.~\ref{fig:seet_selection}, for numerical examples see Table~\ref{tab:h10}.

In the spatial domain, the localized orbitals can be formally obtained by localizing NOs using the well-known Pipek--Mezey~\cite{Pipek:jcp/90/4916} and Boys~\cite{Boys:rmp/32/296} localization schemes. Note that Boys orbitals are a molecular analogue of Wannier orbitals~\cite{Marzari:prb/56/12847}. 
In this work, we use the L\"{o}wdin orthogonalized AO (SAO) basis when dealing with a minimal basis set, while the so-called regional natural orbital (RNO) basis~\cite{Desilva:pccp/14/546,Gu:jpc2006-RNO} is employed for larger basis sets than the minimal one.
The construction of RNO basis for SEET can be briefly described as follows.
Starting from AO density matrix $\textbf{P}^{\mbox{\scriptsize{AO}}}$, the density matrix in SAO basis $\textbf{P}^{\mbox{\scriptsize{SAO}}}$ can be obtained by the L\"{o}wdin orthogonalization,
 \begin{align}
  \textbf{P}^{\mbox{\scriptsize{SAO}}} = \textbf{S}^{1/2} \textbf{P}^{\mbox{\scriptsize{AO}}} \textbf{S}^{1/2},
\end{align}
where $\textbf{S}$ is an AO overlap matrix.
In the next step of RNO basis construction, we separately diagonalize a block density matrix of predefined fragments $i$
\begin{align}
  \textbf{P}^{\mbox{\scriptsize{RNO}}}_i =  \textbf{U}^\dagger_i \textbf{P}^{\mbox{\scriptsize{SAO}}}_i \textbf{U}_i.
\end{align}
The transformation matrix from SAO to RNO basis for a  molecule consisting of $n$ molecular fragments is a direct summation of all block eigenvectors $\textbf{U}_i$
\begin{align}
  \textbf{U} = \textbf{U}_1 \oplus \textbf{U}_2 \oplus ... \oplus \textbf{U}_n.
\end{align}
The density matrix in RNO basis can be obtained as follows
\begin{align}
  \textbf{P}^{\mbox{\scriptsize{RNO}}} =  \textbf{U}^\dagger \textbf{P}^{\mbox{\scriptsize{SAO}}} \textbf{U}. \label{eq:prno}
\end{align}
Finally, the RNO coefficients present in the non-orthogonal AO basis are obtained by a back transformation
\begin{align}
  \textbf{C}^{\mbox{\scriptsize{RNO}}} = \textbf{S}^{-1/2} \textbf{U}.
\end{align}
The active orbitals are then chosen from RNOs belonging to particular molecular fragments.

It is worth mentioning that in the original description of the RNO construction \cite{Desilva:pccp/14/546,Gu:jpc2006-RNO}, the RNO density matrix (Eq.~\ref{eq:prno}) is further diagonalized using the Jacobi rotation to obtain the bonding between fragments.
However, since our purpose is to approximately disentangle the bonding between fragments, we will not proceed according to the original description.

Generally, depending on a system under study, we can use either NO or SAO/RNO bases.
If the entanglement between molecular fragments is large, the NO basis should be used to correctly describe the bonding between fragments.
On the other hand, if molecular fragments are only weakly coupled, we approximately can separate them and employ SAO/RNO bases.
The advantages and disadvantages of each orbital basis will be carefully demonstrated using numerical results in subsection~\ref{numres:energy_spatial_domains}. 

\subsection{Implications for impurity solvers to reach chemical accuracy}\label{realistic_impurity_solvers}

For SEET procedure to be computationally well behaved and accurate one has to fulfill multiple requirements: {\bf (i)} orthogonality of the orbital basis to properly carry out the embedding procedure when an explicit bath representation for a CI solver is necessary, {\bf (ii)} including a realistic Hamiltonian in the impurity problem, {\bf (iii)} hybridizations that simplify the bath fitting procedure,  {\bf (iv)} locality to make the self-energy decay fast with respect to the distance, and {\bf (v)} a possibility of treating many strongly correlated orbitals by an impurity solver. All of these requirements have implications for the possible quantum impurity solvers that can be used in SEET or DMFT procedures. 

For molecules, if chemical accuracy is desired, any modification of the realistic Hamiltonian containing full one- and two-body interactions to a simplified Hamiltonian including only a subset of all interactions may have a detrimental effect~\cite{Lin:prl2011-dmft}. Many of the off-diagonal elements of the realistic Hamiltonian can have a similar magnitude to the diagonal elements, thus presenting no justification for neglecting them and keeping only the diagonal elements. Moreover, solutions obtained by such modifications usually cannot be improved in a  systematic manner.
Consequently, for molecular calculations the safest option for preserving chemical accuracy is to use a class of quantum impurity solvers that employ a full realistic Hamiltonian for the impurity orbitals. 
This requirement allows us to use two classes of solvers: either based on CI expansions or continuous time hybridization expansion quantum Monte Carlo (CT-HYB QMC~\cite{Werner:prl/97/076405,Werner:prb/74/155107,triqs_ctqmc_gull,Gull:rmp2011}).

Solvers based on CI expansions such as restricted active space CI (RASCI) allow us to comfortably include around 8 orbitals in the impurity and around 16-24 bath orbitals. 
They usually work in the zero (or low) temperature formulation and require a bath discretization that can be a potential source of errors. 
The hybridization expansion solvers can handle an infinite bath but face difficulties with Monte Carlo sign problem that gets especially pronounced for non-diagonal hybridizations and
 low temperatures. 
Consequently, for both classes of solvers non-diagonal hybridizations lead to problems.  For CI solvers non-diagonal hybridizations lead to large number of bath orbitals necessary to fit the discrete bath accurately. For hybridization expansion solvers non-diagonal hybridizations lead to long computation times due to sign problem. Thus, a crucial challenge is to define a set of orbitals that either diagonalizes hybridization or minimizes its off-diagonal elements for a range of frequencies. 

\begin{figure} [h]
  \includegraphics[width=\columnwidth]{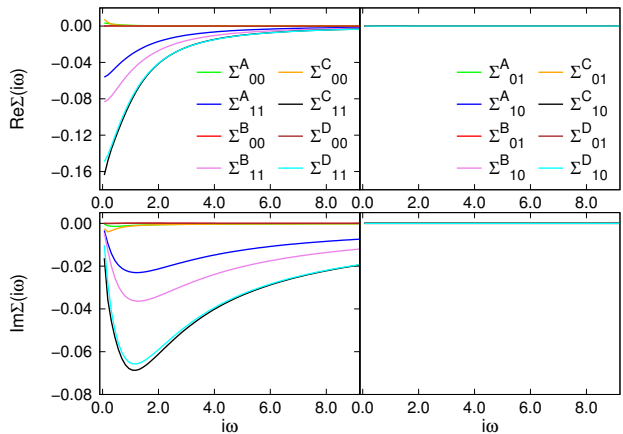}
  \caption{\normalsize Real and imaginary parts of diagonal (left) and off-diagonal (right) hybridization matrix elements $[\mathbf{\Delta}(i\omega)]_{ij}$  in NOs for all the impurities present in H$_8$ chain in the 6-31G basis set at $\beta = 50$. Note that the off-diagonal hybridization elements are zero.}
  \label{fig:hyb_dm}
\end{figure}

\begin{figure} [h]
  \includegraphics[width=\columnwidth]{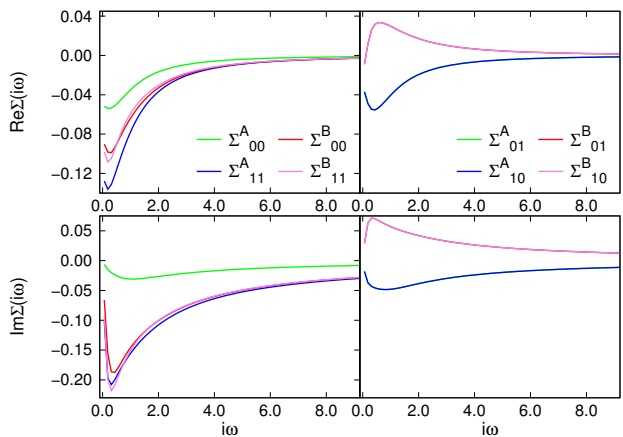}%
  \caption{ \normalsize Real and imaginary parts of diagonal (left) and off-diagonal (right) hybridization matrix elements $[\mathbf{\Delta}(i\omega)]_{ij}$ in Boys orbitals for the impurities A and B present in H$_8$ chain in the  6-31G basis set at $\beta = 50$. Note that the off-diagonal hybridization elements are of a similar magnitude as the diagonal elements.}
  \label{fig:hyb_boys}
\end{figure}

To gain insight into the possibility of making  the hybridization as diagonal as possible, we plot in Figs.~\ref{fig:hyb_dm} and \ref{fig:hyb_boys} hybridizations for different impurities in the H$_8$ chain in the 6-31G basis~\cite{Hehre:jcp/56/2257} using different selection schemes presented in Fig.~\ref{fig:seet_selection}. 
Spatially localized orbitals are obtained using the Boys localization method implemented in \textsc{GAUSSIAN 09}~\cite{g09}. It is evident that NOs yield a diagonal hybridization while Boys orbitals yield hybridization containing off-diagonal elements of large magnitudes. This insight explains some of our observations that NOs are quite advantageous for SEET calculations since the bath fitting procedure can be done using relatively few bath sites. 

We also consider how different orbitals, by influencing hybridization behavior, can change the performance of CT-HYB QMC, in Tables~\ref{tab:hyb_exp_STO3G} and ~\ref{tab:hyb_exp_631G}.
Let us note in passing that for a minimal basis (STO-3G \cite{sto_minimal}), SAOs are very similar to Boys orbitals and the off-diagonal elements of hybridization are significant for SAOs. It is evident that CT-HYB QMC has a much  worse average sign and larger perturbation order in orbital bases that yield non-diagonal hybridizations, thus making such calculations particularly challenging especially at low temperatures since the average perturbation order in CT-HYB QMC calculations grows when temperature decreases.  We observe that even for high temperatures the difference in CT-HYB QMC performance is present and NOs are still favored. 
Moreover, this difference in performance is independent of the number of orbitals present in a given basis and the trend is maintained when a larger orbital basis such as 6-31G is employed. 
Our observations indicate that for molecular examples NOs are advantageous when used with the CT-HYB QMC solver since due to the minimization of the sign problem they lead to a significant computational speed up. A similar conclusion was reached in CT-HYB QMC study of Hubbard model~\cite{Semon:prb/85/201101} where the Monte Carlo sign error was minimized by a canonical transformation that brought the hybridization matrix to a diagonal form. 

In CI calculations, NOs lead to fewer bath orbitals - also resulting in an increased computational efficiency; however, here the savings are not so dramatic since including additional bath orbitals in truncated CI schemes does not result in a significant cost increase.

It is important to showcase one more aspect necessary for a high accuracy quantum impurity solver for realistic problems. A quantum impurity solver capable of treating multiple regimes from weakly to strongly correlated ones, should be able to treat multiple partially occupied orbitals present in the impurity+bath system. To illustrate this statement, in Table~\ref{tab:h10_occ_imp}, we list impurity+bath occupations (8 impurity and 8 bath orbitals) for the H$_10$ chain in the cc-pVDZ basis (with a total of 50 orbitals). It is evident that for short interatomic distances this impurity+bath system has very few partial occupations, however for the stretched geometries, the number of partially occupied orbitals is significant. Consequently, if quantum chemistry methods are used as solvers for the impurity problems, they should be capable of treating many partially occupied orbitals - prompting us to conclude that single reference methods will generally not be capable of treating these impurity problems successfully.

\begin{table*}
  \normalsize
  \caption{\normalsize \label{tab:hyb_exp_STO3G} Different values of the average sign and perturbation order (PO) for four possible impurities present in SAO and NO bases for H$_8$ molecule in the STO-3G basis set. Fig.~\ref{fig:seet_selection} illustrates our impurity selection scheme using SAOs or Boys orbitals (spatial selection) and NOs (occupation selection).  TRIQS~\cite{Seth:cpc/200/274} program was used to perform CT-HYB calculations.}
  \begin{ruledtabular}
  \begin{tabular}{lllccccccccccccc}
    \multirow{2}{*}{$\beta$}          &\multirow{2}{*}{Basis}          &\multicolumn{2}{c}{Impurity A} & \multicolumn{2}{c}{Impurity B}  &  \multicolumn{2}{c}{Impurity C} & \multicolumn{2}{c}{Impurity D}  \\
    \cline{3-4}\cline{5-6}\cline{7-8}\cline{9-10}
     & & $\langle \rm sign \rangle$ & $\langle \rm PO \rangle$ &  $\langle \rm sign \rangle$ & $\langle \rm PO \rangle$ & $\langle \rm sign \rangle$ & $\langle \rm PO \rangle$ &$\langle \rm sign \rangle$ & $\langle \rm PO \rangle$          \\
      \hline
      
      10              & SAO   &0.688 & 4        &0.653 & 6          &0.653 & 6    &0.688 & 4      \\          
      10              & NO    &0.845 & 0        &0.705 & 1          &0.478 & 3    &0.500 & 2      \\
      50              & SAO   &0.232 & 16       &0.262 & 28         &0.262 & 28   &0.232 & 16     \\
      50              & NO    &0.977 & 2        &0.982 & 3          &0.971  & 4   &0.980 & 4      \\
      100             & SAO   &0.050 & 32       &0.065 & 55         &0.065 & 55   &0.055 & 32     \\
      100             & NO    &0.914 & 4        &0.946 & 6          &0.970 & 8    &0.978 & 9      \\
  \end{tabular}
  \end{ruledtabular}
\end{table*}

\begin{table*}
  \normalsize
  \caption{\label{tab:hyb_exp_631G} \normalsize Different values of the average sign and perturbation order (PO) for four possible impurities present in Boys and NO bases for H$_8$ molecule in 6-31G basis set. Fig.~\ref{fig:seet_selection} illustrates our impurity selection scheme using SAOs or Boys orbitals (spatial selection) and NOs (occupation selection).  TRIQS~\cite{Seth:cpc/200/274} program was used to perform CT-HYB calculations.}
  \begin{ruledtabular}
  \begin{tabular}{lllccccccccccccccc}
    \multirow{2}{*}{$\beta$}          &\multirow{2}{*}{Basis}          &\multicolumn{2}{c}{Impurity A} & \multicolumn{2}{c}{Impurity B}  &  \multicolumn{2}{c}{Impurity C} & \multicolumn{2}{c}{Impurity D}  \\
    \cline{3-4}\cline{5-6}\cline{7-8}\cline{9-10}
    & & $\langle \rm sign \rangle$ & $\langle \rm PO \rangle$ &  $\langle \rm sign \rangle$ & $\langle \rm PO \rangle$ & $\langle \rm sign \rangle$ & $\langle \rm PO \rangle$ &$\langle \rm sign \rangle$ & $\langle \rm PO \rangle$          \\
    \hline
      
      10              & Boys   &0.652 & 3        &0.730 & 5          &0.730 & 5    &0.652 & 3      \\          
      10              & NO     &0.633 & 1        &0.468 & 3          &0.377 & 3    &0.370 & 4      \\
      50              & Boys   &0.009 & 21       &0.072 & 28         &0.072 & 28   &0.009 & 21     \\
      50              & NO     &0.919 & 2        &0.922 & 3          &0.931 & 5    &0.964 & 8      \\
      100             & Boys   &0.000 & 42       &0.004 & 56         &0.004 & 56   &0.000 & 42     \\
      100             & NO     &0.722 & 5        &0.741 & 7          &0.823 & 10   &0.907 & 16     \\    
  \end{tabular}
  \end{ruledtabular}
\end{table*}

\begin{table}
  \normalsize
  \caption{\label{tab:h10_occ_imp} \normalsize Orbital occupations from RASCI calculations for an impurity+bath problem consisting of 8 active (impurity) and 8 bath orbitals at $R$ = 1.8 and 3.6 a.u for the H$_10$ chain in the cc-pVDZ basis. The total number of electrons in the impurity+bath problems at $R$ = 1.8 and 3.6 a.u. are 8 and 16, respectively.\\}
  \begin{ruledtabular}
  \begin{tabular}{cccccccccc}
    Orbitals       &R(H-H) = 1.8 a.u.         &R(H-H) = 3.6.a.u \\ 	
    \hline
		1	 		 &1.9827					&2.0000 \\
		2		  	 &1.9693					&2.0000 \\
		3			 &1.9412					&2.0000 \\
		4			 &1.8699					&2.0000 \\
		5			 &0.1377					&1.8220 \\
		6			 &0.0577					&1.7618 \\
		7	 		 &0.0275					&1.6206 \\
		8			 &0.0139					&1.3374 \\
		9			 &0.0001					&0.6679 \\
		10			 &0.0000					&0.3804 \\
		11			 &0.0000					&0.2375 \\
		12			 &0.0000					&0.1723 \\
		13			 &0.0000					&0.0000 \\
		14			 &0.0000					&0.0000 \\
		15		     &0.0000					&0.0000 \\
		16			 &0.0000					&0.0000 \\      
  \end{tabular}
  \end{ruledtabular}
\end{table}

\section{Numerical illustrations}\label{results}

Unless otherwise noted, the \textsc{ORCA} program \cite{orca} was used for all calculations using standard methods, e.g. FCI, CASSCF~\cite{Roos:cp/48/157,Roos:ijqc/18/175,Siegbahn:jcp/74/2384}, and NEVPT2.
The local modified version of the \textsc{DALTON} code \cite{dalton} was employed to generate an RHF input necessary for GF2 and to evaluate (full) CI active space Green's functions~\cite{Zgid:prb/86/165128}.
Throughout this article, to compute potential energy curves accurately and to converge  the electronic energy to $5\times10^{-4}$ a.u. with respect to the inverse temperature $\beta$ and the number of frequencies $w_{max}$, we employ different values of $\beta$ and $w_{max}$ along a single potential energy curve that yield a converged energy for different geometry points. 
We denote our method as SEET(CI/GF2)[$m\times n$o], where $m$ and $n$ are number of active spaces (impurity+bath problems) and number of orbitals in active spaces (number of impurity orbitals), respectively. 
The RASCI solver will be used when the impurity+bath problem is intractable with the FCI solver.

\subsection{Non-local self-energy and screened interactions}\label{numres:screening}

First, let us show that SEET, where the non-local GF2 self-energy is included in the DMFT-like iterations, yields numerically very similar results to a DMFT procedure with screened interactions.
As an example, we consider the H$_6$ chain in the STO-6G basis \cite{sto_minimal}.
All SEET calculations were performed in the SAO basis.
In this work, we employ a statically screened impurity model, in which the screened interaction was obtained using GF2 and is defined via Eq.~\ref{eq:Uscr} at several chosen $\tau$ points.
The effective impurity model present during the DMFT self-consistency is then solved using the FCI impurity solver.

\begin{figure} [h] 
  \includegraphics[width=8.0cm,height=5.5cm]{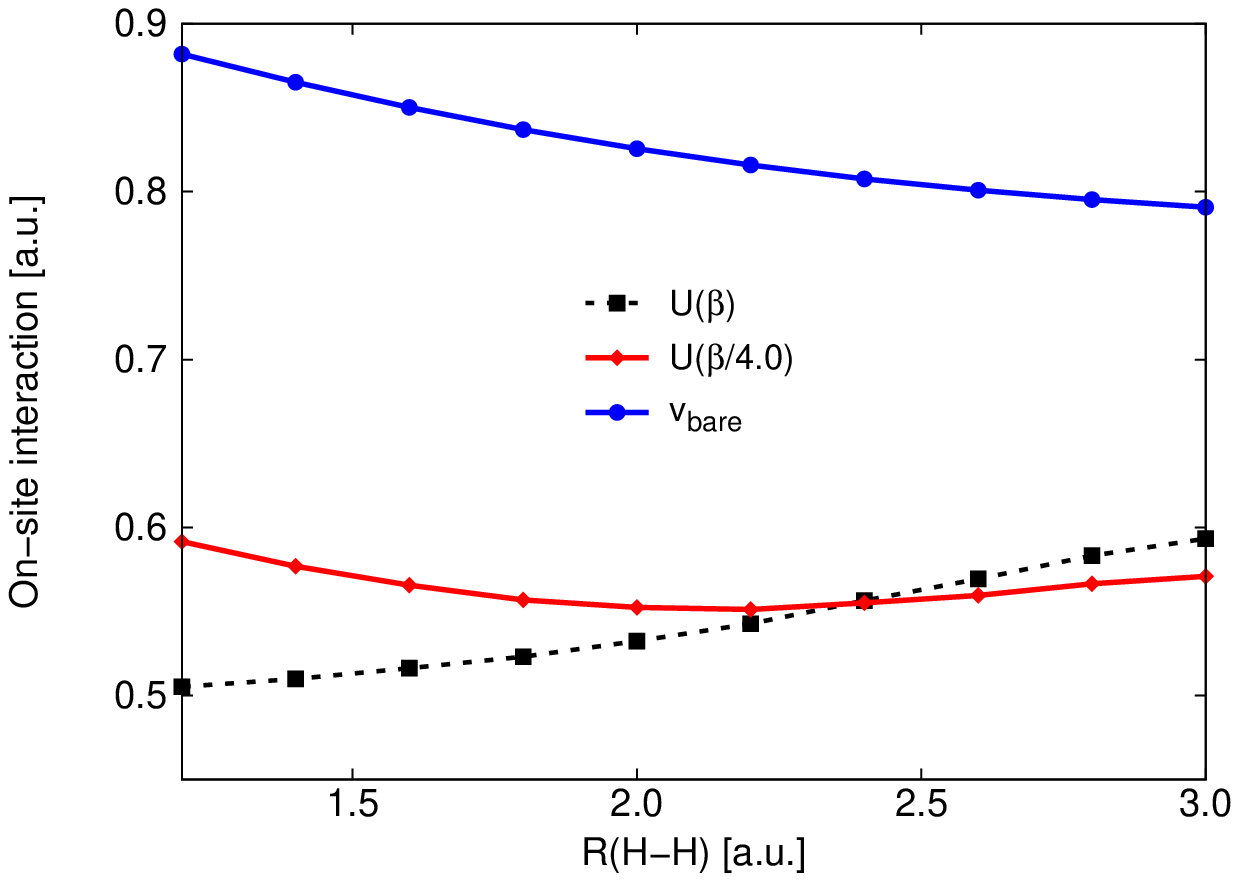}
  \includegraphics[width=8.0cm,height=5.5cm]{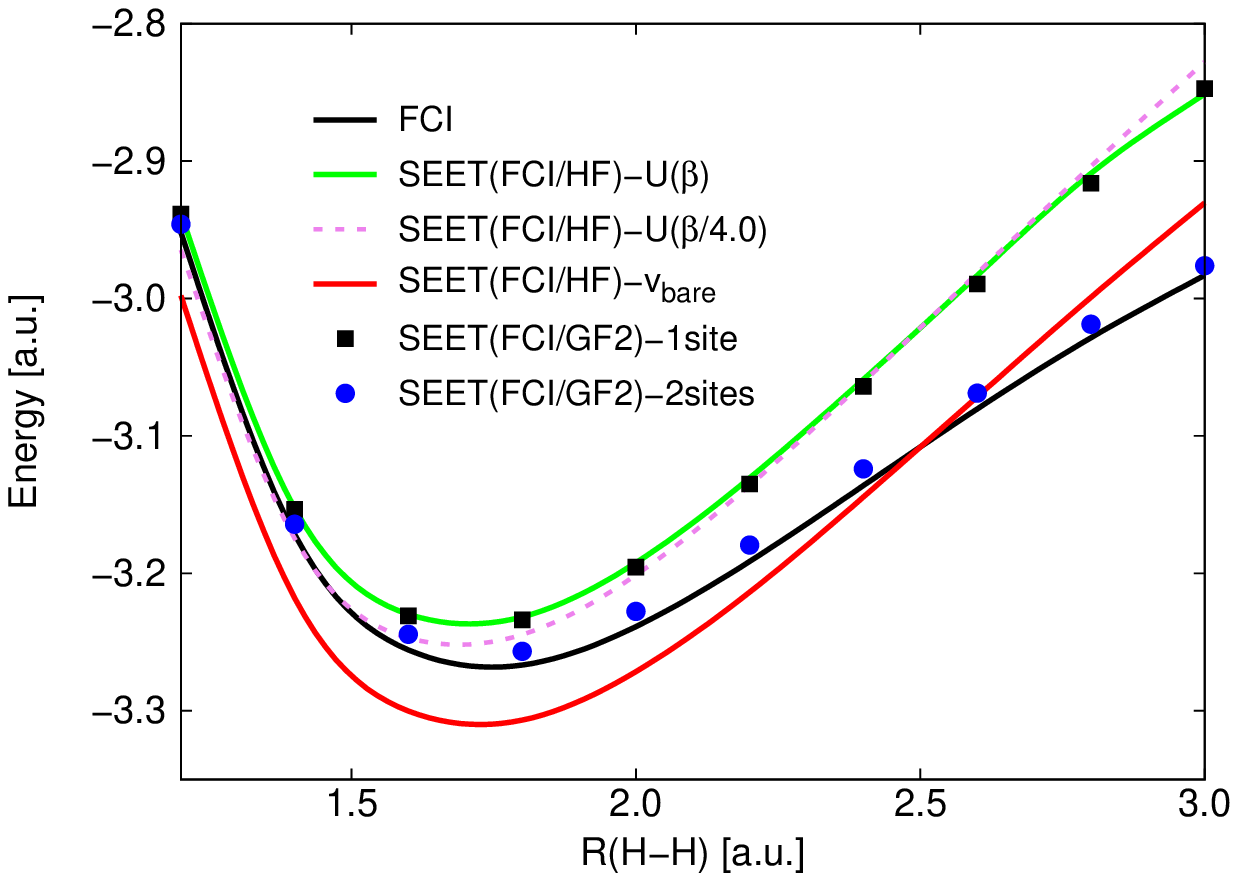}
  \caption{ \normalsize Upper panel: The on-site screened interaction $U(\tau)$, with $\tau = \beta \mbox{ and } \beta/4.0$, and bare Coulomb interaction as functions of the interatomic distance for H$_6$ chain in the STO-6G basis. Lower panel: Potential energy curves of H$_6$ chain in the STO-6G basis calculated using on-site bare and screened interactions. Results from FCI, SEET(FCI/GF2) with a single site and two sites impurities are also provided for comparison.}
  \label{h6c_scr}
\end{figure}

The upper panel of Fig.~\ref{h6c_scr} shows the on-site bare Coulomb interaction and screened interaction $U(\tau)$ at $\tau = \beta$ and $\beta/4.0$ as a function of the interatomic distance $R$.
Note that the on-site bare Coulomb interaction in the SAO basis is not constant and changes with the distance.
The on-site screened interaction should be equal to the bare interaction at large distances when the non-local interactions between atoms vanish. For short interatomic distances, due to the presence of large non-local interactions, the screened interaction is about twice smaller than the bare interaction.

The potential energy curves calculated with the on-site bare interaction, screened interactions, and SEET(FCI/GF2) are shown in the lower panel of Fig.~\ref{h6c_scr}.
Note that SEET(FCI/HF)-$v_{\mbox{\scriptsize{bare}}}$, SEET(FCI/HF)-$U(\beta)$, and SEET(FCI/HF)-$U(\beta/4.0)$  corresponds to DMFT on top of HF with bare interactions $v_{\mbox{\scriptsize{bare}}}$, with on-site  screened interaction $U(\beta)$, and with on-site  screened interaction $U(\beta/4.0)$.
The curves from FCI and SEET(FCI/GF2) with one site and two sites impurities are also plotted for comparison.
It is well-known that the DMFT curve with an on-site bare interaction (SEET(FCI/HF)-$v_{\mbox{\scriptsize{bare}}}$) falls bellow the FCI result around the equilibrium geometry~\cite{Lin:prl2011-dmft}.
When the on-site screened interactions $U(\beta)$ or $U(\beta/4.0)$ are used, the curves move up closer to the FCI result around the equilibrium. 

Interestingly, SEET(FCI/GF2)-1site gives a very similar curve to that of SEET(FCI/HF)-$U(\beta)$, indicating the equivalence between using the non-local self-energy and screened interactions.
To get more accurate results, a larger number of sites than a single site treated by accurate many-body solvers are required.
While, as mentioned previously, the extension of the on-site screened interaction to multiple-orbital spaces may not  be straightforward, 
SEET(FCI/GF2) with larger number of impurity orbitals, for example two sites, can be performed trivially and, as indicated by our results, it significantly improves upon the results of SEET with a single site.

Since the difference between $U(\beta)$ and $U(\beta/4.0)$ curves is relatively small but not negligible, using statically screened interactions  may not be very robust if chemical accuracy is desired.
To fully reach chemical accuracy, a more complicated procedure needs to be carried out to evaluate the dynamically screened interaction $U(\tau)$ for all $\tau$ points, as is done in condensed matter physics~\cite{werner2016dynamical}.
However, this procedure leads to a computational bottleneck for realistic molecules where the active space is usually large, thus a significant  memory would be necessary to store the $U_{ijkl}(\tau)$ matrix for larger orbital spaces.
In contrast, realistic molecular SEET with the non-local self-energy instead of screened interactions requires storing only $[\mathbf{\Sigma}^{GF2}(\tau)]_{ij}$ matrix thus making it less memory demanding.

\subsection{SEET outer loop: non-local self-energy relaxation}\label{numres:seet_loop}
As we have mentioned in the SEET self-consistency description, SEET can be done as a {\it one-shot} procedure where $\mathbf{\Sigma}^{GF2}_{non-local}(i\omega)$ coming from the initial self-consistent GF2 is not updated, or can be performed fully self-consistently where the non-local GF2 self-energy is updated during subsequent outer iterations involving GF2.
Here, we would like to compare the one-shot and self-consistent procedures.

In systems where the entanglement between atoms is equivalent (for instance, hydrogen ring), separating the whole system into spatial fragments in the stretched regime will give rise to an incompatibility between local self-energy from high level theory and non-local self-energy from low level theory.
Moreover, GF2 is not providing the self-energy that is accurate enough since the correlations are strong in the stretched regime.
Consequently, in this regime, it is essential to carry out the outer iterations and update $\mathbf{\Sigma}^{GF2}_{non-local}(\tau)$ while starting from the zeroth order Green's function that contains the strong correlation effects included due to the DMFT-like inner SEET iterations.


\begin{figure} [h]
  \includegraphics[width=8.0cm,height=5.5cm]{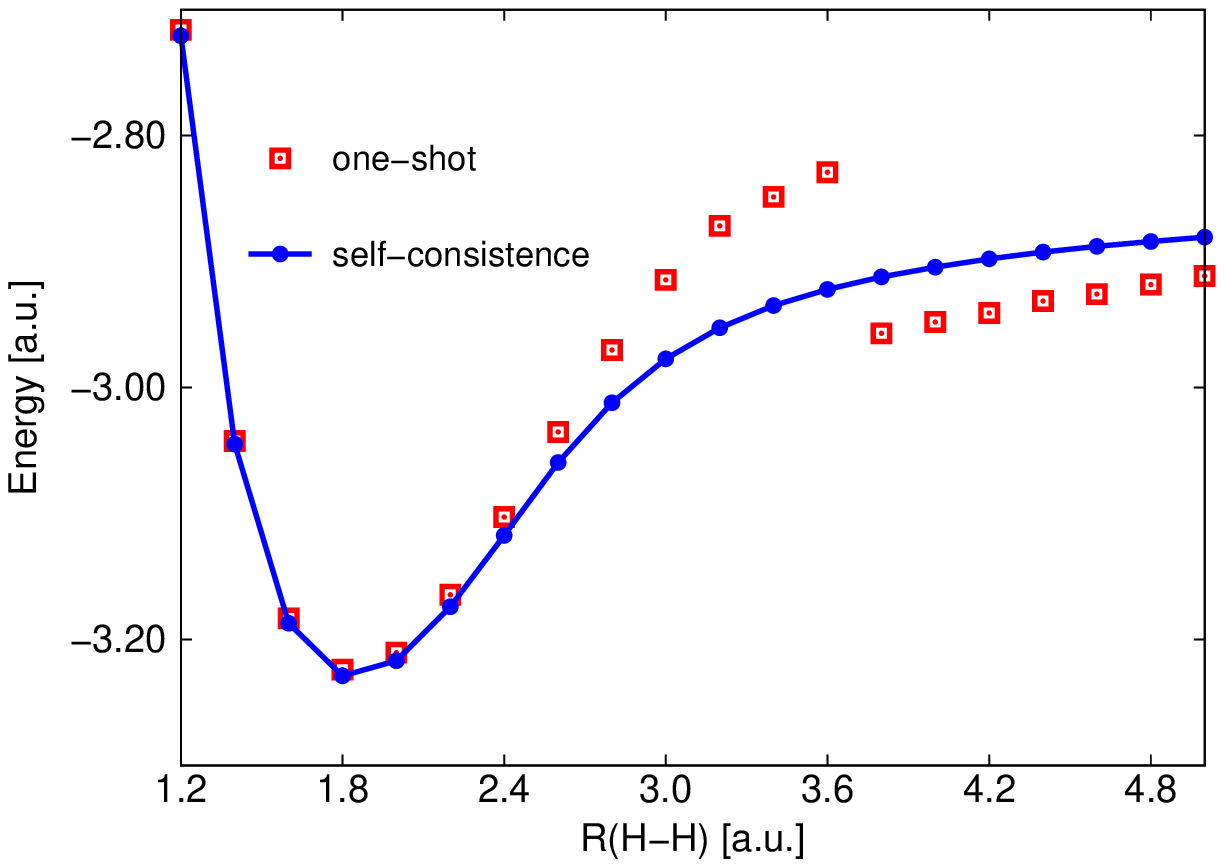}
  \includegraphics[width=8.0cm,height=5.5cm]{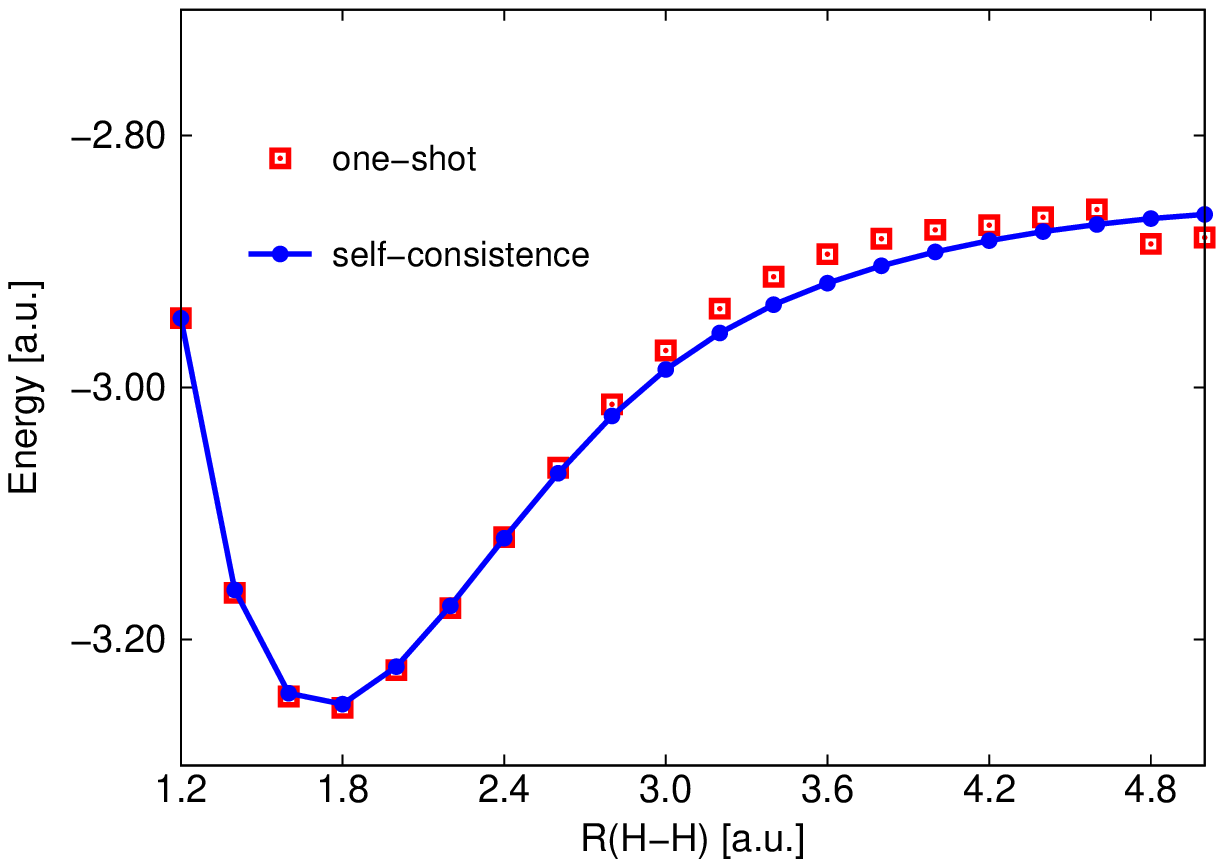}
  \caption{\normalsize Potential energy curves for the H$_6$ ring (upper) and H$_6$ chain (lower) from the one-shot and self-consistent SEET(FCI/GF2)[3$\times$2o]/SAO calculations. The STO-6G basis set was used.}
  \label{h6_nlsigma}
\end{figure}

\begin{figure} 
  \includegraphics[width=\columnwidth]{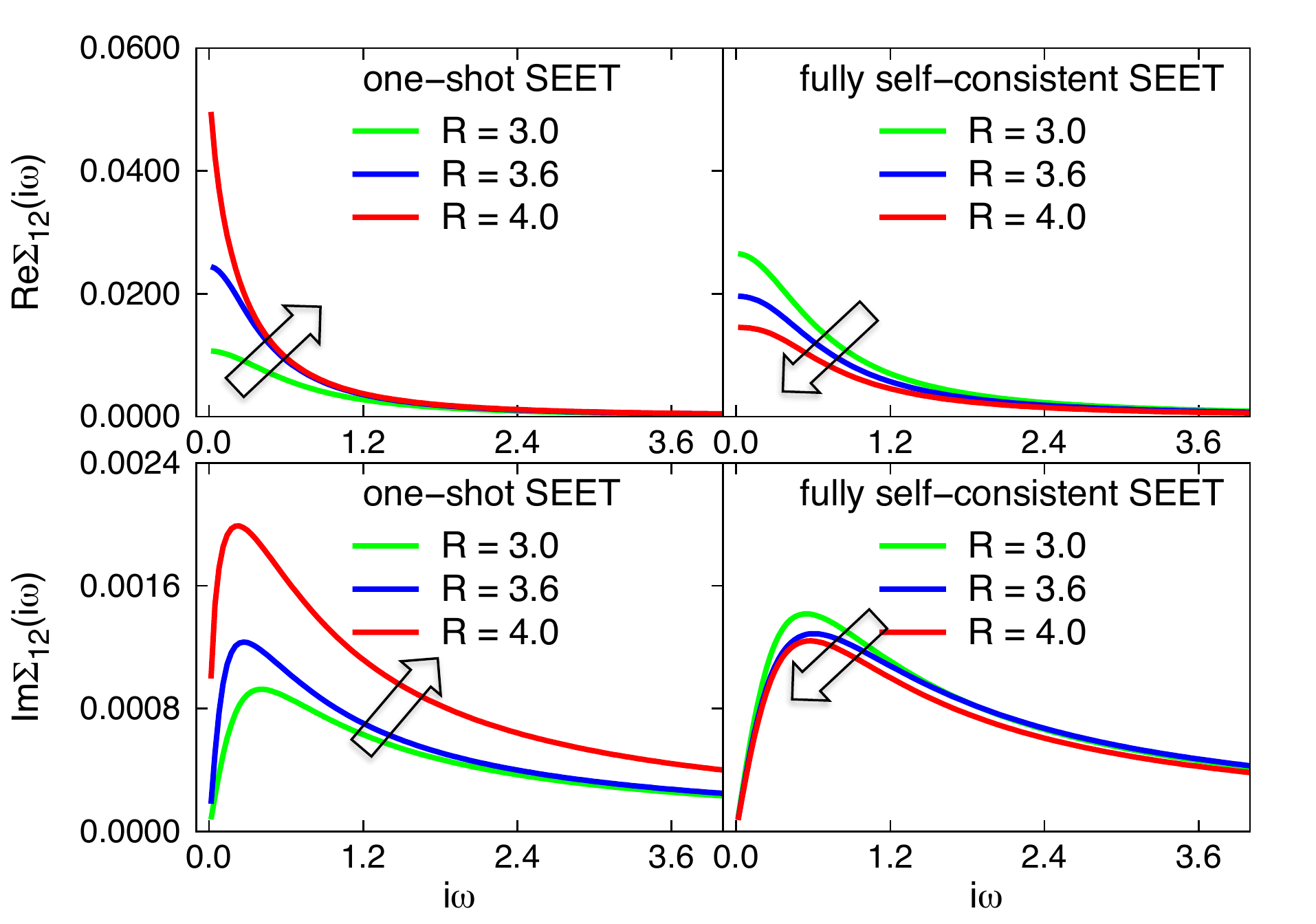}
  \caption{\normalsize Real (upper panel) and imaginary (lower panel) parts of the non-local self-energy $\Sigma_{12}(i\omega)$ after the one shot (left) and fully self-consistent (right) SEET(FCI/GF2)[3$\times$2o]/SAO calculation for the H$_6$ ring in the STO-6G basis.
    Arrows indicate the overall trend of the non-local self-energy with the increasing H-H bond length.}
  \label{sigma12}
\end{figure}

To reveal the importance of the non-local self-energy relaxation, especially, for the spatial fragment based embedding scheme, we plot the potential curves from SEET(FCI/GF2)[3$\times$2o]/SAO calculations for the H$_6$ ring and the H$_6$ chain in the STO-6G basis set.
These results are summarized in Fig.~\ref{h6_nlsigma}.

Around the equilibrium, for both the H$_6$ ring and H$_6$ chain, one-shot and self-consistent GF2-DMFT schemes give almost identical results since the effect of non-local self-energy relaxation is negligible in the weakly correlated regime.
When the interatomic distance is large, the one-shot and self-consistent curves significantly differ from each other and the former breaks down at the dissociation limit.
A smooth dissociation curve is achieved when the GF2 non-local self-energy is relaxed in the presence of the FCI local self-energy.
Obviously, the effect of the non-local self-energy relaxation is more pronounced for the H$_6$ ring than for the H$_6$ chain since the entanglement between unit cells in the former is much stronger than that in the latter.

The relaxation of the non-local self-energy is further explored in Fig.~\ref{sigma12}, where we examine the case of H$_6$ ring.
In principle, the non-local self-energy should decrease with the increasing  inter-fragment distance since the non-local two-electron integrals $v_{ijkl}$ where $i,j,k,l$ belong to different fragments/orbitals should be vanishing. In practice, however, the non-local integrals do not decay fast enough giving rise to quantitatively inaccurate GF2 self-energy in the stretched atoms regime.
Consequently, the non-local GF2 self-energy erroneously increases as a function of  bond distance instead of decreasing, see Fig.~\ref{sigma12}.
The gradual decrease can only  be observed after employing the fully self-consistent SEET when the non-local GF2 self-energy is relaxed in the presence of the FCI self-energy accounting correctly for the strong correlations present at large bond distances. The fully self-consistent SEET yields a smooth transition from the weakly to strongly correlated regime as shown in Fig.~\ref{h6_nlsigma}.
 
\subsection{Energy and spatial domain based embedding}\label{numres:energy_spatial_domains}

The embedding procedure can be performed either in the energy or spatial domain.
In the former the active orbitals are chosen as the most important NOs, whereas in the latter the whole system is split into different spatial fragments (SAOs/RNOs) that are physically/chemically meaningful.

\begin{figure} [h]
  \includegraphics[width=8cm,height=5.5cm]{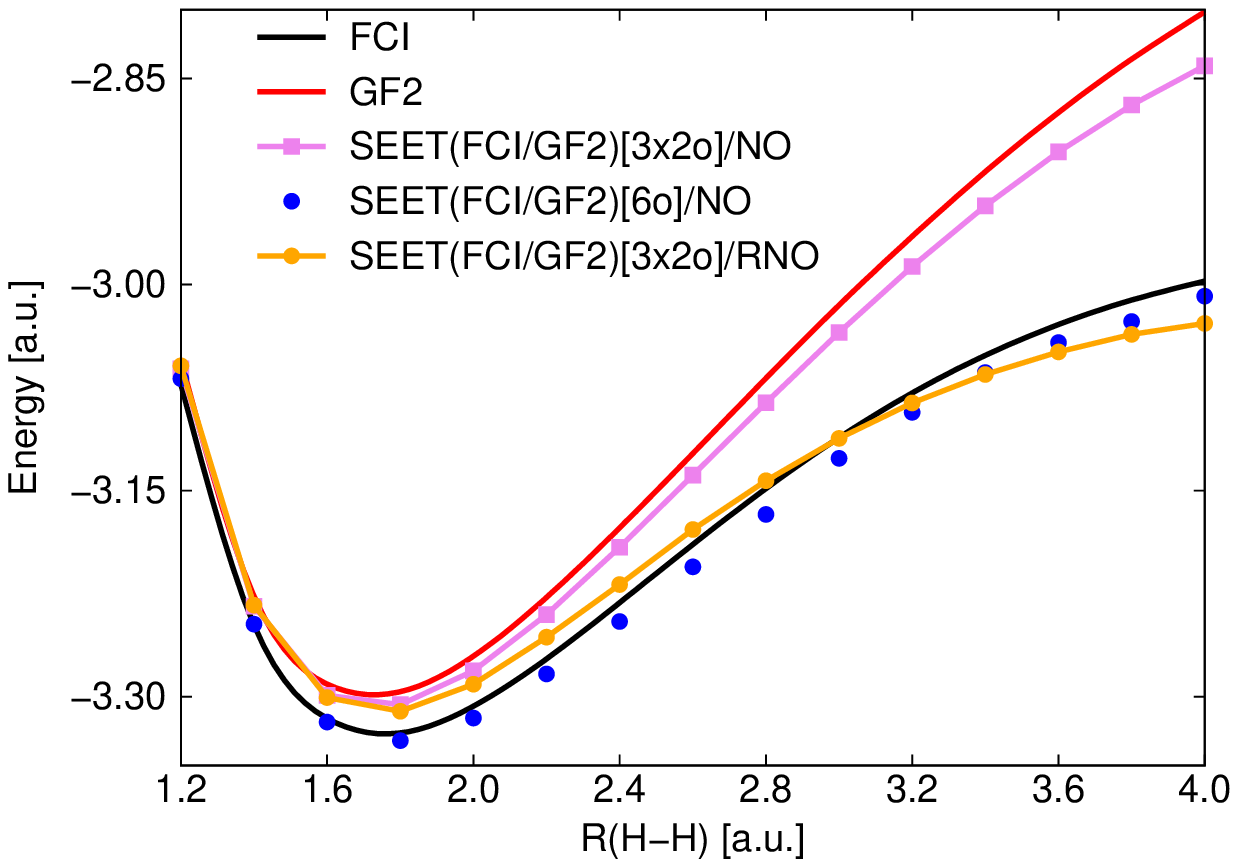}
  \includegraphics[width=8cm,height=5.5cm]{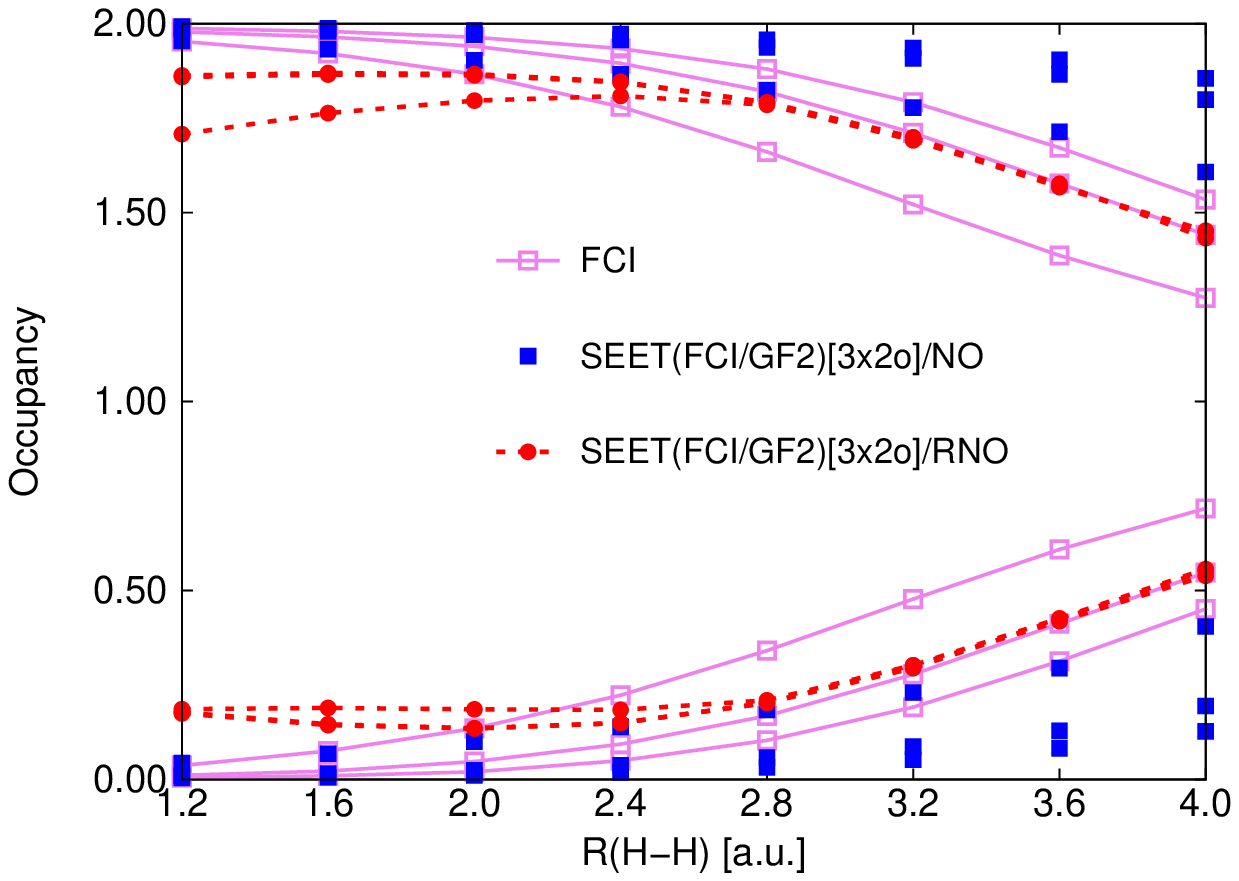}
  \caption{\normalsize Upper panel: potential energy curve for the  H$_6$ chain in the TZ basis. Lower panel: Orbital occupations as a function of bond length.}
  \label{h6tz}
\end{figure}

At first, we consider the H$_6$ chain in the TZ basis set~\cite{Dunning:jcp/55/716} as shown in the upper panel of Fig.~\ref{h6tz}.
An excellent agreement between SEET in NO basis and FCI curve can be observed when a full active space composed of six orbitals, SEET(FCI/GF2)[6o]/NO, is used.
Now, let us consider how splitting the full active space into different groups of NOs and RNOs (3$\times$2o means three groups of two orbitals) is influencing the results.
At short distances, both SEET(FCI/GF2)[3$\times$2o] in NO and RNO bases give very similar results.
However, in the stretched regime,  energies from the calculation in the RNO basis are much closer to the FCI curve than these from the calculation in the NO basis.

For the H$_6$ chain in the TZ basis, to get insight into the behavior of different orbital bases at dissociation, the occupations of six valence orbitals are plotted in the lower panel of Fig.~\ref{h6tz}.
We compare the orbital occupations from SEET(FCI/GF2)[3$\times$2o] in both RNO and NO bases to the exact ones calculated using FCI.

Note that there is a degeneracy of two pairs of active orbitals in the RNO basis corresponding to two ends of the chain.
FCI yields occupation numbers that smoothly transit from weakly to  strongly correlated regime as the bond length increases.
At short distances, i.e. $R \le 2.0$ a.u., NO occupancies are quite close to the FCI ones, while those in the RNO basis have slightly overestimated partial occupations.
This difference can be understood in the following way.
Since the correlation in the short bond regime is weak and the orbitals are mostly unoccupied or doubly occupied, SEET in the NO basis with the active space that is split into several groups is able to capture all the correlation effects well. In fact, even GF2 alone is  good enough to describe correlation effects in this regime.
However, since in the short bond regime in the RNO basis the coupling between fragments is non-negligible,  splitting the molecule into several fragments leads to missing the inter-fragment bonding, thus leading to less accurate results.

Upon bond stretching, fragments become nearly isolated and this is reflected by the degeneracy of all pairs of active orbitals in the RNO basis.
Evidently,  in the RNO basis the occupation numbers in the strongly correlated regime are almost parallel to the FCI ones.
Thus, SEET in the RNO basis can yield a dissociation curve closely following the FCI one at long distances where the correlations are strong.
In the NO basis, splitting the full active space into orbital groups is not sufficient to correctly describe the physics at dissociation limit.
In this case, only the highest occupied molecular orbital (HOMO) and the lowest unoccupied molecular orbital (LUMO) have reasonable occupations, while all the other orbitals have significantly different occupations from the FCI ones.

\begin{figure} [h]
  \includegraphics[width=8.0cm,height=5.5cm]{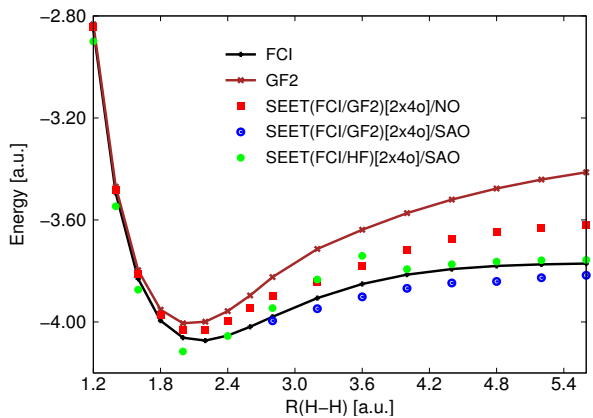}
  \caption{\normalsize Potential energy curves for 2$\times$4 hydrogen lattice in the STO-6G basis.}
  \label{2x4sto}
\end{figure}

We now investigate a 2$\times$4 hydrogen lattice, which is a more complicated example than the H$_6$ chain.
The minimal STO-6G basis was used, so that dividing the whole system into spatial fragments can be done in the SAO basis.
Fig.~\ref{2x4sto} shows the potential energy curves of 2$\times$4 hydrogen lattice from SEET(FCI/GF2)[2$\times$4o] in NO and SAO bases along with GF2 and FCI dissociation curves.
For comparison, we also plot the curve from SEET(FCI/HF)[2$\times$4o] in the SAO basis, i.e. non-local self-energy $\mathbf{\Sigma}^{GF2}_{non-local}(i\omega)$ is not taken into account for this case.
As mentioned previously, due to the lack of the self-energy terms describing non-local correlations, around equilibrium the energy from SEET(FCI/HF) with bare interactions is lower than FCI one, while in the intermediate regime the SEET(FCI/HF) energy is above the FCI one.
Beyond the dissociation limit, the lattice is separated into isolated atoms, thus the electron interaction in such a system is actually just the on-site bare interaction.
Therefore, in the dissociation limit, SEET(FCI/HF) gives energies in a good agreement with FCI results, however, the SEET(FCI/HF) curve is not smooth and has a discontinuity when a transition to the stretched regime happens.
In the short bond region, energies from SEET(FCI/GF2)[2$\times$4o] in the NO basis are better than GF2 energies and quite close to FCI ones, whereas SEET(FCI/GF2)[2$\times$4o] in the SAO basis has difficulty converging because of the large entanglement between fragments. 
In contrast, far away from the equilibrium, SEET(FCI/GF2) energies in the SAO basis are much closer to the exact ones than SEET(FCI/GF2) energies in the NO basis.
However, since GF2 overestimates the non-local self-energy between unit cells at long distances, the SAO curve is below than that from FCI.

\subsection{Comparison with ground-state wavefunction methods}\label{numres:comparison}

After discussing technical aspects of SEET, now we turn to showcase quantitative accuracy of SEET as compared to the standard quantum chemistry ground-state  wavefunction methods.
As mentioned previously, SEET(FCI/GF2) when a full active space is used can be directly compared to methods such as CASPT2 or NEVPT2 since in both of these methods, in an analogous manner to SEET, the strongly correlated (or active) orbitals are described by a higher level method than the weakly correlated (inactive) orbitals. In cases of CASPT2, NEVPT2, and SEET(FCI/GF2) the perturbative description involves a low level perturbative expansion. 
Therefore, it is necessary and interesting to make a numerical comparison between SEET(FCI/GF2) and CASPT2 or NEVPT2.
We have  done  such comparison in our earlier work~\cite{Lan:jcp/143/241102} but only for SEET in the NO basis.

\subsubsection{ LiH$_2$ and LiH$_4$ chains in TZ basis}

\begin{figure} [t]
  \includegraphics[width=8.0cm,height=5.5cm]{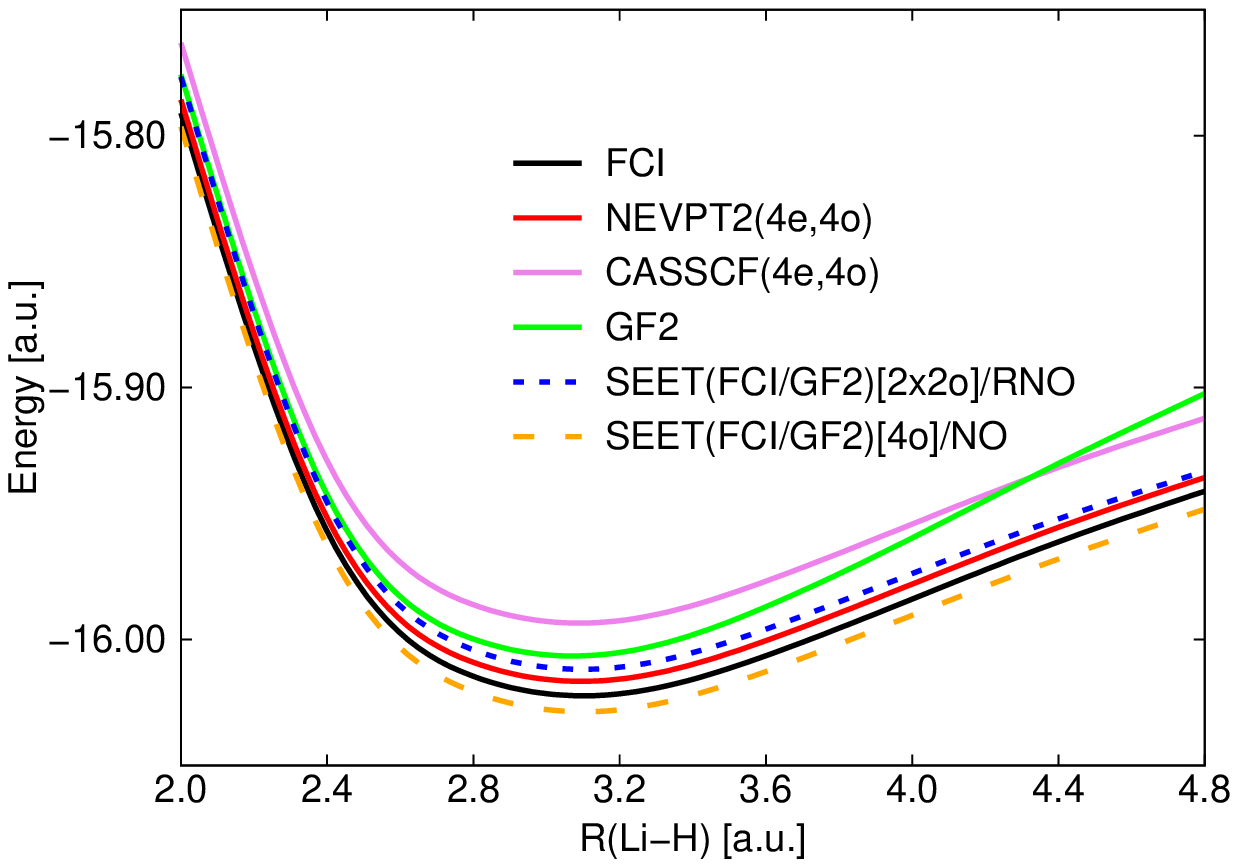}
  \includegraphics[width=8.0cm,height=5.5cm]{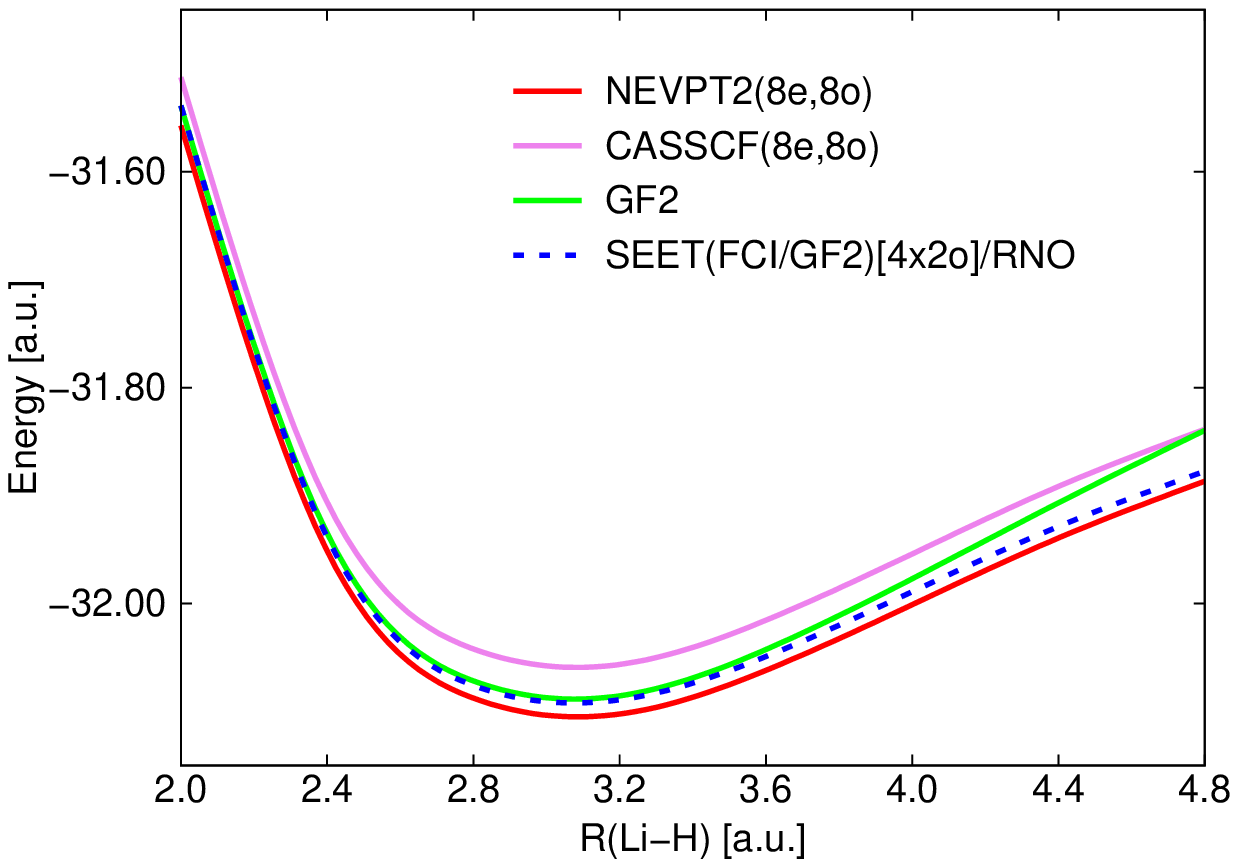}
  \caption{\normalsize Potential energy curves evaluated using different methods for LiH$_2$ (upper panel) and LiH$_4$ (lower panel) chains in the TZ basis. }
  \label{lih_tz}
\end{figure}

\begin{figure} [t]
  \includegraphics[width=8.0cm,height=5.5cm]{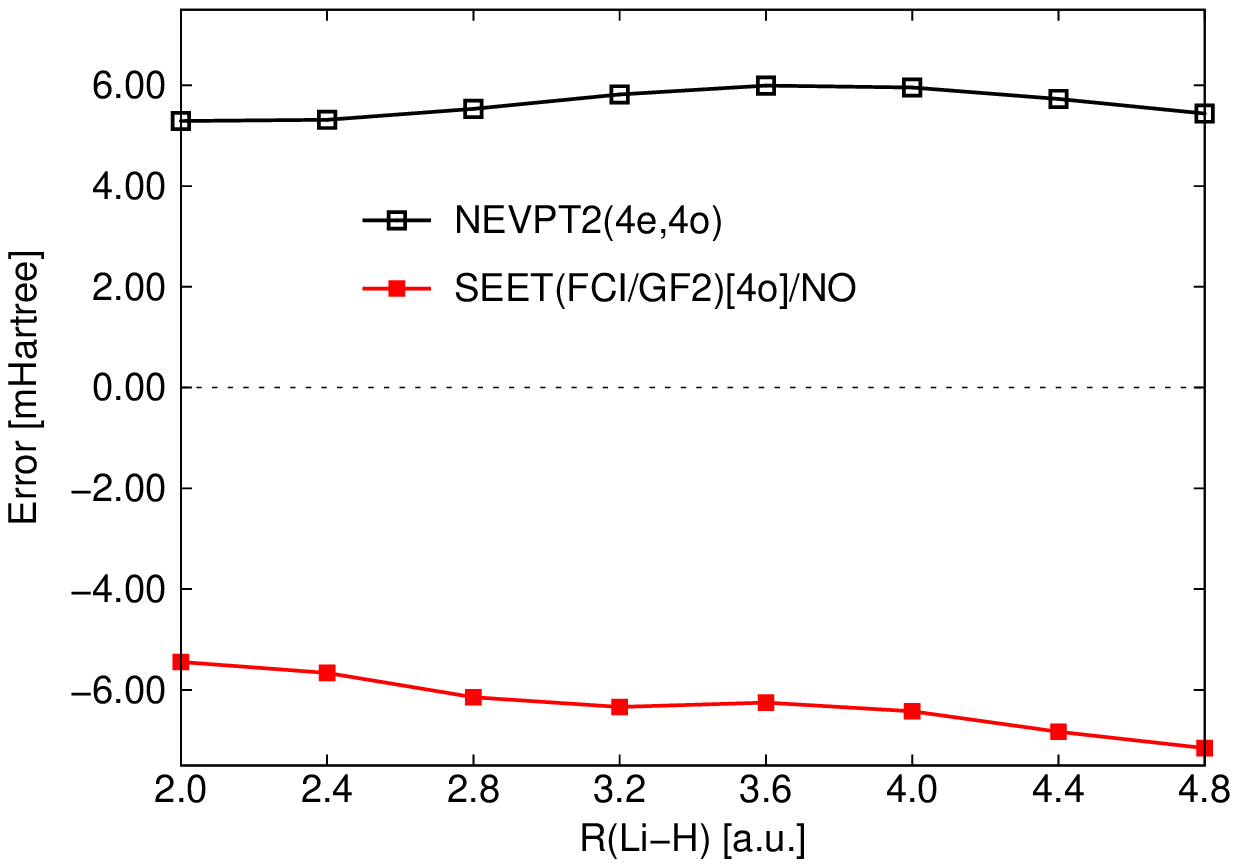}
  \includegraphics[width=8.0cm,height=5.5cm]{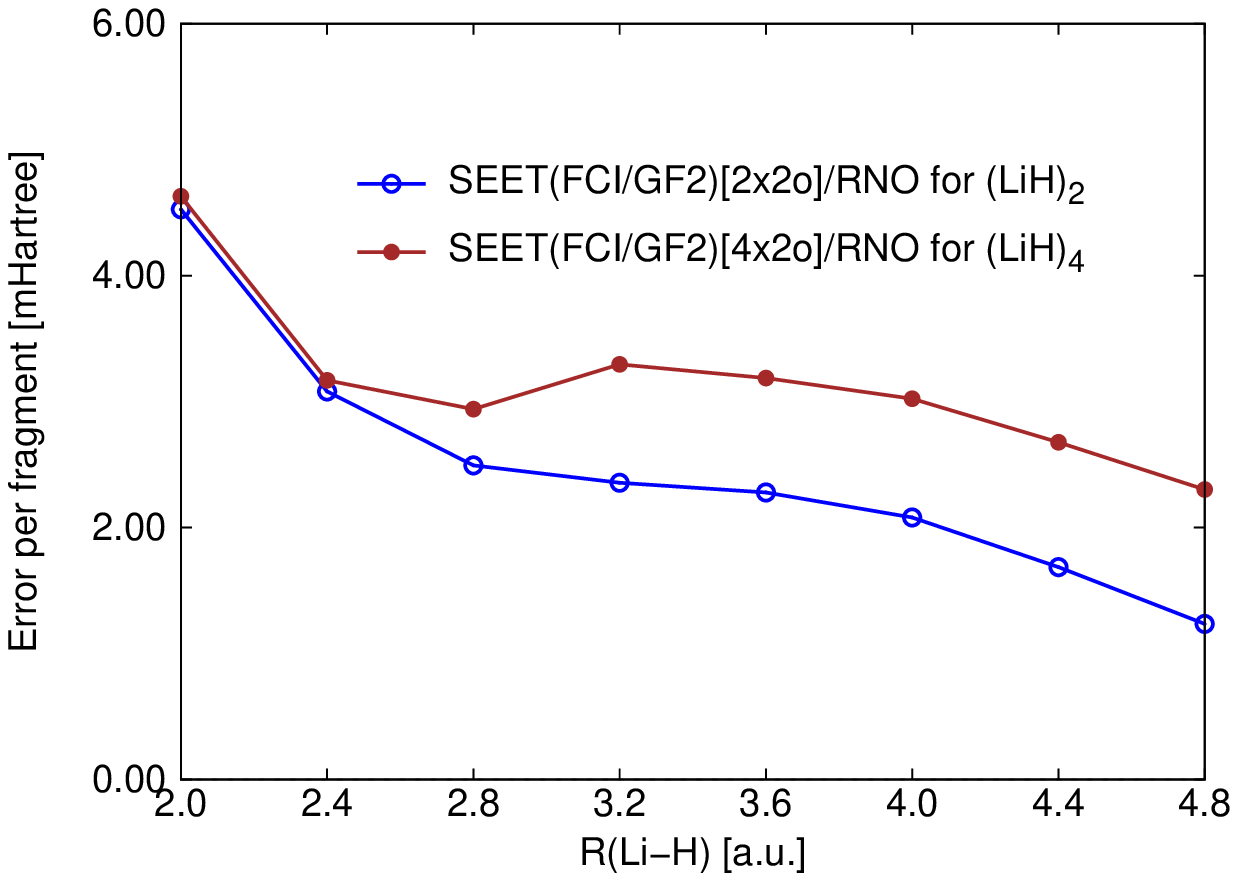}
  \caption{\normalsize Upper panel: LiH$_2$ chain in the TZ basis. Energy errors (in mHartree) $E_{\rm FCI}-E_{\rm X}$ for X = NEVPT2(4e,4o) and SEET(FCI/GF2)[4o]/NO as a function of bond distances. Lower panel: Energy errors per fragment (in mHartree) $(E_{\rm NEVPT2}-E_{X})/N$  for SEET in the RNO basis for $N=2,4$ for both LiH$_2$ and LiH$_4$ in the TZ basis.}
\label{lih_tz_errors}
\end{figure}

Here, we further analyze SEET results in the RNO basis.
In Fig.~\ref{lih_tz}, we present potential energy curves of LiH$_2$ and LiH$_4$ chains in the TZ basis.
For the RNO basis, (LiH)$_n$ chain (with $n = 2, 4$) is divided into $n$ LiH fragments and the active orbitals that are used to construct the Anderson impurity models are constructed from two valence orbitals of each fragment.
Both CASSCF and NEVPT2 correctly describe the dissociation.
While GF2 yields accurate energies around the equilibrium, its curve is not parallel to the FCI one at long distances.
However, when static correlation is properly treated using SEET(FCI/GF2), in both bases NO and RNO, the dissociation regime is described correctly.
In particular for LiH$_2$, as seen in the upper panel of Fig.~\ref{lih_tz_errors}, although the SEET(FCI/GF2)[4o] in the NO basis yields a curve below the exact one, errors are  of the same order as those of NEVPT2(4e,4o).
SEET(FCI/GF2)[2$\times$2o] in the RNO basis closely follows the NEVPT2 curve for all the distances considered here, thus having an error essentially equivalent to NEVPT2.

For LiH$_4$, we cannot evaluate FCI, thus we compare our results against NEVPT2. In the lower panel of Fig.~\ref{lih_tz_errors}, we plot the error in the energy per fragment for LiH$_2$ and LiH$_4$ molecules when compared to NEVPT2 energies per fragment.
Interestingly, the difference between errors per fragment of LiH$_2$ and LiH$_4$ cases are very small ($\le 1.0$ mHartree), indicating that the correct description of SEET(FCI/GF2) using LiH fragment as a repeating unit holds true regardless of the length of system.

\subsubsection{H$_{50}$ chain in STO-6G basis}

\begin{figure} [t]
  \includegraphics[width=8.0cm,height=5.5cm]{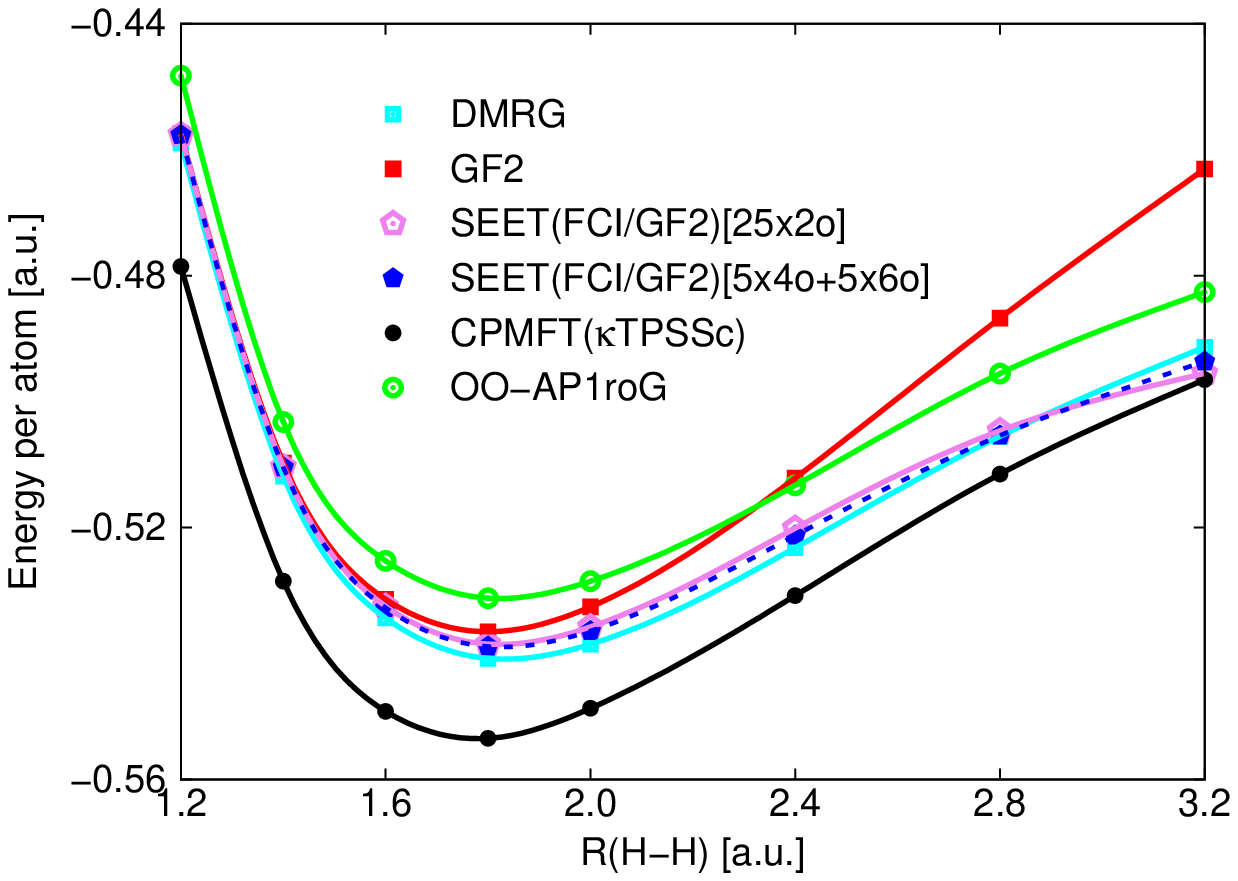}
  \includegraphics[width=8.0cm,height=5.5cm]{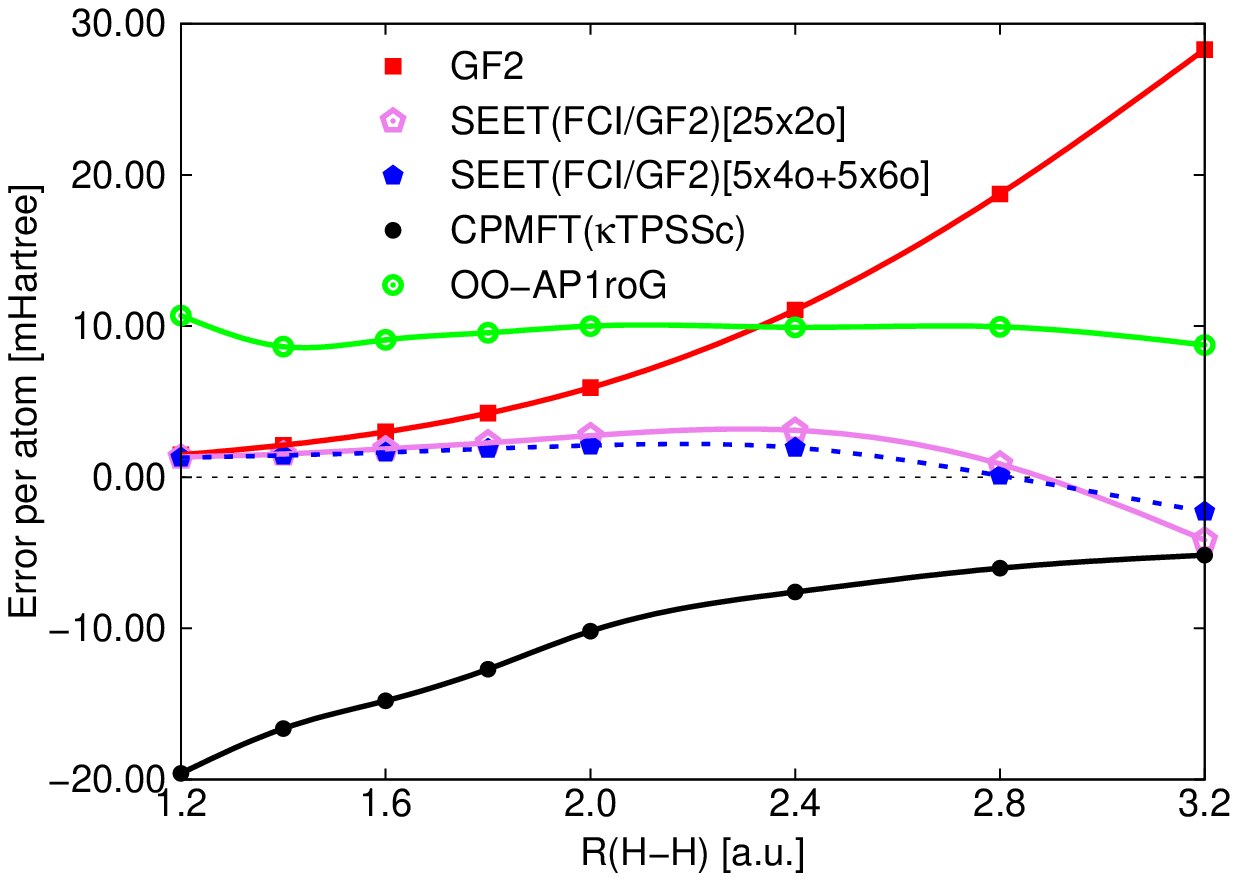}
  \caption{\normalsize Upper panel: potential energy curves for H$_{50}$ chain in the STO-6G basis. DMRG, CPMFT($\kappa$TPSSc), and OO-AP1roG data are taken from Refs.~\onlinecite{Hachmann:jcp2006-h50dmrg,Tsuchimochi:jcp2009-cpmft,Boguslawski2014}.
   Lower panel: energy error per atom (in mHartree) with respect to  the DMRG data.
  }
  \label{h50sto}
\end{figure}

Now we test SEET(FCI/GF2) in the SAO basis on a well-known, non-trivial benchmark system, H$_{50}$ chain in the STO-6G basis. The exact solution is available from DMRG calculation~\cite{Hachmann:jcp2006-h50dmrg}.
For a full comparison, we also present results from established theories capable of targeting strongly correlated molecules such as orbital-optimized antisymmetric product of 1-reference-orbital geminals (OO-AP1roG)~\cite{Boguslawski2014} and constrained-pairing mean-field theory combined with $\kappa$TPSSc functional [CPMFT($\kappa$TPSSc)]~\cite{Tsuchimochi:jcp2009-cpmft}.
Potential energy curves and errors relative to DMRG energies are displayed in the upper and lower panels of Fig.~\ref{h50sto}.

OO-AP1roG and CPMFT($\kappa$TPSSc) curves are far above and far below the DMRG reference, respectively. While the CPMFT($\kappa$TPSSc) curve displays huge non-parallelity errors near the equilibrium geometry, the OO-AP1roG curve remains nicely parallel to the DMRG curve.
GF2 gives very good energies for short distances; however, it  largely deviates from DMRG at long distances.
SEET(FCI/GF2)[25$\times$2o], where 25 Anderson impurities containing two impurity orbitals are embedded in the GF2 self-energy, yields a significantly improved energies at long distances when compared to GF2 alone.
The errors can be further minimized when SEET(FCI/GF2)[5$\times$4o+5$\times$6o] calculation is carried out.

\subsubsection{H$_{10}$ chain in cc-pVDZ basis}

\begin{figure} [h]
  \includegraphics[width=8.0cm,height=5.5cm]{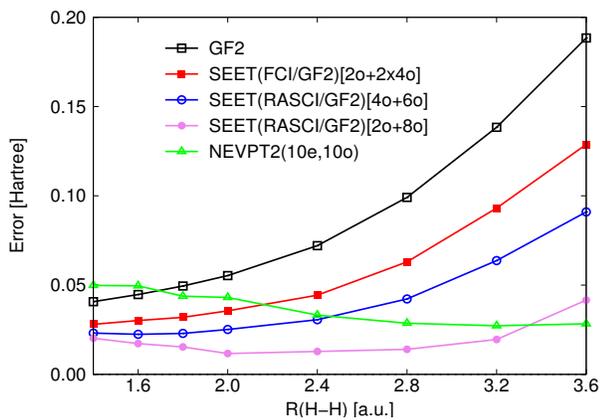}
  \caption{\normalsize Errors in energy (in a.u.) with respect to DMRG reference as a function of R(H-H) for GF2, SEET with different active spaces in the NO basis, and NEVPT2(10e,10o). All energies (in a.u.) are presented in Table~\ref{tab:h10}.}
  \label{h10err}
\end{figure}

Here, we explore the concept of active space splitting where the full number of active orbitals is divided into several groups of orbitals used to build Anderson impurity models. To demonstrate that SEET is systematically improvable, when the number of orbitals used to build the impurities is increased, we performed calculations with different number of orbitals in the impurities for H$_{10}$ chain in the cc-pVDZ basis \cite{dunning_ccpvdz}. The total number of orbitals in this basis set is 50 while the size of full active space is 10 orbitals.
SEET results are summarized in Table~\ref{tab:h10} along with GF2, NEVPT2(10e,10o) and DMRG \cite{h10ccpvdz_dmrg} energies for comparison.
The DMRG data were computed using the \textsc{BLOCK} program~\cite{dmrg_block_2012,dmrg_block_2015}.
The errors relative to DMRG are shown in Fig~\ref{h10err} as a function of bond length.
At short distances ($R < 2.0$ a.u.), GF2 energies are comparable to those from the NEVPT2(10e,10o) calculation.
Upon bond stretching, when compared to DMRG, the GF2 error strongly increases, while NEVPT2(10e,10o) one slowly decreases.
We can see that the errors of GF2 are significantly reduced when SEET(FCI/GF2)[2o+2$\times$4o] calculation is carried out.
For stretched distances, a systematic reduction of errors can be observed when the number of impurity  orbitals is systematically enlarged starting from (2o+2$\times$4o), (4o+6o), to (2o+8o). Interestingly, this error reduction with an increasing number of impurity orbitals is very systematic and is independent of the distances on the potential energy curve.

\begin{table*}
  \normalsize
  \caption{\label{tab:h10} \normalsize Potential energies (in a.u.) as a function of bond distance (in a.u.) for H$_{10}$ chain in the cc-pVDZ basis. Energies from GF2 and SEET with different number of impurity orbitals building the active space are compared to NEVPT2(10e,10o) and DMRG energies~\cite{h10ccpvdz_dmrg}.  FCI solver is used to treat [2o+2$\times$4o] and [4o+6o] impurities, while RASCI solver is used for [2o+8o] impurity. Non-parallelity error [NPE] (a.u.) which is the difference between the largest and smallest errors with respect to DMRG references are also provided.\\}
  \begin{ruledtabular}
  \begin{tabular}{cccccccccc}
    \multirow{2}{*}{R(H-H)}          &\multirow{2}{*}{GF2}          &\multicolumn{3}{c}{SEET(CI/GF2)} &\multirow{2}{*}{NEVPT2(10e,10o)}  &\multirow{2}{*}{DMRG}   \\
    \cline{3-5}
                                     &                                &[2o+2$\times$4o] &[4o+6o] &[2o+8o] &  &           \\
      \hline
      
      1.4             &--5.367 9    &--5.380 6                  &--5.385 6         &--5.388 5         &--5.358 9           &--5.408 7      \\          
      1.6             &--5.525 5    &--5.540 0                  &--5.547 7         &--5.552 9         &--5.520 6           &--5.570 2      \\
      1.8             &--5.564 6    &--5.582 1                  &--5.593 6         &--5.598 8         &--5.570 3           &--5.614 1      \\
      2.0             &--5.539 5    &--5.559 3                  &--5.569 7         &--5.583 2         &--5.551 8           &--5.594 9      \\
      2.4             &--5.403 9    &--5.431 7                  &--5.445 4         &--5.463 2         &--5.442 9           &--5.476 1      \\
      2.8             &--5.235 2    &--5.271 4                  &--5.292 1         &--5.320 4         &--5.305 7           &--5.334 4      \\
      3.2             &--5.074 1    &--5.119 3                  &--5.148 7         &--5.192 9         &--5.185 3           &--5.212 5      \\
      3.6             &--4.935 3    &--4.995 0                  &--5.032 8         &--5.082 2         &--5.095 4           &--5.123 8      \\
      \hline
     NPE               &\,\,\,0.147 7	    &\,\,\,0.100 8	                 &\,\,\,0.068 6	       &\,\,\,0.028 8	      &\,\,\,0.021 5             &               \end{tabular}
  \end{ruledtabular}
\end{table*}

Consequently, one can expect that in cases where the number of active space orbitals is too large to be treated within one impurity, it is possible to split the active space orbitals among several impurities and systematically improve the answer. 
We would like to stress that this systematic improvement will become crucial for systems where the exact answer is unknown, thus the only way of assessing if the level of accuracy given by SEET is sufficient will be coming from internal SEET criteria and checking if the answer obtained does not change drastically upon enlarging the number of impurity orbitals. 

\section{Conclusions}\label{conclusions}

In this paper, we have presented a detailed discussion of the molecular Green's function quantum embedding scheme called SEET.
The self-energy separation characteristic for SEET onto strongly correlated/active/subsystem and weakly correlated/inactive/environment parts is completely general and does not specify how the self-energies for these fragments will be evaluated in practice. While many schemes are possible, in this paper we used a scheme where first the whole molecule is treated by the perturbative self-consistent GF2 approach and then selected strongly correlated orbitals are used to build impurity+bath models that are solved in DMFT-like procedure in the presence of self-energy coming from the weakly correlated/inactive/environment orbitals. 

We aim for SEET to be systematically improvable, without empirical parameters, and reaching chemical accuracy. Consequently, we discussed many aspects of SEET that were developed by us to fulfill these strict demands. SEET is a Green's function method capable not only of delivering ground state energies but also many more physically relevant quantities such as free energies, ionization potentials (IP) and electron affinities (EA), or temperature-dependent magnetic susceptibility; however, here we focused on analyzing SEET results for small molecular examples where multiple ground state methods are known to give excellent results and ground state energies can be easily used to asses the SEET performance.

We started our considerations by explaining the double self-consistency loop present in SEET, where in the inner DMFT-like loop the active space/impurity self-energy is updated using an accurate many-body solver. The outer loop requires an update of the self-energy for inactive orbitals performed by us at the GF2 level; however other inexpensive \textit{ab initio} methods could also be used for the calculation of the inactive orbitals self-energy. We observed, when analyzing numerical results, that this outer loop's self-energy update is crucial for classes of systems where the initial perturbative description was not quantitative. 

Next, we have analyzed different schemes for selecting the strongly correlated/active/subsystem orbitals either based on the energy (occupations) or spatial criteria. 
We stress that in the energy (or occupation) scheme the strongly correlated orbital selection is done mainly based on the occupations of correlated one-body density matrix, thus not only relying on intuitive means. 
Moreover,  in the result section for H$_{10}$ chain, we have demonstrated that the results of such a selection scheme can be systematically improved.
We have also analyzed how different orbital bases, SAO, RNO, and NO influence SEET results. As expected, we found that NO basis, which allows us easily describe a large inter-orbital entanglement, yields very good results for equilibrium geometries; however, in the limit of separated fragment bases that localize orbitals, such as SAOs/RNOs, yield better energies. 

We discussed optimal impurity solvers that can be used with SEET, stressing that a full realistic Hamiltonian should be employed for strongly correlated orbitals if one desires systematically improvable and accurate results. Moreover, we have observed that suitable solvers (to be applicable in multiple regimes from weakly to strongly correlated) must be able to deal with near degeneracy and have to be able to treat multiple strongly correlated impurity orbitals. Finally, we highlighted the influence of the orbital basis onto the performance of impurity solvers by noticing that for molecular examples the NO basis seems to be particularly advantageous for the CT-HYB and RASCI solvers.

In SEET the influence of the non-local interactions on the strongly correlated/active/embedded system is contained in the non-local self-energy evaluated at the perturbative level. We have compared SEET results with DMFT results where the non-local interactions are accounted for by employing the effective on-site interactions $U(\tau)$ present in the impurity model. We found that for molecular examples both methods give almost identical results thus showing that SEET with bare Coulomb interactions present in the impurity model solved in the presence of self-energy coming from the inactive orbitals allows us to avoid the downfolding procedure and constructing an effective Hamiltonian for the impurity model.

Finally, we have performed multiple total energy comparisons against established quantum chemistry methods such as NEVPT2 and DMRG. For calculations in the NO basis and when a full active space is used, SEET can be considered as a Green's function analog of CASPT2 or NEVPT2 procedure. We have demonstrated that when employing a full set of active space orbitals SEET(CI/GF2) yields energies that are very close to NEVPT2. However, unlike CASPT2 or NEVPT2, SEET(CI/GF2)  is a perturb and diagonalize scheme that does not require storing or evaluating of two-, three-, or four-body reduced density matrices and avoids the intruder state problem present in CASPT2. Here, we focused on analyzing how splitting of the full set of active orbitals into several impurities can be done in a systematic manner leading to an improvable and controlled procedure. We have shown that both in the spatial and energy domains, SEET results can be systematically improved by using multi-site impurities. We analyzed the spatial domain using the example of H$_{50}$ chain in STO-6G basis, while the energy domain was examined on the example of H$_{10}$ chain in cc-pVDZ basis. Moreover, we have shown that SEET results in a small non-parallelity error when compared against DMRG and other established multi-reference methods. 
These results that indicate systematic trends provide us with a number of important self-contained assessment tools that in the future can be used to help us rigorously analyze the accuracy of SEET in the absence of known results. 

We believe that the considerations and results presented here are further establishing SEET as a quantum embedding Green's function method that is controlled, systematically improvable, and not only able to reach accuracy comparable to currently established active space quantum chemistry methods but also   flexible enough to yield other spectral and thermodynamic quantities than the ground state energies. 

\section*{Acknowledgments}

T.N.L, A.A.K, and D.Z. would like to acknowledge support from the US Department of Energy (DOE) grant No. ER16391. Authors also thank Ushnish Ray and Garnet Kin-Lic Chan for providing the DMRG data for the H$_{10}$ chain in the cc-pVDZ basis set.


\begin{thebibliography}{83}%
\makeatletter
\providecommand \@ifxundefined [1]{%
 \@ifx{#1\undefined}
}%
\providecommand \@ifnum [1]{%
 \ifnum #1\expandafter \@firstoftwo
 \else \expandafter \@secondoftwo
 \fi
}%
\providecommand \@ifx [1]{%
 \ifx #1\expandafter \@firstoftwo
 \else \expandafter \@secondoftwo
 \fi
}%
\providecommand \natexlab [1]{#1}%
\providecommand \enquote  [1]{``#1''}%
\providecommand \bibnamefont  [1]{#1}%
\providecommand \bibfnamefont [1]{#1}%
\providecommand \citenamefont [1]{#1}%
\providecommand \href@noop [0]{\@secondoftwo}%
\providecommand \href [0]{\begingroup \@sanitize@url \@href}%
\providecommand \@href[1]{\@@startlink{#1}\@@href}%
\providecommand \@@href[1]{\endgroup#1\@@endlink}%
\providecommand \@sanitize@url [0]{\catcode `\\12\catcode `\$12\catcode
  `\&12\catcode `\#12\catcode `\^12\catcode `\_12\catcode `\%12\relax}%
\providecommand \@@startlink[1]{}%
\providecommand \@@endlink[0]{}%
\providecommand \url  [0]{\begingroup\@sanitize@url \@url }%
\providecommand \@url [1]{\endgroup\@href {#1}{\urlprefix }}%
\providecommand \urlprefix  [0]{URL }%
\providecommand \Eprint [0]{\href }%
\providecommand \doibase [0]{http://dx.doi.org/}%
\providecommand \selectlanguage [0]{\@gobble}%
\providecommand \bibinfo  [0]{\@secondoftwo}%
\providecommand \bibfield  [0]{\@secondoftwo}%
\providecommand \translation [1]{[#1]}%
\providecommand \BibitemOpen [0]{}%
\providecommand \bibitemStop [0]{}%
\providecommand \bibitemNoStop [0]{.\EOS\space}%
\providecommand \EOS [0]{\spacefactor3000\relax}%
\providecommand \BibitemShut  [1]{\csname bibitem#1\endcsname}%
\let\auto@bib@innerbib\@empty
\bibitem [{\citenamefont {Ghigo}\ \emph {et~al.}(2004)\citenamefont {Ghigo},
  \citenamefont {Roos},\ and\ \citenamefont {Malmqvist}}]{Ghigo:cpl/396/142}%
  \BibitemOpen
  \bibfield  {author} {\bibinfo {author} {\bibfnamefont {G.}~\bibnamefont
  {Ghigo}}, \bibinfo {author} {\bibfnamefont {B.~O.}\ \bibnamefont {Roos}}, \
  and\ \bibinfo {author} {\bibfnamefont {P.~A.}\ \bibnamefont {Malmqvist}},\
  }\href {\doibase http://dx.doi.org/10.1016/j.cplett.2004.08.032} {\bibfield
  {journal} {\bibinfo  {journal} {Chem. Phys. Lett.}\ }\textbf {\bibinfo
  {volume} {396}},\ \bibinfo {pages} {142} (\bibinfo {year}
  {2004})}\BibitemShut {NoStop}%
\bibitem [{\citenamefont {Andersson}\ \emph {et~al.}(1990)\citenamefont
  {Andersson}, \citenamefont {Malmqvist}, \citenamefont {Roos}, \citenamefont
  {Sadlej},\ and\ \citenamefont {Wolinski}}]{Kerstin:jpc/94/5483}%
  \BibitemOpen
  \bibfield  {author} {\bibinfo {author} {\bibfnamefont {K.}~\bibnamefont
  {Andersson}}, \bibinfo {author} {\bibfnamefont {P.~A.}\ \bibnamefont
  {Malmqvist}}, \bibinfo {author} {\bibfnamefont {B.~O.}\ \bibnamefont {Roos}},
  \bibinfo {author} {\bibfnamefont {A.~J.}\ \bibnamefont {Sadlej}}, \ and\
  \bibinfo {author} {\bibfnamefont {K.}~\bibnamefont {Wolinski}},\ }\href@noop
  {} {\bibfield  {journal} {\bibinfo  {journal} {J. Phys. Chem.}\ }\textbf
  {\bibinfo {volume} {94}},\ \bibinfo {pages} {5483} (\bibinfo {year}
  {1990})}\BibitemShut {NoStop}%
\bibitem [{\citenamefont {Angeli}\ \emph {et~al.}(2001)\citenamefont {Angeli},
  \citenamefont {Cimiraglia}, \citenamefont {Evangelisti}, \citenamefont
  {Leininger},\ and\ \citenamefont {Malrieu}}]{Angeli:jcp/114/10252}%
  \BibitemOpen
  \bibfield  {author} {\bibinfo {author} {\bibfnamefont {C.}~\bibnamefont
  {Angeli}}, \bibinfo {author} {\bibfnamefont {R.}~\bibnamefont {Cimiraglia}},
  \bibinfo {author} {\bibfnamefont {S.}~\bibnamefont {Evangelisti}}, \bibinfo
  {author} {\bibfnamefont {T.}~\bibnamefont {Leininger}}, \ and\ \bibinfo
  {author} {\bibfnamefont {J.-P.}\ \bibnamefont {Malrieu}},\ }\href
  {http://scitation.aip.org/content/aip/journal/jcp/114/23/10.1063/1.1361246}
  {\bibfield  {journal} {\bibinfo  {journal} {J. Chem. Phys.}\ }\textbf
  {\bibinfo {volume} {114}},\ \bibinfo {pages} {10252} (\bibinfo {year}
  {2001})}\BibitemShut {NoStop}%
\bibitem [{\citenamefont {Angeli}\ \emph {et~al.}(2002)\citenamefont {Angeli},
  \citenamefont {Cimiraglia},\ and\ \citenamefont
  {Malrieu}}]{Angeli:jcp/117/9138}%
  \BibitemOpen
  \bibfield  {author} {\bibinfo {author} {\bibfnamefont {C.}~\bibnamefont
  {Angeli}}, \bibinfo {author} {\bibfnamefont {R.}~\bibnamefont {Cimiraglia}},
  \ and\ \bibinfo {author} {\bibfnamefont {J.-P.}\ \bibnamefont {Malrieu}},\
  }\href
  {http://scitation.aip.org/content/aip/journal/jcp/117/20/10.1063/1.1515317}
  {\bibfield  {journal} {\bibinfo  {journal} {J. Chem. Phys.}\ }\textbf
  {\bibinfo {volume} {117}},\ \bibinfo {pages} {9138} (\bibinfo {year}
  {2002})}\BibitemShut {NoStop}%
\bibitem [{\citenamefont {Bartlett}\ and\ \citenamefont
  {Musia\l{}}(2007)}]{Bartlett:rmp/79/291}%
  \BibitemOpen
  \bibfield  {author} {\bibinfo {author} {\bibfnamefont {R.~J.}\ \bibnamefont
  {Bartlett}}\ and\ \bibinfo {author} {\bibfnamefont {M.}~\bibnamefont
  {Musia\l{}}},\ }\href {\doibase 10.1103/RevModPhys.79.291} {\bibfield
  {journal} {\bibinfo  {journal} {Rev. Mod. Phys.}\ }\textbf {\bibinfo {volume}
  {79}},\ \bibinfo {pages} {291} (\bibinfo {year} {2007})}\BibitemShut
  {NoStop}%
\bibitem [{\citenamefont {Jeziorski}\ and\ \citenamefont
  {Monkhorst}(1981)}]{Jeziorski:pra/24/1668}%
  \BibitemOpen
  \bibfield  {author} {\bibinfo {author} {\bibfnamefont {B.}~\bibnamefont
  {Jeziorski}}\ and\ \bibinfo {author} {\bibfnamefont {H.~J.}\ \bibnamefont
  {Monkhorst}},\ }\href {\doibase 10.1103/PhysRevA.24.1668} {\bibfield
  {journal} {\bibinfo  {journal} {Phys. Rev. A}\ }\textbf {\bibinfo {volume}
  {24}},\ \bibinfo {pages} {1668} (\bibinfo {year} {1981})}\BibitemShut
  {NoStop}%
\bibitem [{\citenamefont {Jeziorski}\ and\ \citenamefont
  {Paldus}(1988)}]{Jeziorski:jcp/88/5673}%
  \BibitemOpen
  \bibfield  {author} {\bibinfo {author} {\bibfnamefont {B.}~\bibnamefont
  {Jeziorski}}\ and\ \bibinfo {author} {\bibfnamefont {J.}~\bibnamefont
  {Paldus}},\ }\href {\doibase http://dx.doi.org/10.1063/1.454528} {\bibfield
  {journal} {\bibinfo  {journal} {J. Chem. Phys.}\ }\textbf {\bibinfo {volume}
  {88}},\ \bibinfo {pages} {5673} (\bibinfo {year} {1988})}\BibitemShut
  {NoStop}%
\bibitem [{\citenamefont {Musia\l{}}\ \emph {et~al.}(2011)\citenamefont
  {Musia\l{}}, \citenamefont {Perera},\ and\ \citenamefont
  {Bartlett}}]{Musial:jcp/134/114108}%
  \BibitemOpen
  \bibfield  {author} {\bibinfo {author} {\bibfnamefont {M.}~\bibnamefont
  {Musia\l{}}}, \bibinfo {author} {\bibfnamefont {A.}~\bibnamefont {Perera}}, \
  and\ \bibinfo {author} {\bibfnamefont {R.~J.}\ \bibnamefont {Bartlett}},\
  }\href
  {http://scitation.aip.org/content/aip/journal/jcp/134/11/10.1063/1.3567115}
  {\bibfield  {journal} {\bibinfo  {journal} {J. Chem. Phys.}\ }\textbf
  {\bibinfo {volume} {134}},\ \bibinfo {pages} {114108} (\bibinfo {year}
  {2011})}\BibitemShut {NoStop}%
\bibitem [{\citenamefont {Lindgren}(1978)}]{Lindgren:ijqc/14/1097}%
  \BibitemOpen
  \bibfield  {author} {\bibinfo {author} {\bibfnamefont {I.}~\bibnamefont
  {Lindgren}},\ }\href {\doibase 10.1002/qua.560140804} {\bibfield  {journal}
  {\bibinfo  {journal} {Int. J. Quant. Chem.}\ }\textbf {\bibinfo {volume}
  {14}},\ \bibinfo {pages} {33} (\bibinfo {year} {1978})}\BibitemShut {NoStop}%
\bibitem [{\citenamefont {Mukherjee}\ \emph {et~al.}(1975)\citenamefont
  {Mukherjee}, \citenamefont {Moitra},\ and\ \citenamefont
  {Mukhopadhyay}}]{Mukherjee:mp/30/1861}%
  \BibitemOpen
  \bibfield  {author} {\bibinfo {author} {\bibfnamefont {D.}~\bibnamefont
  {Mukherjee}}, \bibinfo {author} {\bibfnamefont {R.~K.}\ \bibnamefont
  {Moitra}}, \ and\ \bibinfo {author} {\bibfnamefont {A.}~\bibnamefont
  {Mukhopadhyay}},\ }\href {\doibase 10.1080/00268977500103351} {\bibfield
  {journal} {\bibinfo  {journal} {Mol. Phys.}\ }\textbf {\bibinfo {volume}
  {30}},\ \bibinfo {pages} {1861} (\bibinfo {year} {1975})}\BibitemShut
  {NoStop}%
\bibitem [{\citenamefont {Szalay}\ \emph {et~al.}(2012)\citenamefont {Szalay},
  \citenamefont {M\"{u}ller}, \citenamefont {Gidofalvi}, \citenamefont
  {Lischka},\ and\ \citenamefont {Shepard}}]{Szalay:cr/112/108}%
  \BibitemOpen
  \bibfield  {author} {\bibinfo {author} {\bibfnamefont {P.~G.}\ \bibnamefont
  {Szalay}}, \bibinfo {author} {\bibfnamefont {T.}~\bibnamefont {M\"{u}ller}},
  \bibinfo {author} {\bibfnamefont {G.}~\bibnamefont {Gidofalvi}}, \bibinfo
  {author} {\bibfnamefont {H.}~\bibnamefont {Lischka}}, \ and\ \bibinfo
  {author} {\bibfnamefont {R.}~\bibnamefont {Shepard}},\ }\href {\doibase
  10.1021/cr200137a} {\bibfield  {journal} {\bibinfo  {journal} {Chem. Rev.}\
  }\textbf {\bibinfo {volume} {112}},\ \bibinfo {pages} {108} (\bibinfo {year}
  {2012})}\BibitemShut {NoStop}%
\bibitem [{\citenamefont {Gr\"{a}fenstein}\ and\ \citenamefont
  {Cremer}(2000)}]{Grafenstein:cpl/316/569}%
  \BibitemOpen
  \bibfield  {author} {\bibinfo {author} {\bibfnamefont {J.}~\bibnamefont
  {Gr\"{a}fenstein}}\ and\ \bibinfo {author} {\bibfnamefont {D.}~\bibnamefont
  {Cremer}},\ }\href@noop {} {\bibfield  {journal} {\bibinfo  {journal} {Chem.
  Phys. Lett.}\ }\textbf {\bibinfo {volume} {316}},\ \bibinfo {pages} {569}
  (\bibinfo {year} {2000})}\BibitemShut {NoStop}%
\bibitem [{\citenamefont {Gr\"{a}fenstein}\ and\ \citenamefont
  {Cremer}(2005)}]{Grafenstein:mp/103/279}%
  \BibitemOpen
  \bibfield  {author} {\bibinfo {author} {\bibfnamefont {J.}~\bibnamefont
  {Gr\"{a}fenstein}}\ and\ \bibinfo {author} {\bibfnamefont {D.}~\bibnamefont
  {Cremer}},\ }\href@noop {} {\bibfield  {journal} {\bibinfo  {journal} {Mol.
  Phys.}\ }\textbf {\bibinfo {volume} {103}},\ \bibinfo {pages} {279} (\bibinfo
  {year} {2005})}\BibitemShut {NoStop}%
\bibitem [{\citenamefont {Pollet}\ \emph {et~al.}(2002)\citenamefont {Pollet},
  \citenamefont {Savin}, \citenamefont {Leininger},\ and\ \citenamefont
  {Stoll}}]{Pollet:jcp/116/1250}%
  \BibitemOpen
  \bibfield  {author} {\bibinfo {author} {\bibfnamefont {R.}~\bibnamefont
  {Pollet}}, \bibinfo {author} {\bibfnamefont {A.}~\bibnamefont {Savin}},
  \bibinfo {author} {\bibfnamefont {T.}~\bibnamefont {Leininger}}, \ and\
  \bibinfo {author} {\bibfnamefont {H.}~\bibnamefont {Stoll}},\ }\href@noop {}
  {\bibfield  {journal} {\bibinfo  {journal} {J. Chem. Phys.}\ }\textbf
  {\bibinfo {volume} {116}},\ \bibinfo {pages} {1250} (\bibinfo {year}
  {2002})}\BibitemShut {NoStop}%
\bibitem [{\citenamefont {Manni}\ \emph {et~al.}(2014)\citenamefont {Manni},
  \citenamefont {Carlson}, \citenamefont {Luo}, \citenamefont {Ma},
  \citenamefont {Olsen}, \citenamefont {Truhlar},\ and\ \citenamefont
  {Gagliardi}}]{Limanni:jctc/10/3669}%
  \BibitemOpen
  \bibfield  {author} {\bibinfo {author} {\bibfnamefont {G.~L.}\ \bibnamefont
  {Manni}}, \bibinfo {author} {\bibfnamefont {R.~K.}\ \bibnamefont {Carlson}},
  \bibinfo {author} {\bibfnamefont {S.}~\bibnamefont {Luo}}, \bibinfo {author}
  {\bibfnamefont {D.}~\bibnamefont {Ma}}, \bibinfo {author} {\bibfnamefont
  {J.}~\bibnamefont {Olsen}}, \bibinfo {author} {\bibfnamefont {D.~G.}\
  \bibnamefont {Truhlar}}, \ and\ \bibinfo {author} {\bibfnamefont
  {L.}~\bibnamefont {Gagliardi}},\ }\href@noop {} {\bibfield  {journal}
  {\bibinfo  {journal} {J. Chem. Theory Comput.}\ }\textbf {\bibinfo {volume}
  {10}},\ \bibinfo {pages} {3669} (\bibinfo {year} {2014})}\BibitemShut
  {NoStop}%
\bibitem [{\citenamefont {Carlson}\ \emph {et~al.}(2015)\citenamefont
  {Carlson}, \citenamefont {Truhlar},\ and\ \citenamefont
  {Gagliardi}}]{Carlson:jctc/11/4077}%
  \BibitemOpen
  \bibfield  {author} {\bibinfo {author} {\bibfnamefont {R.~K.}\ \bibnamefont
  {Carlson}}, \bibinfo {author} {\bibfnamefont {D.~G.}\ \bibnamefont
  {Truhlar}}, \ and\ \bibinfo {author} {\bibfnamefont {L.}~\bibnamefont
  {Gagliardi}},\ }\href@noop {} {\bibfield  {journal} {\bibinfo  {journal} {J.
  Chem. Theory Comput.}\ }\textbf {\bibinfo {volume} {11}},\ \bibinfo {pages}
  {4077} (\bibinfo {year} {2015})}\BibitemShut {NoStop}%
\bibitem [{\citenamefont {Georges}\ \emph {et~al.}(1996)\citenamefont
  {Georges}, \citenamefont {Kotliar}, \citenamefont {Krauth},\ and\
  \citenamefont {Rozenberg}}]{Georges:rmp/68/13}%
  \BibitemOpen
  \bibfield  {author} {\bibinfo {author} {\bibfnamefont {A.}~\bibnamefont
  {Georges}}, \bibinfo {author} {\bibfnamefont {G.}~\bibnamefont {Kotliar}},
  \bibinfo {author} {\bibfnamefont {W.}~\bibnamefont {Krauth}}, \ and\ \bibinfo
  {author} {\bibfnamefont {M.~J.}\ \bibnamefont {Rozenberg}},\ }\href {\doibase
  10.1103/RevModPhys.68.13} {\bibfield  {journal} {\bibinfo  {journal} {Rev.
  Mod. Phys.}\ }\textbf {\bibinfo {volume} {68}},\ \bibinfo {pages} {13}
  (\bibinfo {year} {1996})}\BibitemShut {NoStop}%
\bibitem [{\citenamefont {Kotliar}\ \emph {et~al.}(2006)\citenamefont
  {Kotliar}, \citenamefont {Savrasov}, \citenamefont {Haule}, \citenamefont
  {Oudovenko}, \citenamefont {Parcollet},\ and\ \citenamefont
  {Marianetti}}]{Kotliar:rmp/78/865}%
  \BibitemOpen
  \bibfield  {author} {\bibinfo {author} {\bibfnamefont {G.}~\bibnamefont
  {Kotliar}}, \bibinfo {author} {\bibfnamefont {S.~Y.}\ \bibnamefont
  {Savrasov}}, \bibinfo {author} {\bibfnamefont {K.}~\bibnamefont {Haule}},
  \bibinfo {author} {\bibfnamefont {V.~S.}\ \bibnamefont {Oudovenko}}, \bibinfo
  {author} {\bibfnamefont {O.}~\bibnamefont {Parcollet}}, \ and\ \bibinfo
  {author} {\bibfnamefont {C.~A.}\ \bibnamefont {Marianetti}},\ }\href@noop {}
  {\bibfield  {journal} {\bibinfo  {journal} {Rev. Mod. Phys.}\ }\textbf
  {\bibinfo {volume} {78}},\ \bibinfo {pages} {865} (\bibinfo {year}
  {2006})}\BibitemShut {NoStop}%
\bibitem [{\citenamefont {Maier}\ \emph {et~al.}(2005)\citenamefont {Maier},
  \citenamefont {Jarrell}, \citenamefont {Pruschke},\ and\ \citenamefont
  {Hettler}}]{Maier:rmp/77/1027}%
  \BibitemOpen
  \bibfield  {author} {\bibinfo {author} {\bibfnamefont {T.}~\bibnamefont
  {Maier}}, \bibinfo {author} {\bibfnamefont {M.}~\bibnamefont {Jarrell}},
  \bibinfo {author} {\bibfnamefont {T.}~\bibnamefont {Pruschke}}, \ and\
  \bibinfo {author} {\bibfnamefont {M.~H.}\ \bibnamefont {Hettler}},\ }\href
  {\doibase 10.1103/RevModPhys.77.1027} {\bibfield  {journal} {\bibinfo
  {journal} {Rev. Mod. Phys.}\ }\textbf {\bibinfo {volume} {77}},\ \bibinfo
  {pages} {1027} (\bibinfo {year} {2005})}\BibitemShut {NoStop}%
\bibitem [{\citenamefont {Hedin}(1965)}]{Hedin:pr/139/A796}%
  \BibitemOpen
  \bibfield  {author} {\bibinfo {author} {\bibfnamefont {L.}~\bibnamefont
  {Hedin}},\ }\href {\doibase 10.1103/PhysRev.139.A796} {\bibfield  {journal}
  {\bibinfo  {journal} {Phys. Rev.}\ }\textbf {\bibinfo {volume} {139}},\
  \bibinfo {pages} {A796} (\bibinfo {year} {1965})}\BibitemShut {NoStop}%
\bibitem [{\citenamefont {Biermann}\ \emph {et~al.}(2003)\citenamefont
  {Biermann}, \citenamefont {Aryasetiawan},\ and\ \citenamefont
  {Georges}}]{Biermann:prl/90/086402}%
  \BibitemOpen
  \bibfield  {author} {\bibinfo {author} {\bibfnamefont {S.}~\bibnamefont
  {Biermann}}, \bibinfo {author} {\bibfnamefont {F.}~\bibnamefont
  {Aryasetiawan}}, \ and\ \bibinfo {author} {\bibfnamefont {A.}~\bibnamefont
  {Georges}},\ }\href {\doibase 10.1103/PhysRevLett.90.086402} {\bibfield
  {journal} {\bibinfo  {journal} {Phys. Rev. Lett.}\ }\textbf {\bibinfo
  {volume} {90}},\ \bibinfo {pages} {086402} (\bibinfo {year}
  {2003})}\BibitemShut {NoStop}%
\bibitem [{\citenamefont {Tomczak}\ \emph {et~al.}(2012)\citenamefont
  {Tomczak}, \citenamefont {Casula}, \citenamefont {Miyake}, \citenamefont
  {Aryasetiawan},\ and\ \citenamefont {Biermann}}]{Tomczak:epl/100/67001}%
  \BibitemOpen
  \bibfield  {author} {\bibinfo {author} {\bibfnamefont {J.~M.}\ \bibnamefont
  {Tomczak}}, \bibinfo {author} {\bibfnamefont {M.}~\bibnamefont {Casula}},
  \bibinfo {author} {\bibfnamefont {T.}~\bibnamefont {Miyake}}, \bibinfo
  {author} {\bibfnamefont {F.}~\bibnamefont {Aryasetiawan}}, \ and\ \bibinfo
  {author} {\bibfnamefont {S.}~\bibnamefont {Biermann}},\ }\href
  {http://stacks.iop.org/0295-5075/100/i=6/a=67001} {\bibfield  {journal}
  {\bibinfo  {journal} {EPL}\ }\textbf {\bibinfo {volume} {100}},\ \bibinfo
  {pages} {67001} (\bibinfo {year} {2012})}\BibitemShut {NoStop}%
\bibitem [{\citenamefont {Kananenka}\ \emph {et~al.}(2015)\citenamefont
  {Kananenka}, \citenamefont {Gull},\ and\ \citenamefont
  {Zgid}}]{Kananenka:prb/91/121111}%
  \BibitemOpen
  \bibfield  {author} {\bibinfo {author} {\bibfnamefont {A.~A.}\ \bibnamefont
  {Kananenka}}, \bibinfo {author} {\bibfnamefont {E.}~\bibnamefont {Gull}}, \
  and\ \bibinfo {author} {\bibfnamefont {D.}~\bibnamefont {Zgid}},\ }\href
  {\doibase 10.1103/PhysRevB.91.121111} {\bibfield  {journal} {\bibinfo
  {journal} {Phys. Rev. B}\ }\textbf {\bibinfo {volume} {91}},\ \bibinfo
  {pages} {121111} (\bibinfo {year} {2015})}\BibitemShut {NoStop}%
\bibitem [{\citenamefont {Lan}\ \emph {et~al.}(2015)\citenamefont {Lan},
  \citenamefont {Kananenka},\ and\ \citenamefont {Zgid}}]{Lan:jcp/143/241102}%
  \BibitemOpen
  \bibfield  {author} {\bibinfo {author} {\bibfnamefont {T.~N.}\ \bibnamefont
  {Lan}}, \bibinfo {author} {\bibfnamefont {A.~A.}\ \bibnamefont {Kananenka}},
  \ and\ \bibinfo {author} {\bibfnamefont {D.}~\bibnamefont {Zgid}},\ }\href
  {http://scitation.aip.org/content/aip/journal/jcp/143/24/10.1063/1.4938562}
  {\bibfield  {journal} {\bibinfo  {journal} {J. Chem. Phys.}\ }\textbf
  {\bibinfo {volume} {143}},\ \bibinfo {pages} {241102} (\bibinfo {year}
  {2015})}\BibitemShut {NoStop}%
\bibitem [{\citenamefont {Phillips}\ and\ \citenamefont
  {Zgid}(2014)}]{Phillips:jcp/140/241101}%
  \BibitemOpen
  \bibfield  {author} {\bibinfo {author} {\bibfnamefont {J.~J.}\ \bibnamefont
  {Phillips}}\ and\ \bibinfo {author} {\bibfnamefont {D.}~\bibnamefont
  {Zgid}},\ }\href
  {http://scitation.aip.org/content/aip/journal/jcp/140/24/10.1063/1.4884951}
  {\bibfield  {journal} {\bibinfo  {journal} {J. Chem. Phys.}\ }\textbf
  {\bibinfo {volume} {140}},\ \bibinfo {pages} {241101} (\bibinfo {year}
  {2014})}\BibitemShut {NoStop}%
\bibitem [{\citenamefont {Dahlen}\ and\ \citenamefont {van
  Leeuwen}(2005)}]{Dahlen:jcp/122/164102}%
  \BibitemOpen
  \bibfield  {author} {\bibinfo {author} {\bibfnamefont {N.~E.}\ \bibnamefont
  {Dahlen}}\ and\ \bibinfo {author} {\bibfnamefont {R.}~\bibnamefont {van
  Leeuwen}},\ }\href@noop {} {\bibfield  {journal} {\bibinfo  {journal} {J.
  Chem. Phys.}\ }\textbf {\bibinfo {volume} {122}},\ \bibinfo {pages} {164102}
  (\bibinfo {year} {2005})}\BibitemShut {NoStop}%
\bibitem [{\citenamefont {Rusakov}\ and\ \citenamefont
  {Zgid}(2016)}]{Rusakov:jcp/144/054106}%
  \BibitemOpen
  \bibfield  {author} {\bibinfo {author} {\bibfnamefont {A.~A.}\ \bibnamefont
  {Rusakov}}\ and\ \bibinfo {author} {\bibfnamefont {D.}~\bibnamefont {Zgid}},\
  }\href
  {http://scitation.aip.org/content/aip/journal/jcp/144/5/10.1063/1.4940900}
  {\bibfield  {journal} {\bibinfo  {journal} {J. Chem. Phys.}\ }\textbf
  {\bibinfo {volume} {144}},\ \bibinfo {pages} {054106} (\bibinfo {year}
  {2016})}\BibitemShut {NoStop}%
\bibitem [{\citenamefont {Welden}\ \emph {et~al.}()\citenamefont {Welden},
  \citenamefont {Rusakov},\ and\ \citenamefont {Zgid}}]{Welden_arxiv}%
  \BibitemOpen
  \bibfield  {author} {\bibinfo {author} {\bibfnamefont {A.~R.}\ \bibnamefont
  {Welden}}, \bibinfo {author} {\bibfnamefont {A.}~\bibnamefont {Rusakov}}, \
  and\ \bibinfo {author} {\bibfnamefont {D.}~\bibnamefont {Zgid}},\ }\href@noop
  {} {\ }\bibinfo {note} {{e}-print arXiv:1605.06563
  [physics.chem-ph]}\BibitemShut {NoStop}%
\bibitem [{\citenamefont {Zgid}\ and\ \citenamefont
  {Chan}(2011)}]{Zgid:jcp/134/094115}%
  \BibitemOpen
  \bibfield  {author} {\bibinfo {author} {\bibfnamefont {D.}~\bibnamefont
  {Zgid}}\ and\ \bibinfo {author} {\bibfnamefont {G.~K.-L.}\ \bibnamefont
  {Chan}},\ }\href@noop {} {\bibfield  {journal} {\bibinfo  {journal} {J. Chem.
  Phys.}\ }\textbf {\bibinfo {volume} {134}},\ \bibinfo {pages} {094115}
  (\bibinfo {year} {2011})}\BibitemShut {NoStop}%
\bibitem [{\citenamefont {Zgid}\ \emph {et~al.}(2012)\citenamefont {Zgid},
  \citenamefont {Gull},\ and\ \citenamefont {Chan}}]{Zgid:prb/86/165128}%
  \BibitemOpen
  \bibfield  {author} {\bibinfo {author} {\bibfnamefont {D.}~\bibnamefont
  {Zgid}}, \bibinfo {author} {\bibfnamefont {E.}~\bibnamefont {Gull}}, \ and\
  \bibinfo {author} {\bibfnamefont {G.~K.-L.}\ \bibnamefont {Chan}},\ }\href
  {\doibase 10.1103/PhysRevB.86.165128} {\bibfield  {journal} {\bibinfo
  {journal} {Phys. Rev. B}\ }\textbf {\bibinfo {volume} {86}},\ \bibinfo
  {pages} {165128} (\bibinfo {year} {2012})}\BibitemShut {NoStop}%
\bibitem [{\citenamefont {Gull}\ \emph {et~al.}(2008)\citenamefont {Gull},
  \citenamefont {Werner}, \citenamefont {Parcollet},\ and\ \citenamefont
  {Troyer}}]{Gull:el/82/57003}%
  \BibitemOpen
  \bibfield  {author} {\bibinfo {author} {\bibfnamefont {E.}~\bibnamefont
  {Gull}}, \bibinfo {author} {\bibfnamefont {P.}~\bibnamefont {Werner}},
  \bibinfo {author} {\bibfnamefont {O.}~\bibnamefont {Parcollet}}, \ and\
  \bibinfo {author} {\bibfnamefont {M.}~\bibnamefont {Troyer}},\ }\href
  {http://stacks.iop.org/0295-5075/82/i=5/a=57003} {\bibfield  {journal}
  {\bibinfo  {journal} {Europhys. Lett.}\ }\textbf {\bibinfo {volume} {82}},\
  \bibinfo {pages} {57003} (\bibinfo {year} {2008})}\BibitemShut {NoStop}%
\bibitem [{\citenamefont {Gull}\ \emph
  {et~al.}(2011{\natexlab{a}})\citenamefont {Gull}, \citenamefont {Staar},
  \citenamefont {Fuchs}, \citenamefont {Nukala}, \citenamefont {Summers},
  \citenamefont {Pruschke}, \citenamefont {Schulthess},\ and\ \citenamefont
  {Maier}}]{Gull:prb/83/075122}%
  \BibitemOpen
  \bibfield  {author} {\bibinfo {author} {\bibfnamefont {E.}~\bibnamefont
  {Gull}}, \bibinfo {author} {\bibfnamefont {P.}~\bibnamefont {Staar}},
  \bibinfo {author} {\bibfnamefont {S.}~\bibnamefont {Fuchs}}, \bibinfo
  {author} {\bibfnamefont {P.}~\bibnamefont {Nukala}}, \bibinfo {author}
  {\bibfnamefont {M.~S.}\ \bibnamefont {Summers}}, \bibinfo {author}
  {\bibfnamefont {T.}~\bibnamefont {Pruschke}}, \bibinfo {author}
  {\bibfnamefont {T.~C.}\ \bibnamefont {Schulthess}}, \ and\ \bibinfo {author}
  {\bibfnamefont {T.}~\bibnamefont {Maier}},\ }\href {\doibase
  10.1103/PhysRevB.83.075122} {\bibfield  {journal} {\bibinfo  {journal} {Phys.
  Rev. B}\ }\textbf {\bibinfo {volume} {83}},\ \bibinfo {pages} {075122}
  (\bibinfo {year} {2011}{\natexlab{a}})}\BibitemShut {NoStop}%
\bibitem [{\citenamefont {White}(1992)}]{White:prl/69/2863}%
  \BibitemOpen
  \bibfield  {author} {\bibinfo {author} {\bibfnamefont {S.~R.}\ \bibnamefont
  {White}},\ }\href {\doibase 10.1103/PhysRevLett.69.2863} {\bibfield
  {journal} {\bibinfo  {journal} {Phys. Rev. Lett.}\ }\textbf {\bibinfo
  {volume} {69}},\ \bibinfo {pages} {2863} (\bibinfo {year}
  {1992})}\BibitemShut {NoStop}%
\bibitem [{\citenamefont {White}\ and\ \citenamefont
  {Martin}(1999)}]{White:jcp/110/4127}%
  \BibitemOpen
  \bibfield  {author} {\bibinfo {author} {\bibfnamefont {S.~R.}\ \bibnamefont
  {White}}\ and\ \bibinfo {author} {\bibfnamefont {R.~L.}\ \bibnamefont
  {Martin}},\ }\href {\doibase http://dx.doi.org/10.1063/1.478295} {\bibfield
  {journal} {\bibinfo  {journal} {J. Chem. Phys.}\ }\textbf {\bibinfo {volume}
  {110}},\ \bibinfo {pages} {4127} (\bibinfo {year} {1999})}\BibitemShut
  {NoStop}%
\bibitem [{\citenamefont {Chan}\ and\ \citenamefont
  {Head-Gordon}(2003)}]{Chan_dmrg_jcp2003}%
  \BibitemOpen
  \bibfield  {author} {\bibinfo {author} {\bibfnamefont {G.~K.-L.}\
  \bibnamefont {Chan}}\ and\ \bibinfo {author} {\bibfnamefont {M.}~\bibnamefont
  {Head-Gordon}},\ }\href {\doibase http://dx.doi.org/10.1063/1.1574318}
  {\bibfield  {journal} {\bibinfo  {journal} {J. Chem. Phys.}\ }\textbf
  {\bibinfo {volume} {118}},\ \bibinfo {pages} {8551} (\bibinfo {year}
  {2003})}\BibitemShut {NoStop}%
\bibitem [{\citenamefont {Chan}(2004)}]{Chan_dmrg_jcp2004}%
  \BibitemOpen
  \bibfield  {author} {\bibinfo {author} {\bibfnamefont {G.~K.-L.}\
  \bibnamefont {Chan}},\ }\href {\doibase http://dx.doi.org/10.1063/1.1638734}
  {\bibfield  {journal} {\bibinfo  {journal} {J. Chem. Phys.}\ }\textbf
  {\bibinfo {volume} {120}},\ \bibinfo {pages} {3172} (\bibinfo {year}
  {2004})}\BibitemShut {NoStop}%
\bibitem [{\citenamefont {Zgid}\ and\ \citenamefont
  {Nooijen}(2008)}]{Zgid:jcp/128/144116}%
  \BibitemOpen
  \bibfield  {author} {\bibinfo {author} {\bibfnamefont {D.}~\bibnamefont
  {Zgid}}\ and\ \bibinfo {author} {\bibfnamefont {M.}~\bibnamefont {Nooijen}},\
  }\href
  {http://scitation.aip.org/content/aip/journal/jcp/128/14/10.1063/1.2883981}
  {\bibfield  {journal} {\bibinfo  {journal} {J. Chem. Phys.}\ }\textbf
  {\bibinfo {volume} {128}},\ \bibinfo {pages} {144116} (\bibinfo {year}
  {2008})}\BibitemShut {NoStop}%
\bibitem [{\citenamefont {Kurashige}\ and\ \citenamefont
  {Yanai}(2009)}]{Kurashige:jcp/130/234114}%
  \BibitemOpen
  \bibfield  {author} {\bibinfo {author} {\bibfnamefont {Y.}~\bibnamefont
  {Kurashige}}\ and\ \bibinfo {author} {\bibfnamefont {T.}~\bibnamefont
  {Yanai}},\ }\href
  {http://scitation.aip.org/content/aip/journal/jcp/130/23/10.1063/1.3152576}
  {\bibfield  {journal} {\bibinfo  {journal} {J. Chem. Phys.}\ }\textbf
  {\bibinfo {volume} {130}},\ \bibinfo {pages} {234114} (\bibinfo {year}
  {2009})}\BibitemShut {NoStop}%
\bibitem [{\citenamefont {Marti}\ and\ \citenamefont
  {Reiher}(2011)}]{Marti:pccp/13/6750}%
  \BibitemOpen
  \bibfield  {author} {\bibinfo {author} {\bibfnamefont {K.~H.}\ \bibnamefont
  {Marti}}\ and\ \bibinfo {author} {\bibfnamefont {M.}~\bibnamefont {Reiher}},\
  }\href {\doibase 10.1039/C0CP01883J} {\bibfield  {journal} {\bibinfo
  {journal} {Phys. Chem. Chem. Phys.}\ }\textbf {\bibinfo {volume} {13}},\
  \bibinfo {pages} {6750} (\bibinfo {year} {2011})}\BibitemShut {NoStop}%
\bibitem [{\citenamefont {Bickers}\ and\ \citenamefont
  {Scalapino}(1989)}]{Bickers:ap/193/206}%
  \BibitemOpen
  \bibfield  {author} {\bibinfo {author} {\bibfnamefont {N.}~\bibnamefont
  {Bickers}}\ and\ \bibinfo {author} {\bibfnamefont {D.}~\bibnamefont
  {Scalapino}},\ }\href {\doibase
  http://dx.doi.org/10.1016/0003-4916(89)90359-X} {\bibfield  {journal}
  {\bibinfo  {journal} {Ann. Phys.}\ }\textbf {\bibinfo {volume} {193}},\
  \bibinfo {pages} {206} (\bibinfo {year} {1989})}\BibitemShut {NoStop}%
\bibitem [{\citenamefont {Bickers}\ \emph {et~al.}(1989)\citenamefont
  {Bickers}, \citenamefont {Scalapino},\ and\ \citenamefont
  {White}}]{Bickers:prl/62/961}%
  \BibitemOpen
  \bibfield  {author} {\bibinfo {author} {\bibfnamefont {N.~E.}\ \bibnamefont
  {Bickers}}, \bibinfo {author} {\bibfnamefont {D.~J.}\ \bibnamefont
  {Scalapino}}, \ and\ \bibinfo {author} {\bibfnamefont {S.~R.}\ \bibnamefont
  {White}},\ }\href {\doibase 10.1103/PhysRevLett.62.961} {\bibfield  {journal}
  {\bibinfo  {journal} {Phys. Rev. Lett.}\ }\textbf {\bibinfo {volume} {62}},\
  \bibinfo {pages} {961} (\bibinfo {year} {1989})}\BibitemShut {NoStop}%
\bibitem [{\citenamefont {Anderson}(1961)}]{Anderson:pr/124/41}%
  \BibitemOpen
  \bibfield  {author} {\bibinfo {author} {\bibfnamefont {P.~W.}\ \bibnamefont
  {Anderson}},\ }\href {\doibase 10.1103/PhysRev.124.41} {\bibfield  {journal}
  {\bibinfo  {journal} {Phys. Rev.}\ }\textbf {\bibinfo {volume} {124}},\
  \bibinfo {pages} {41} (\bibinfo {year} {1961})}\BibitemShut {NoStop}%
\bibitem [{\citenamefont {Davidson}(1975)}]{Davidson:jcp/17/87}%
  \BibitemOpen
  \bibfield  {author} {\bibinfo {author} {\bibfnamefont {E.~R.}\ \bibnamefont
  {Davidson}},\ }\href {\doibase
  http://dx.doi.org/10.1016/0021-9991(75)90065-0} {\bibfield  {journal}
  {\bibinfo  {journal} {J. Comput. Phys.}\ }\textbf {\bibinfo {volume} {17}},\
  \bibinfo {pages} {87} (\bibinfo {year} {1975})}\BibitemShut {NoStop}%
\bibitem [{\citenamefont {Capone}\ \emph {et~al.}(2007)\citenamefont {Capone},
  \citenamefont {de' Medici},\ and\ \citenamefont
  {Georges}}]{Capone:prb/76/245116}%
  \BibitemOpen
  \bibfield  {author} {\bibinfo {author} {\bibfnamefont {M.}~\bibnamefont
  {Capone}}, \bibinfo {author} {\bibfnamefont {L.}~\bibnamefont {de' Medici}},
  \ and\ \bibinfo {author} {\bibfnamefont {A.}~\bibnamefont {Georges}},\ }\href
  {\doibase 10.1103/PhysRevB.76.245116} {\bibfield  {journal} {\bibinfo
  {journal} {Phys. Rev. B}\ }\textbf {\bibinfo {volume} {76}},\ \bibinfo
  {pages} {245116} (\bibinfo {year} {2007})}\BibitemShut {NoStop}%
\bibitem [{\citenamefont {Lin}\ \emph {et~al.}(1993)\citenamefont {Lin},
  \citenamefont {Gubernatis}, \citenamefont {Gould},\ and\ \citenamefont
  {Tobochnik}}]{Lin:cp/7/400}%
  \BibitemOpen
  \bibfield  {author} {\bibinfo {author} {\bibfnamefont {H.}~\bibnamefont
  {Lin}}, \bibinfo {author} {\bibfnamefont {J.}~\bibnamefont {Gubernatis}},
  \bibinfo {author} {\bibfnamefont {H.}~\bibnamefont {Gould}}, \ and\ \bibinfo
  {author} {\bibfnamefont {J.}~\bibnamefont {Tobochnik}},\ }\href {\doibase
  http://dx.doi.org/10.1063/1.4823192} {\bibfield  {journal} {\bibinfo
  {journal} {Comput. Phys.}\ }\textbf {\bibinfo {volume} {7}},\ \bibinfo
  {pages} {400} (\bibinfo {year} {1993})}\BibitemShut {NoStop}%
\bibitem [{\citenamefont {Liebsch}\ and\ \citenamefont
  {Ishida}(2012)}]{Liebsch:jpcm/24/053201}%
  \BibitemOpen
  \bibfield  {author} {\bibinfo {author} {\bibfnamefont {A.}~\bibnamefont
  {Liebsch}}\ and\ \bibinfo {author} {\bibfnamefont {H.}~\bibnamefont
  {Ishida}},\ }\href {http://stacks.iop.org/0953-8984/24/i=5/a=053201}
  {\bibfield  {journal} {\bibinfo  {journal} {J. Phys.: Condens. Matter}\
  }\textbf {\bibinfo {volume} {24}},\ \bibinfo {pages} {053201} (\bibinfo
  {year} {2012})}\BibitemShut {NoStop}%
\bibitem [{\citenamefont {Kananenka}\ \emph
  {et~al.}(2016{\natexlab{a}})\citenamefont {Kananenka}, \citenamefont
  {Phillips},\ and\ \citenamefont {Zgid}}]{Kananenka:jctc/12/564}%
  \BibitemOpen
  \bibfield  {author} {\bibinfo {author} {\bibfnamefont {A.~A.}\ \bibnamefont
  {Kananenka}}, \bibinfo {author} {\bibfnamefont {J.~J.}\ \bibnamefont
  {Phillips}}, \ and\ \bibinfo {author} {\bibfnamefont {D.}~\bibnamefont
  {Zgid}},\ }\href {\doibase 10.1021/acs.jctc.5b00884} {\bibfield  {journal}
  {\bibinfo  {journal} {J. Chem. Theory Comput.}\ }\textbf {\bibinfo {volume}
  {12}},\ \bibinfo {pages} {564} (\bibinfo {year}
  {2016}{\natexlab{a}})}\BibitemShut {NoStop}%
\bibitem [{\citenamefont {Galitskii}(1958)}]{Galitskii:zetp/34/151}%
  \BibitemOpen
  \bibfield  {author} {\bibinfo {author} {\bibfnamefont {V.~M.}\ \bibnamefont
  {Galitskii}},\ }\href@noop {} {\bibfield  {journal} {\bibinfo  {journal} {Zh.
  Eksp. Teor. Fiz.}\ }\textbf {\bibinfo {volume} {34}},\ \bibinfo {pages} {151}
  (\bibinfo {year} {1958})}\BibitemShut {NoStop}%
\bibitem [{\citenamefont {Gukelberger}\ \emph {et~al.}(2015)\citenamefont
  {Gukelberger}, \citenamefont {Huang},\ and\ \citenamefont
  {Werner}}]{Gukelberger:prb/91/235114}%
  \BibitemOpen
  \bibfield  {author} {\bibinfo {author} {\bibfnamefont {J.}~\bibnamefont
  {Gukelberger}}, \bibinfo {author} {\bibfnamefont {L.}~\bibnamefont {Huang}},
  \ and\ \bibinfo {author} {\bibfnamefont {P.}~\bibnamefont {Werner}},\ }\href
  {\doibase 10.1103/PhysRevB.91.235114} {\bibfield  {journal} {\bibinfo
  {journal} {Phys. Rev. B}\ }\textbf {\bibinfo {volume} {91}},\ \bibinfo
  {pages} {235114} (\bibinfo {year} {2015})}\BibitemShut {NoStop}%
\bibitem [{\citenamefont {Kananenka}\ \emph
  {et~al.}(2016{\natexlab{b}})\citenamefont {Kananenka}, \citenamefont
  {Welden}, \citenamefont {Lan}, \citenamefont {Gull},\ and\ \citenamefont
  {Zgid}}]{Kananenka:jctc/12/2250}%
  \BibitemOpen
  \bibfield  {author} {\bibinfo {author} {\bibfnamefont {A.~A.}\ \bibnamefont
  {Kananenka}}, \bibinfo {author} {\bibfnamefont {A.~R.}\ \bibnamefont
  {Welden}}, \bibinfo {author} {\bibfnamefont {T.~N.}\ \bibnamefont {Lan}},
  \bibinfo {author} {\bibfnamefont {E.}~\bibnamefont {Gull}}, \ and\ \bibinfo
  {author} {\bibfnamefont {D.}~\bibnamefont {Zgid}},\ }\href {\doibase
  10.1021/acs.jctc.6b00178} {\bibfield  {journal} {\bibinfo  {journal} {J.
  Chem. Theory Comput.}\ }\textbf {\bibinfo {volume} {12}},\ \bibinfo {pages}
  {2250} (\bibinfo {year} {2016}{\natexlab{b}})}\BibitemShut {NoStop}%
\bibitem [{\citenamefont {Cococcioni}\ and\ \citenamefont
  {de~Gironcoli}(2005)}]{Cococcioni:prb/71/035105}%
  \BibitemOpen
  \bibfield  {author} {\bibinfo {author} {\bibfnamefont {M.}~\bibnamefont
  {Cococcioni}}\ and\ \bibinfo {author} {\bibfnamefont {S.}~\bibnamefont
  {de~Gironcoli}},\ }\href {\doibase 10.1103/PhysRevB.71.035105} {\bibfield
  {journal} {\bibinfo  {journal} {Phys. Rev. B}\ }\textbf {\bibinfo {volume}
  {71}},\ \bibinfo {pages} {035105} (\bibinfo {year} {2005})}\BibitemShut
  {NoStop}%
\bibitem [{\citenamefont {Aryasetiawan}\ \emph {et~al.}(2004)\citenamefont
  {Aryasetiawan}, \citenamefont {Imada}, \citenamefont {Georges}, \citenamefont
  {Kotliar}, \citenamefont {Biermann},\ and\ \citenamefont
  {Lichtenstein}}]{Aryasetiawan:prb/70195104}%
  \BibitemOpen
  \bibfield  {author} {\bibinfo {author} {\bibfnamefont {F.}~\bibnamefont
  {Aryasetiawan}}, \bibinfo {author} {\bibfnamefont {M.}~\bibnamefont {Imada}},
  \bibinfo {author} {\bibfnamefont {A.}~\bibnamefont {Georges}}, \bibinfo
  {author} {\bibfnamefont {G.}~\bibnamefont {Kotliar}}, \bibinfo {author}
  {\bibfnamefont {S.}~\bibnamefont {Biermann}}, \ and\ \bibinfo {author}
  {\bibfnamefont {A.~I.}\ \bibnamefont {Lichtenstein}},\ }\href {\doibase
  10.1103/PhysRevB.70.195104} {\bibfield  {journal} {\bibinfo  {journal} {Phys.
  Rev. B}\ }\textbf {\bibinfo {volume} {70}},\ \bibinfo {pages} {195104}
  (\bibinfo {year} {2004})}\BibitemShut {NoStop}%
\bibitem [{\citenamefont {Werner}\ and\ \citenamefont
  {Millis}(2010)}]{werner:prl2010}%
  \BibitemOpen
  \bibfield  {author} {\bibinfo {author} {\bibfnamefont {P.}~\bibnamefont
  {Werner}}\ and\ \bibinfo {author} {\bibfnamefont {A.~J.}\ \bibnamefont
  {Millis}},\ }\href {\doibase 10.1103/PhysRevLett.104.146401} {\bibfield
  {journal} {\bibinfo  {journal} {Phys. Rev. Lett.}\ }\textbf {\bibinfo
  {volume} {104}},\ \bibinfo {pages} {146401} (\bibinfo {year}
  {2010})}\BibitemShut {NoStop}%
\bibitem [{\citenamefont {Werner}\ and\ \citenamefont
  {Casula}()}]{werner2016dynamical}%
  \BibitemOpen
  \bibfield  {author} {\bibinfo {author} {\bibfnamefont {P.}~\bibnamefont
  {Werner}}\ and\ \bibinfo {author} {\bibfnamefont {M.}~\bibnamefont
  {Casula}},\ }\href@noop {} {\ }\bibinfo {note} {{e}-print arXiv:1602.00584
  [cond-mat.str-el]}\BibitemShut {NoStop}%
\bibitem [{\citenamefont {Rusakov}\ \emph {et~al.}(2014)\citenamefont
  {Rusakov}, \citenamefont {Phillips},\ and\ \citenamefont
  {Zgid}}]{Rusakov:jcp/141/194105}%
  \BibitemOpen
  \bibfield  {author} {\bibinfo {author} {\bibfnamefont {A.~A.}\ \bibnamefont
  {Rusakov}}, \bibinfo {author} {\bibfnamefont {J.~J.}\ \bibnamefont
  {Phillips}}, \ and\ \bibinfo {author} {\bibfnamefont {D.}~\bibnamefont
  {Zgid}},\ }\href@noop {} {\bibfield  {journal} {\bibinfo  {journal} {J. Chem.
  Phys.}\ }\textbf {\bibinfo {volume} {141}},\ \bibinfo {pages} {194105}
  (\bibinfo {year} {2014})}\BibitemShut {NoStop}%
\bibitem [{\citenamefont {Pipek}\ and\ \citenamefont
  {Mezey}(1989)}]{Pipek:jcp/90/4916}%
  \BibitemOpen
  \bibfield  {author} {\bibinfo {author} {\bibfnamefont {J.}~\bibnamefont
  {Pipek}}\ and\ \bibinfo {author} {\bibfnamefont {P.~G.}\ \bibnamefont
  {Mezey}},\ }\href@noop {} {\bibfield  {journal} {\bibinfo  {journal} {J.
  Chem. Phys.}\ }\textbf {\bibinfo {volume} {90}},\ \bibinfo {pages} {4916}
  (\bibinfo {year} {1989})}\BibitemShut {NoStop}%
\bibitem [{\citenamefont {Boys}(1960)}]{Boys:rmp/32/296}%
  \BibitemOpen
  \bibfield  {author} {\bibinfo {author} {\bibfnamefont {S.~F.}\ \bibnamefont
  {Boys}},\ }\href {\doibase 10.1103/RevModPhys.32.296} {\bibfield  {journal}
  {\bibinfo  {journal} {Rev. Mod. Phys.}\ }\textbf {\bibinfo {volume} {32}},\
  \bibinfo {pages} {296} (\bibinfo {year} {1960})}\BibitemShut {NoStop}%
\bibitem [{\citenamefont {Marzari}\ and\ \citenamefont
  {Vanderbilt}(1997)}]{Marzari:prb/56/12847}%
  \BibitemOpen
  \bibfield  {author} {\bibinfo {author} {\bibfnamefont {N.}~\bibnamefont
  {Marzari}}\ and\ \bibinfo {author} {\bibfnamefont {D.}~\bibnamefont
  {Vanderbilt}},\ }\href {\doibase 10.1103/PhysRevB.56.12847} {\bibfield
  {journal} {\bibinfo  {journal} {Phys. Rev. B}\ }\textbf {\bibinfo {volume}
  {56}},\ \bibinfo {pages} {12847} (\bibinfo {year} {1997})}\BibitemShut
  {NoStop}%
\bibitem [{\citenamefont {de~Silva}\ \emph {et~al.}(2012)\citenamefont
  {de~Silva}, \citenamefont {Giebultowski},\ and\ \citenamefont
  {Korchowiec}}]{Desilva:pccp/14/546}%
  \BibitemOpen
  \bibfield  {author} {\bibinfo {author} {\bibfnamefont {P.}~\bibnamefont
  {de~Silva}}, \bibinfo {author} {\bibfnamefont {M.}~\bibnamefont
  {Giebultowski}}, \ and\ \bibinfo {author} {\bibfnamefont {J.}~\bibnamefont
  {Korchowiec}},\ }\href@noop {} {\bibfield  {journal} {\bibinfo  {journal}
  {Phys. Chem. Chem. Phys.}\ }\textbf {\bibinfo {volume} {14}},\ \bibinfo
  {pages} {546} (\bibinfo {year} {2012})}\BibitemShut {NoStop}%
\bibitem [{\citenamefont {Gu}\ \emph {et~al.}(2004)\citenamefont {Gu},
  \citenamefont {Aoki}, \citenamefont {Korchowiec}, \citenamefont {Imamura},\
  and\ \citenamefont {Kirtman}}]{Gu:jpc2006-RNO}%
  \BibitemOpen
  \bibfield  {author} {\bibinfo {author} {\bibfnamefont {F.~L.}\ \bibnamefont
  {Gu}}, \bibinfo {author} {\bibfnamefont {Y.}~\bibnamefont {Aoki}}, \bibinfo
  {author} {\bibfnamefont {J.}~\bibnamefont {Korchowiec}}, \bibinfo {author}
  {\bibfnamefont {A.}~\bibnamefont {Imamura}}, \ and\ \bibinfo {author}
  {\bibfnamefont {B.}~\bibnamefont {Kirtman}},\ }\href {\doibase
  http://dx.doi.org/10.1063/1.1812736} {\bibfield  {journal} {\bibinfo
  {journal} {J. Chem. Phys.}\ }\textbf {\bibinfo {volume} {121}},\ \bibinfo
  {pages} {10385} (\bibinfo {year} {2004})}\BibitemShut {NoStop}%
\bibitem [{\citenamefont {Lin}\ \emph {et~al.}(2011)\citenamefont {Lin},
  \citenamefont {Marianetti}, \citenamefont {Millis},\ and\ \citenamefont
  {Reichman}}]{Lin:prl2011-dmft}%
  \BibitemOpen
  \bibfield  {author} {\bibinfo {author} {\bibfnamefont {N.}~\bibnamefont
  {Lin}}, \bibinfo {author} {\bibfnamefont {C.~A.}\ \bibnamefont {Marianetti}},
  \bibinfo {author} {\bibfnamefont {A.~J.}\ \bibnamefont {Millis}}, \ and\
  \bibinfo {author} {\bibfnamefont {D.~R.}\ \bibnamefont {Reichman}},\ }\href
  {\doibase 10.1103/PhysRevLett.106.096402} {\bibfield  {journal} {\bibinfo
  {journal} {Phys. Rev. Lett.}\ }\textbf {\bibinfo {volume} {106}},\ \bibinfo
  {pages} {096402} (\bibinfo {year} {2011})}\BibitemShut {NoStop}%
\bibitem [{\citenamefont {Werner}\ \emph {et~al.}(2006)\citenamefont {Werner},
  \citenamefont {Comanac}, \citenamefont {de' Medici}, \citenamefont {Troyer},\
  and\ \citenamefont {Millis}}]{Werner:prl/97/076405}%
  \BibitemOpen
  \bibfield  {author} {\bibinfo {author} {\bibfnamefont {P.}~\bibnamefont
  {Werner}}, \bibinfo {author} {\bibfnamefont {A.}~\bibnamefont {Comanac}},
  \bibinfo {author} {\bibfnamefont {L.}~\bibnamefont {de' Medici}}, \bibinfo
  {author} {\bibfnamefont {M.}~\bibnamefont {Troyer}}, \ and\ \bibinfo {author}
  {\bibfnamefont {A.~J.}\ \bibnamefont {Millis}},\ }\href {\doibase
  10.1103/PhysRevLett.97.076405} {\bibfield  {journal} {\bibinfo  {journal}
  {Phys. Rev. Lett.}\ }\textbf {\bibinfo {volume} {97}},\ \bibinfo {pages}
  {076405} (\bibinfo {year} {2006})}\BibitemShut {NoStop}%
\bibitem [{\citenamefont {Werner}\ and\ \citenamefont
  {Millis}(2006)}]{Werner:prb/74/155107}%
  \BibitemOpen
  \bibfield  {author} {\bibinfo {author} {\bibfnamefont {P.}~\bibnamefont
  {Werner}}\ and\ \bibinfo {author} {\bibfnamefont {A.~J.}\ \bibnamefont
  {Millis}},\ }\href {\doibase 10.1103/PhysRevB.74.155107} {\bibfield
  {journal} {\bibinfo  {journal} {Phys. Rev. B}\ }\textbf {\bibinfo {volume}
  {74}},\ \bibinfo {pages} {155107} (\bibinfo {year} {2006})}\BibitemShut
  {NoStop}%
\bibitem [{\citenamefont {Gull}()}]{triqs_ctqmc_gull}%
  \BibitemOpen
  \bibfield  {author} {\bibinfo {author} {\bibfnamefont {E.}~\bibnamefont
  {Gull}},\ }\href {\doibase 10.3929/ethz-a-005722583} {\
  10.3929/ethz-a-005722583},\ \bibinfo {note} {ph.D. thesis, ETH {Z\"{u}rich},
  {\bf{2008}}}\BibitemShut {NoStop}%
\bibitem [{\citenamefont {Gull}\ \emph
  {et~al.}(2011{\natexlab{b}})\citenamefont {Gull}, \citenamefont {Millis},
  \citenamefont {Lichtenstein}, \citenamefont {Rubtsov}, \citenamefont
  {Troyer},\ and\ \citenamefont {Werner}}]{Gull:rmp2011}%
  \BibitemOpen
  \bibfield  {author} {\bibinfo {author} {\bibfnamefont {E.}~\bibnamefont
  {Gull}}, \bibinfo {author} {\bibfnamefont {A.~J.}\ \bibnamefont {Millis}},
  \bibinfo {author} {\bibfnamefont {A.~I.}\ \bibnamefont {Lichtenstein}},
  \bibinfo {author} {\bibfnamefont {A.~N.}\ \bibnamefont {Rubtsov}}, \bibinfo
  {author} {\bibfnamefont {M.}~\bibnamefont {Troyer}}, \ and\ \bibinfo {author}
  {\bibfnamefont {P.}~\bibnamefont {Werner}},\ }\href {\doibase
  10.1103/RevModPhys.83.349} {\bibfield  {journal} {\bibinfo  {journal} {Rev.
  Mod. Phys.}\ }\textbf {\bibinfo {volume} {83}},\ \bibinfo {pages} {349}
  (\bibinfo {year} {2011}{\natexlab{b}})}\BibitemShut {NoStop}%
\bibitem [{\citenamefont {Hehre}\ \emph {et~al.}(1972)\citenamefont {Hehre},
  \citenamefont {Ditchfield},\ and\ \citenamefont {Pople}}]{Hehre:jcp/56/2257}%
  \BibitemOpen
  \bibfield  {author} {\bibinfo {author} {\bibfnamefont {W.~J.}\ \bibnamefont
  {Hehre}}, \bibinfo {author} {\bibfnamefont {R.}~\bibnamefont {Ditchfield}}, \
  and\ \bibinfo {author} {\bibfnamefont {J.~A.}\ \bibnamefont {Pople}},\ }\href
  {\doibase http://dx.doi.org/10.1063/1.1677527} {\bibfield  {journal}
  {\bibinfo  {journal} {J. Chem. Phys.}\ }\textbf {\bibinfo {volume} {56}},\
  \bibinfo {pages} {2257} (\bibinfo {year} {1972})}\BibitemShut {NoStop}%
\bibitem [{\citenamefont {Frisch}\ \emph {et~al.}()\citenamefont {Frisch},
  \citenamefont {Trucks}, \citenamefont {Schlegel}, \citenamefont {Scuseria},
  \citenamefont {Robb}, \citenamefont {Cheeseman}, \citenamefont {Scalmani},
  \citenamefont {Barone}, \citenamefont {Mennucci}, \citenamefont {Petersson},
  \citenamefont {Nakatsuji}, \citenamefont {Caricato}, \citenamefont {Li},
  \citenamefont {Hratchian}, \citenamefont {Izmaylov}, \citenamefont {Bloino},
  \citenamefont {Zheng}, \citenamefont {Sonnenberg}, \citenamefont {Hada},
  \citenamefont {Ehara}, \citenamefont {Toyota}, \citenamefont {Fukuda},
  \citenamefont {Hasegawa}, \citenamefont {Ishida}, \citenamefont {Nakajima},
  \citenamefont {Honda}, \citenamefont {Kitao}, \citenamefont {Nakai},
  \citenamefont {Vreven}, \citenamefont {Montgomery}, \citenamefont {Peralta},
  \citenamefont {Ogliaro}, \citenamefont {Bearpark}, \citenamefont {Heyd},
  \citenamefont {Brothers}, \citenamefont {Kudin}, \citenamefont {Staroverov},
  \citenamefont {Kobayashi}, \citenamefont {Normand}, \citenamefont
  {Raghavachari}, \citenamefont {Rendell}, \citenamefont {Burant},
  \citenamefont {Iyengar}, \citenamefont {Tomasi}, \citenamefont {Cossi},
  \citenamefont {Rega}, \citenamefont {Millam}, \citenamefont {Klene},
  \citenamefont {Knox}, \citenamefont {Cross}, \citenamefont {Bakken},
  \citenamefont {Adamo}, \citenamefont {Jaramillo}, \citenamefont {Gomperts},
  \citenamefont {Stratmann}, \citenamefont {Yazyev}, \citenamefont {Austin},
  \citenamefont {Cammi}, \citenamefont {Pomelli}, \citenamefont {Ochterski},
  \citenamefont {Martin}, \citenamefont {Morokuma}, \citenamefont {Zakrzewski},
  \citenamefont {Voth}, \citenamefont {Salvador}, \citenamefont {Dannenberg},
  \citenamefont {Dapprich}, \citenamefont {Daniels}, \citenamefont {Farkas},
  \citenamefont {Foresman}, \citenamefont {Ortiz}, \citenamefont {Cioslowski},\
  and\ \citenamefont {Fox}}]{g09}%
  \BibitemOpen
  \bibfield  {author} {\bibinfo {author} {\bibfnamefont {M.~J.}\ \bibnamefont
  {Frisch}}, \bibinfo {author} {\bibfnamefont {G.~W.}\ \bibnamefont {Trucks}},
  \bibinfo {author} {\bibfnamefont {H.~B.}\ \bibnamefont {Schlegel}}, \bibinfo
  {author} {\bibfnamefont {G.~E.}\ \bibnamefont {Scuseria}}, \bibinfo {author}
  {\bibfnamefont {M.~A.}\ \bibnamefont {Robb}}, \bibinfo {author}
  {\bibfnamefont {J.~R.}\ \bibnamefont {Cheeseman}}, \bibinfo {author}
  {\bibfnamefont {G.}~\bibnamefont {Scalmani}}, \bibinfo {author}
  {\bibfnamefont {V.}~\bibnamefont {Barone}}, \bibinfo {author} {\bibfnamefont
  {B.}~\bibnamefont {Mennucci}}, \bibinfo {author} {\bibfnamefont {G.~A.}\
  \bibnamefont {Petersson}}, \bibinfo {author} {\bibfnamefont {H.}~\bibnamefont
  {Nakatsuji}}, \bibinfo {author} {\bibfnamefont {M.}~\bibnamefont {Caricato}},
  \bibinfo {author} {\bibfnamefont {X.}~\bibnamefont {Li}}, \bibinfo {author}
  {\bibfnamefont {H.~P.}\ \bibnamefont {Hratchian}}, \bibinfo {author}
  {\bibfnamefont {A.~F.}\ \bibnamefont {Izmaylov}}, \bibinfo {author}
  {\bibfnamefont {J.}~\bibnamefont {Bloino}}, \bibinfo {author} {\bibfnamefont
  {G.}~\bibnamefont {Zheng}}, \bibinfo {author} {\bibfnamefont {J.~L.}\
  \bibnamefont {Sonnenberg}}, \bibinfo {author} {\bibfnamefont
  {M.}~\bibnamefont {Hada}}, \bibinfo {author} {\bibfnamefont {M.}~\bibnamefont
  {Ehara}}, \bibinfo {author} {\bibfnamefont {K.}~\bibnamefont {Toyota}},
  \bibinfo {author} {\bibfnamefont {R.}~\bibnamefont {Fukuda}}, \bibinfo
  {author} {\bibfnamefont {J.}~\bibnamefont {Hasegawa}}, \bibinfo {author}
  {\bibfnamefont {M.}~\bibnamefont {Ishida}}, \bibinfo {author} {\bibfnamefont
  {T.}~\bibnamefont {Nakajima}}, \bibinfo {author} {\bibfnamefont
  {Y.}~\bibnamefont {Honda}}, \bibinfo {author} {\bibfnamefont
  {O.}~\bibnamefont {Kitao}}, \bibinfo {author} {\bibfnamefont
  {H.}~\bibnamefont {Nakai}}, \bibinfo {author} {\bibfnamefont
  {T.}~\bibnamefont {Vreven}}, \bibinfo {author} {\bibfnamefont {J.~A.}\
  \bibnamefont {Montgomery}, \bibfnamefont {{Jr.}}}, \bibinfo {author}
  {\bibfnamefont {J.~E.}\ \bibnamefont {Peralta}}, \bibinfo {author}
  {\bibfnamefont {F.}~\bibnamefont {Ogliaro}}, \bibinfo {author} {\bibfnamefont
  {M.}~\bibnamefont {Bearpark}}, \bibinfo {author} {\bibfnamefont {J.~J.}\
  \bibnamefont {Heyd}}, \bibinfo {author} {\bibfnamefont {E.}~\bibnamefont
  {Brothers}}, \bibinfo {author} {\bibfnamefont {K.~N.}\ \bibnamefont {Kudin}},
  \bibinfo {author} {\bibfnamefont {V.~N.}\ \bibnamefont {Staroverov}},
  \bibinfo {author} {\bibfnamefont {R.}~\bibnamefont {Kobayashi}}, \bibinfo
  {author} {\bibfnamefont {J.}~\bibnamefont {Normand}}, \bibinfo {author}
  {\bibfnamefont {K.}~\bibnamefont {Raghavachari}}, \bibinfo {author}
  {\bibfnamefont {A.}~\bibnamefont {Rendell}}, \bibinfo {author} {\bibfnamefont
  {J.~C.}\ \bibnamefont {Burant}}, \bibinfo {author} {\bibfnamefont {S.~S.}\
  \bibnamefont {Iyengar}}, \bibinfo {author} {\bibfnamefont {J.}~\bibnamefont
  {Tomasi}}, \bibinfo {author} {\bibfnamefont {M.}~\bibnamefont {Cossi}},
  \bibinfo {author} {\bibfnamefont {N.}~\bibnamefont {Rega}}, \bibinfo {author}
  {\bibfnamefont {J.~M.}\ \bibnamefont {Millam}}, \bibinfo {author}
  {\bibfnamefont {M.}~\bibnamefont {Klene}}, \bibinfo {author} {\bibfnamefont
  {J.~E.}\ \bibnamefont {Knox}}, \bibinfo {author} {\bibfnamefont {J.~B.}\
  \bibnamefont {Cross}}, \bibinfo {author} {\bibfnamefont {V.}~\bibnamefont
  {Bakken}}, \bibinfo {author} {\bibfnamefont {C.}~\bibnamefont {Adamo}},
  \bibinfo {author} {\bibfnamefont {J.}~\bibnamefont {Jaramillo}}, \bibinfo
  {author} {\bibfnamefont {R.}~\bibnamefont {Gomperts}}, \bibinfo {author}
  {\bibfnamefont {R.~E.}\ \bibnamefont {Stratmann}}, \bibinfo {author}
  {\bibfnamefont {O.}~\bibnamefont {Yazyev}}, \bibinfo {author} {\bibfnamefont
  {A.~J.}\ \bibnamefont {Austin}}, \bibinfo {author} {\bibfnamefont
  {R.}~\bibnamefont {Cammi}}, \bibinfo {author} {\bibfnamefont
  {C.}~\bibnamefont {Pomelli}}, \bibinfo {author} {\bibfnamefont {J.~W.}\
  \bibnamefont {Ochterski}}, \bibinfo {author} {\bibfnamefont {R.~L.}\
  \bibnamefont {Martin}}, \bibinfo {author} {\bibfnamefont {K.}~\bibnamefont
  {Morokuma}}, \bibinfo {author} {\bibfnamefont {V.~G.}\ \bibnamefont
  {Zakrzewski}}, \bibinfo {author} {\bibfnamefont {G.~A.}\ \bibnamefont
  {Voth}}, \bibinfo {author} {\bibfnamefont {P.}~\bibnamefont {Salvador}},
  \bibinfo {author} {\bibfnamefont {J.~J.}\ \bibnamefont {Dannenberg}},
  \bibinfo {author} {\bibfnamefont {S.}~\bibnamefont {Dapprich}}, \bibinfo
  {author} {\bibfnamefont {A.~D.}\ \bibnamefont {Daniels}}, \bibinfo {author}
  {\bibfnamefont {{\"{O}}.}~\bibnamefont {Farkas}}, \bibinfo {author}
  {\bibfnamefont {J.~B.}\ \bibnamefont {Foresman}}, \bibinfo {author}
  {\bibfnamefont {J.~V.}\ \bibnamefont {Ortiz}}, \bibinfo {author}
  {\bibfnamefont {J.}~\bibnamefont {Cioslowski}}, \ and\ \bibinfo {author}
  {\bibfnamefont {D.~J.}\ \bibnamefont {Fox}},\ }\href@noop {} {\enquote
  {\bibinfo {title} {Gaussian∼09 {R}evision {A}.02},}\ }\bibinfo {note}
  {Gaussian Inc. Wallingford CT 2009}\BibitemShut {NoStop}%
\bibitem [{\citenamefont {Hehre}\ \emph {et~al.}(1969)\citenamefont {Hehre},
  \citenamefont {Stewart},\ and\ \citenamefont {Pople}}]{sto_minimal}%
  \BibitemOpen
  \bibfield  {author} {\bibinfo {author} {\bibfnamefont {W.~J.}\ \bibnamefont
  {Hehre}}, \bibinfo {author} {\bibfnamefont {R.~F.}\ \bibnamefont {Stewart}},
  \ and\ \bibinfo {author} {\bibfnamefont {J.~A.}\ \bibnamefont {Pople}},\
  }\href {\doibase http://dx.doi.org/10.1063/1.1672392} {\bibfield  {journal}
  {\bibinfo  {journal} {J. Chem. Phys.}\ }\textbf {\bibinfo {volume} {51}},\
  \bibinfo {pages} {2657} (\bibinfo {year} {1969})}\BibitemShut {NoStop}%
\bibitem [{\citenamefont {S\'emon}\ and\ \citenamefont
  {Tremblay}(2012)}]{Semon:prb/85/201101}%
  \BibitemOpen
  \bibfield  {author} {\bibinfo {author} {\bibfnamefont {P.}~\bibnamefont
  {S\'emon}}\ and\ \bibinfo {author} {\bibfnamefont {A.-M.~S.}\ \bibnamefont
  {Tremblay}},\ }\href {\doibase 10.1103/PhysRevB.85.201101} {\bibfield
  {journal} {\bibinfo  {journal} {Phys. Rev. B}\ }\textbf {\bibinfo {volume}
  {85}},\ \bibinfo {pages} {201101} (\bibinfo {year} {2012})}\BibitemShut
  {NoStop}%
\bibitem [{\citenamefont {Seth}\ \emph {et~al.}(2016)\citenamefont {Seth},
  \citenamefont {Krivenko}, \citenamefont {Ferrero},\ and\ \citenamefont
  {Parcollet}}]{Seth:cpc/200/274}%
  \BibitemOpen
  \bibfield  {author} {\bibinfo {author} {\bibfnamefont {P.}~\bibnamefont
  {Seth}}, \bibinfo {author} {\bibfnamefont {I.}~\bibnamefont {Krivenko}},
  \bibinfo {author} {\bibfnamefont {M.}~\bibnamefont {Ferrero}}, \ and\
  \bibinfo {author} {\bibfnamefont {O.}~\bibnamefont {Parcollet}},\ }\href
  {\doibase http://dx.doi.org/10.1016/j.cpc.2015.10.023} {\bibfield  {journal}
  {\bibinfo  {journal} {Comp. Phys. Comm.}\ }\textbf {\bibinfo {volume}
  {200}},\ \bibinfo {pages} {274} (\bibinfo {year} {2016})}\BibitemShut
  {NoStop}%
\bibitem [{\citenamefont {Neese}(2012)}]{orca}%
  \BibitemOpen
  \bibfield  {author} {\bibinfo {author} {\bibfnamefont {F.}~\bibnamefont
  {Neese}},\ }\href {\doibase 10.1002/wcms.81} {\bibfield  {journal} {\bibinfo
  {journal} {Wiley Interdiscip. Rev.: Comput. Mol. Sci.}\ }\textbf {\bibinfo
  {volume} {2}},\ \bibinfo {pages} {73} (\bibinfo {year} {2012})}\BibitemShut
  {NoStop}%
\bibitem [{\citenamefont {Roos}\ \emph {et~al.}(1980)\citenamefont {Roos},
  \citenamefont {Taylor},\ and\ \citenamefont {Si≐gbahn}}]{Roos:cp/48/157}%
  \BibitemOpen
  \bibfield  {author} {\bibinfo {author} {\bibfnamefont {B.~O.}\ \bibnamefont
  {Roos}}, \bibinfo {author} {\bibfnamefont {P.~R.}\ \bibnamefont {Taylor}}, \
  and\ \bibinfo {author} {\bibfnamefont {P.~E.}\ \bibnamefont {Si≐gbahn}},\
  }\href {\doibase http://dx.doi.org/10.1016/0301-0104(80)80045-0} {\bibfield
  {journal} {\bibinfo  {journal} {Chem. Phys.}\ }\textbf {\bibinfo {volume}
  {48}},\ \bibinfo {pages} {157} (\bibinfo {year} {1980})}\BibitemShut
  {NoStop}%
\bibitem [{\citenamefont {Roos}(1980)}]{Roos:ijqc/18/175}%
  \BibitemOpen
  \bibfield  {author} {\bibinfo {author} {\bibfnamefont {B.~O.}\ \bibnamefont
  {Roos}},\ }\href {\doibase 10.1002/qua.560180822} {\bibfield  {journal}
  {\bibinfo  {journal} {Int. J. Quant. Chem.}\ }\textbf {\bibinfo {volume}
  {18}},\ \bibinfo {pages} {175} (\bibinfo {year} {1980})}\BibitemShut
  {NoStop}%
\bibitem [{\citenamefont {Siegbahn}\ \emph {et~al.}(1981)\citenamefont
  {Siegbahn}, \citenamefont {Alml{\"{o}}f}, \citenamefont {Heiberg},\ and\
  \citenamefont {Roos}}]{Siegbahn:jcp/74/2384}%
  \BibitemOpen
  \bibfield  {author} {\bibinfo {author} {\bibfnamefont {P.~E.~M.}\
  \bibnamefont {Siegbahn}}, \bibinfo {author} {\bibfnamefont {J.}~\bibnamefont
  {Alml{\"{o}}f}}, \bibinfo {author} {\bibfnamefont {A.}~\bibnamefont
  {Heiberg}}, \ and\ \bibinfo {author} {\bibfnamefont {B.~O.}\ \bibnamefont
  {Roos}},\ }\href {\doibase http://dx.doi.org/10.1063/1.441359} {\bibfield
  {journal} {\bibinfo  {journal} {J. Chem. Phys.}\ }\textbf {\bibinfo {volume}
  {74}},\ \bibinfo {pages} {2384} (\bibinfo {year} {1981})}\BibitemShut
  {NoStop}%
\bibitem [{\citenamefont {Aidas}\ \emph {et~al.}(2014)\citenamefont {Aidas},
  \citenamefont {Angeli}, \citenamefont {Bak}, \citenamefont {Bakken},
  \citenamefont {Bast}, \citenamefont {Boman}, \citenamefont {Christiansen},
  \citenamefont {Cimiraglia}, \citenamefont {Coriani}, \citenamefont {Dahle},
  \citenamefont {Dalskov}, \citenamefont {Ekström}, \citenamefont
  {Enevoldsen}, \citenamefont {Eriksen}, \citenamefont {Ettenhuber},
  \citenamefont {Fernández}, \citenamefont {Ferrighi}, \citenamefont {Fliegl},
  \citenamefont {Frediani}, \citenamefont {Hald}, \citenamefont {Halkier},
  \citenamefont {Hättig}, \citenamefont {Heiberg}, \citenamefont {Helgaker},
  \citenamefont {Hennum}, \citenamefont {Hettema}, \citenamefont {Hjertenæs},
  \citenamefont {Høst}, \citenamefont {Høyvik}, \citenamefont {Iozzi},
  \citenamefont {Jansík}, \citenamefont {Jensen}, \citenamefont {Jonsson},
  \citenamefont {Jørgensen}, \citenamefont {Kauczor}, \citenamefont
  {Kirpekar}, \citenamefont {Kjærgaard}, \citenamefont {Klopper},
  \citenamefont {Knecht}, \citenamefont {Kobayashi}, \citenamefont {Koch},
  \citenamefont {Kongsted}, \citenamefont {Krapp}, \citenamefont {Kristensen},
  \citenamefont {Ligabue}, \citenamefont {Lutnæs}, \citenamefont {Melo},
  \citenamefont {Mikkelsen}, \citenamefont {Myhre}, \citenamefont {Neiss},
  \citenamefont {Nielsen}, \citenamefont {Norman}, \citenamefont {Olsen},
  \citenamefont {Olsen}, \citenamefont {Osted}, \citenamefont {Packer},
  \citenamefont {Pawlowski}, \citenamefont {Pedersen}, \citenamefont {Provasi},
  \citenamefont {Reine}, \citenamefont {Rinkevicius}, \citenamefont {Ruden},
  \citenamefont {Ruud}, \citenamefont {Rybkin}, \citenamefont {Sałek},
  \citenamefont {Samson}, \citenamefont {de~Merás}, \citenamefont {Saue},
  \citenamefont {Sauer}, \citenamefont {Schimmelpfennig}, \citenamefont
  {Sneskov}, \citenamefont {Steindal}, \citenamefont {Sylvester-Hvid},
  \citenamefont {Taylor}, \citenamefont {Teale}, \citenamefont {Tellgren},
  \citenamefont {Tew}, \citenamefont {Thorvaldsen}, \citenamefont {Thøgersen},
  \citenamefont {Vahtras}, \citenamefont {Watson}, \citenamefont {Wilson},
  \citenamefont {Ziolkowski},\ and\ \citenamefont {Ågren}}]{dalton}%
  \BibitemOpen
  \bibfield  {author} {\bibinfo {author} {\bibfnamefont {K.}~\bibnamefont
  {Aidas}}, \bibinfo {author} {\bibfnamefont {C.}~\bibnamefont {Angeli}},
  \bibinfo {author} {\bibfnamefont {K.~L.}\ \bibnamefont {Bak}}, \bibinfo
  {author} {\bibfnamefont {V.}~\bibnamefont {Bakken}}, \bibinfo {author}
  {\bibfnamefont {R.}~\bibnamefont {Bast}}, \bibinfo {author} {\bibfnamefont
  {L.}~\bibnamefont {Boman}}, \bibinfo {author} {\bibfnamefont
  {O.}~\bibnamefont {Christiansen}}, \bibinfo {author} {\bibfnamefont
  {R.}~\bibnamefont {Cimiraglia}}, \bibinfo {author} {\bibfnamefont
  {S.}~\bibnamefont {Coriani}}, \bibinfo {author} {\bibfnamefont
  {P.}~\bibnamefont {Dahle}}, \bibinfo {author} {\bibfnamefont {E.~K.}\
  \bibnamefont {Dalskov}}, \bibinfo {author} {\bibfnamefont {U.}~\bibnamefont
  {Ekström}}, \bibinfo {author} {\bibfnamefont {T.}~\bibnamefont
  {Enevoldsen}}, \bibinfo {author} {\bibfnamefont {J.~J.}\ \bibnamefont
  {Eriksen}}, \bibinfo {author} {\bibfnamefont {P.}~\bibnamefont {Ettenhuber}},
  \bibinfo {author} {\bibfnamefont {B.}~\bibnamefont {Fernández}}, \bibinfo
  {author} {\bibfnamefont {L.}~\bibnamefont {Ferrighi}}, \bibinfo {author}
  {\bibfnamefont {H.}~\bibnamefont {Fliegl}}, \bibinfo {author} {\bibfnamefont
  {L.}~\bibnamefont {Frediani}}, \bibinfo {author} {\bibfnamefont
  {K.}~\bibnamefont {Hald}}, \bibinfo {author} {\bibfnamefont {A.}~\bibnamefont
  {Halkier}}, \bibinfo {author} {\bibfnamefont {C.}~\bibnamefont {Hättig}},
  \bibinfo {author} {\bibfnamefont {H.}~\bibnamefont {Heiberg}}, \bibinfo
  {author} {\bibfnamefont {T.}~\bibnamefont {Helgaker}}, \bibinfo {author}
  {\bibfnamefont {A.~C.}\ \bibnamefont {Hennum}}, \bibinfo {author}
  {\bibfnamefont {H.}~\bibnamefont {Hettema}}, \bibinfo {author} {\bibfnamefont
  {E.}~\bibnamefont {Hjertenæs}}, \bibinfo {author} {\bibfnamefont
  {S.}~\bibnamefont {Høst}}, \bibinfo {author} {\bibfnamefont {I.-M.}\
  \bibnamefont {Høyvik}}, \bibinfo {author} {\bibfnamefont {M.~F.}\
  \bibnamefont {Iozzi}}, \bibinfo {author} {\bibfnamefont {B.}~\bibnamefont
  {Jansík}}, \bibinfo {author} {\bibfnamefont {H.~J.~A.}\ \bibnamefont
  {Jensen}}, \bibinfo {author} {\bibfnamefont {D.}~\bibnamefont {Jonsson}},
  \bibinfo {author} {\bibfnamefont {P.}~\bibnamefont {Jørgensen}}, \bibinfo
  {author} {\bibfnamefont {J.}~\bibnamefont {Kauczor}}, \bibinfo {author}
  {\bibfnamefont {S.}~\bibnamefont {Kirpekar}}, \bibinfo {author}
  {\bibfnamefont {T.}~\bibnamefont {Kjærgaard}}, \bibinfo {author}
  {\bibfnamefont {W.}~\bibnamefont {Klopper}}, \bibinfo {author} {\bibfnamefont
  {S.}~\bibnamefont {Knecht}}, \bibinfo {author} {\bibfnamefont
  {R.}~\bibnamefont {Kobayashi}}, \bibinfo {author} {\bibfnamefont
  {H.}~\bibnamefont {Koch}}, \bibinfo {author} {\bibfnamefont {J.}~\bibnamefont
  {Kongsted}}, \bibinfo {author} {\bibfnamefont {A.}~\bibnamefont {Krapp}},
  \bibinfo {author} {\bibfnamefont {K.}~\bibnamefont {Kristensen}}, \bibinfo
  {author} {\bibfnamefont {A.}~\bibnamefont {Ligabue}}, \bibinfo {author}
  {\bibfnamefont {O.~B.}\ \bibnamefont {Lutnæs}}, \bibinfo {author}
  {\bibfnamefont {J.~I.}\ \bibnamefont {Melo}}, \bibinfo {author}
  {\bibfnamefont {K.~V.}\ \bibnamefont {Mikkelsen}}, \bibinfo {author}
  {\bibfnamefont {R.~H.}\ \bibnamefont {Myhre}}, \bibinfo {author}
  {\bibfnamefont {C.}~\bibnamefont {Neiss}}, \bibinfo {author} {\bibfnamefont
  {C.~B.}\ \bibnamefont {Nielsen}}, \bibinfo {author} {\bibfnamefont
  {P.}~\bibnamefont {Norman}}, \bibinfo {author} {\bibfnamefont
  {J.}~\bibnamefont {Olsen}}, \bibinfo {author} {\bibfnamefont {J.~M.~H.}\
  \bibnamefont {Olsen}}, \bibinfo {author} {\bibfnamefont {A.}~\bibnamefont
  {Osted}}, \bibinfo {author} {\bibfnamefont {M.~J.}\ \bibnamefont {Packer}},
  \bibinfo {author} {\bibfnamefont {F.}~\bibnamefont {Pawlowski}}, \bibinfo
  {author} {\bibfnamefont {T.~B.}\ \bibnamefont {Pedersen}}, \bibinfo {author}
  {\bibfnamefont {P.~F.}\ \bibnamefont {Provasi}}, \bibinfo {author}
  {\bibfnamefont {S.}~\bibnamefont {Reine}}, \bibinfo {author} {\bibfnamefont
  {Z.}~\bibnamefont {Rinkevicius}}, \bibinfo {author} {\bibfnamefont {T.~A.}\
  \bibnamefont {Ruden}}, \bibinfo {author} {\bibfnamefont {K.}~\bibnamefont
  {Ruud}}, \bibinfo {author} {\bibfnamefont {V.~V.}\ \bibnamefont {Rybkin}},
  \bibinfo {author} {\bibfnamefont {P.}~\bibnamefont {Sałek}}, \bibinfo
  {author} {\bibfnamefont {C.~C.~M.}\ \bibnamefont {Samson}}, \bibinfo {author}
  {\bibfnamefont {A.~S.}\ \bibnamefont {de~Merás}}, \bibinfo {author}
  {\bibfnamefont {T.}~\bibnamefont {Saue}}, \bibinfo {author} {\bibfnamefont
  {S.~P.~A.}\ \bibnamefont {Sauer}}, \bibinfo {author} {\bibfnamefont
  {B.}~\bibnamefont {Schimmelpfennig}}, \bibinfo {author} {\bibfnamefont
  {K.}~\bibnamefont {Sneskov}}, \bibinfo {author} {\bibfnamefont {A.~H.}\
  \bibnamefont {Steindal}}, \bibinfo {author} {\bibfnamefont {K.~O.}\
  \bibnamefont {Sylvester-Hvid}}, \bibinfo {author} {\bibfnamefont {P.~R.}\
  \bibnamefont {Taylor}}, \bibinfo {author} {\bibfnamefont {A.~M.}\
  \bibnamefont {Teale}}, \bibinfo {author} {\bibfnamefont {E.~I.}\ \bibnamefont
  {Tellgren}}, \bibinfo {author} {\bibfnamefont {D.~P.}\ \bibnamefont {Tew}},
  \bibinfo {author} {\bibfnamefont {A.~J.}\ \bibnamefont {Thorvaldsen}},
  \bibinfo {author} {\bibfnamefont {L.}~\bibnamefont {Thøgersen}}, \bibinfo
  {author} {\bibfnamefont {O.}~\bibnamefont {Vahtras}}, \bibinfo {author}
  {\bibfnamefont {M.~A.}\ \bibnamefont {Watson}}, \bibinfo {author}
  {\bibfnamefont {D.~J.~D.}\ \bibnamefont {Wilson}}, \bibinfo {author}
  {\bibfnamefont {M.}~\bibnamefont {Ziolkowski}}, \ and\ \bibinfo {author}
  {\bibfnamefont {H.}~\bibnamefont {Ågren}},\ }\href {\doibase
  10.1002/wcms.1172} {\bibfield  {journal} {\bibinfo  {journal} {Wiley
  Interdiscip. Rev.: Comput. Mol. Sci.}\ }\textbf {\bibinfo {volume} {4}},\
  \bibinfo {pages} {269} (\bibinfo {year} {2014})}\BibitemShut {NoStop}%
\bibitem [{\citenamefont {Dunning}(1971)}]{Dunning:jcp/55/716}%
  \BibitemOpen
  \bibfield  {author} {\bibinfo {author} {\bibfnamefont {T.~H.}\ \bibnamefont
  {Dunning}},\ }\href@noop {} {\bibfield  {journal} {\bibinfo  {journal} {J.
  Chem. Phys.}\ }\textbf {\bibinfo {volume} {55}},\ \bibinfo {pages} {716}
  (\bibinfo {year} {1971})}\BibitemShut {NoStop}%
\bibitem [{\citenamefont {Hachmann}\ \emph {et~al.}(2006)\citenamefont
  {Hachmann}, \citenamefont {Cardoen},\ and\ \citenamefont
  {Chan}}]{Hachmann:jcp2006-h50dmrg}%
  \BibitemOpen
  \bibfield  {author} {\bibinfo {author} {\bibfnamefont {J.}~\bibnamefont
  {Hachmann}}, \bibinfo {author} {\bibfnamefont {W.}~\bibnamefont {Cardoen}}, \
  and\ \bibinfo {author} {\bibfnamefont {G.~K.-L.}\ \bibnamefont {Chan}},\
  }\href@noop {} {\bibfield  {journal} {\bibinfo  {journal} {J. Chem. Phys.}\
  }\textbf {\bibinfo {volume} {125}},\ \bibinfo {pages} {144101} (\bibinfo
  {year} {2006})}\BibitemShut {NoStop}%
\bibitem [{\citenamefont {Boguslawski}\ \emph {et~al.}(2014)\citenamefont
  {Boguslawski}, \citenamefont {Tecmer}, \citenamefont {Ayers}, \citenamefont
  {Bultinck}, \citenamefont {De~Baerdemacker},\ and\ \citenamefont
  {Van~Neck}}]{Boguslawski2014}%
  \BibitemOpen
  \bibfield  {author} {\bibinfo {author} {\bibfnamefont {K.}~\bibnamefont
  {Boguslawski}}, \bibinfo {author} {\bibfnamefont {P.}~\bibnamefont {Tecmer}},
  \bibinfo {author} {\bibfnamefont {P.~W.}\ \bibnamefont {Ayers}}, \bibinfo
  {author} {\bibfnamefont {P.}~\bibnamefont {Bultinck}}, \bibinfo {author}
  {\bibfnamefont {S.}~\bibnamefont {De~Baerdemacker}}, \ and\ \bibinfo {author}
  {\bibfnamefont {D.}~\bibnamefont {Van~Neck}},\ }\href {\doibase
  10.1103/PhysRevB.89.201106} {\bibfield  {journal} {\bibinfo  {journal} {Phys.
  Rev. B}\ }\textbf {\bibinfo {volume} {89}},\ \bibinfo {pages} {201106}
  (\bibinfo {year} {2014})}\BibitemShut {NoStop}%
\bibitem [{\citenamefont {Tsuchimochi}\ and\ \citenamefont
  {Scuseria}(2009)}]{Tsuchimochi:jcp2009-cpmft}%
  \BibitemOpen
  \bibfield  {author} {\bibinfo {author} {\bibfnamefont {T.}~\bibnamefont
  {Tsuchimochi}}\ and\ \bibinfo {author} {\bibfnamefont {G.~E.}\ \bibnamefont
  {Scuseria}},\ }\href@noop {} {\bibfield  {journal} {\bibinfo  {journal} {J.
  Chem. Phys.}\ }\textbf {\bibinfo {volume} {131}},\ \bibinfo {pages} {121102}
  (\bibinfo {year} {2009})}\BibitemShut {NoStop}%
\bibitem [{\citenamefont {Dunning}(1989)}]{dunning_ccpvdz}%
  \BibitemOpen
  \bibfield  {author} {\bibinfo {author} {\bibfnamefont {T.~H.}\ \bibnamefont
  {Dunning}},\ }\href {\doibase http://dx.doi.org/10.1063/1.456153} {\bibfield
  {journal} {\bibinfo  {journal} {J. Chem. Phys.}\ }\textbf {\bibinfo {volume}
  {90}},\ \bibinfo {pages} {1007} (\bibinfo {year} {1989})}\BibitemShut
  {NoStop}%
\bibitem [{\citenamefont {Ray}\ and\ \citenamefont
  {Chan}(2016)}]{h10ccpvdz_dmrg}%
  \BibitemOpen
  \bibfield  {author} {\bibinfo {author} {\bibfnamefont {U.}~\bibnamefont
  {Ray}}\ and\ \bibinfo {author} {\bibfnamefont {G.~K.-L.}\ \bibnamefont
  {Chan}},\ }\href@noop {} {\  (\bibinfo {year} {2016})},\ \bibinfo {note}
  {private communication}\BibitemShut {NoStop}%
\bibitem [{\citenamefont {Sharma}\ and\ \citenamefont
  {Chan}(2012)}]{dmrg_block_2012}%
  \BibitemOpen
  \bibfield  {author} {\bibinfo {author} {\bibfnamefont {S.}~\bibnamefont
  {Sharma}}\ and\ \bibinfo {author} {\bibfnamefont {G.~K.-L.}\ \bibnamefont
  {Chan}},\ }\href {\doibase http://dx.doi.org/10.1063/1.3695642} {\bibfield
  {journal} {\bibinfo  {journal} {J. Chem. Phys.}\ }\textbf {\bibinfo {volume}
  {136}},\ \bibinfo {pages} {124121} (\bibinfo {year} {2012})}\BibitemShut
  {NoStop}%
\bibitem [{\citenamefont {Olivares-Amaya}\ \emph {et~al.}(2015)\citenamefont
  {Olivares-Amaya}, \citenamefont {Hu}, \citenamefont {Nakatani}, \citenamefont
  {Sharma}, \citenamefont {Yang},\ and\ \citenamefont
  {Chan}}]{dmrg_block_2015}%
  \BibitemOpen
  \bibfield  {author} {\bibinfo {author} {\bibfnamefont {R.}~\bibnamefont
  {Olivares-Amaya}}, \bibinfo {author} {\bibfnamefont {W.}~\bibnamefont {Hu}},
  \bibinfo {author} {\bibfnamefont {N.}~\bibnamefont {Nakatani}}, \bibinfo
  {author} {\bibfnamefont {S.}~\bibnamefont {Sharma}}, \bibinfo {author}
  {\bibfnamefont {J.}~\bibnamefont {Yang}}, \ and\ \bibinfo {author}
  {\bibfnamefont {G.~K.-L.}\ \bibnamefont {Chan}},\ }\href {\doibase
  http://dx.doi.org/10.1063/1.4905329} {\bibfield  {journal} {\bibinfo
  {journal} {J. Chem. Phys.}\ }\textbf {\bibinfo {volume} {142}},\ \bibinfo
  {pages} {034102} (\bibinfo {year} {2015})}\BibitemShut {NoStop}%
\end{thebibliography}

%

\end{document}